\documentclass[twocolumn,showpacs,preprintnumbers,amsmath,amssymb,floatfix,nofootinbib]{revtex4}
\usepackage{amsmath,amsfonts,bbm,euscript,hyperref}
\usepackage{graphicx}
\usepackage{dcolumn}
\usepackage{bm}
\usepackage{epsfig}
\epsfclipon

\newcommand{\nco}{\newcommand}
\nco{\beq}{\begin{equation}} \nco{\eeq}{\end{equation}}
\nco{\beqa}{\begin{eqnarray}} \nco{\eeqa}{\end{eqnarray}}
\def\be{\begin{equation}}
\def\ee{\end{equation}}    
\def\baray{\begin{eqnarray}}
\def\earay{\end{eqnarray}}

\nco{\lra}{\leftrightarrow}

\nco{\sss}{\scriptscriptstyle} \nco{\dphi}{\varphi}
\nco{\lsim}{\mbox{\raisebox{-.6ex}{~$\stackrel{<}{\sim}$~}}}
\nco{\gsim}{\mbox{\raisebox{-.6ex}{~$\stackrel{>}{\sim}$~}}}

\def\IK{\relax{\rm I\kern-.20em K}}
\def\IM{\relax{\rm I\kern-.20em M}}

\def\lsim{\mbox{\raisebox{-.6ex}{~$\stackrel{<}{\sim}$~}}}
\def\gsim{\mbox{\raisebox{-.6ex}{~$\stackrel{>}{\sim}$~}}}
\def\sss{\scriptscriptstyle}

\def\sH{\mathcal{H}}

\def\grad{\vec{\nabla}}

\begin{document}


\title{On Features and Nongaussianity from Inflationary Particle Production}

\author{Neil Barnaby}

\affiliation{%
\centerline{Canadian Institute for Theoretical Astrophysics,}
\centerline{University of Toronto, McLennan Physical Laboratories,
60 St.\ George Street, Toronto, Ontario, Canada  M5S 3H8}
e-mail:\ barnaby@cita.utoronto.ca}

\date{June 2010}

\begin{abstract} 
Interactions between the inflaton and any additional fields can lead to isolated bursts of particle production during inflation (for example from parametric 
resonance or a phase transition).  Inflationary particle production leaves localized features in the spectrum and bispectrum of the observable 
cosmological fluctuations, via the Infra-Red (IR) cascading mechanism.  We focus on a simple prototype interaction $g^2 (\phi-\phi_0)^2\chi^2$ between the 
inflaton, $\phi$, and iso-inflaton, $\chi$; extending previous work on this model in two directions.  First, we quantify the magnitude of the produced 
nongaussianity by extracting the moments of the probability distribution function from lattice field theory simulations.  We argue that the bispectrum feature 
from particle production might be observable for reasonable values of the coupling, $g^2$.  Second, we develop a detailed analytical theory of particle 
production and IR cascading during inflation, which is in excellent agreement with numerical simulations.  Our formalism improves significantly on 
previous approaches by consistently incorporating both the expansion of the universe and also metric perturbations.  We use this new formalism to 
estimate the shape of the bispectrum from particle production, showing this to be distinguishable from other mechanisms that predict large nongaussianity.
\end{abstract}


\maketitle


\section{Introduction} 

The inflationary paradigm has become a cornerstone of modern cosmology.  As measurements of the Cosmic Microwave Background (CMB) 
radiation grow increasingly precise, it has become topical to look beyond the simplest single-field, slow-roll inflationary scenario.  In particular, it
is interesting to determine the extent to which non-minimal signatures, such as features in the primordial power spectrum or observable nongaussianities,
can be accommodated by microscopically sensible inflation models.  Efforts in this direction are valuable because they allow us to test our theoretical 
prejudices and provide observers with well-motivated templates for departures from the standard scenario.  Finally, a detection of some non-minimal
features might open a rare observational window into fundamental particle physics at extremely high energy scales.  In this work, we will consider a very 
simple and well-motivated class of models which predict novel observable signatures in the spectrum and bispectrum of the primordial curvature 
fluctuations.

In a variety of inflation models, the motion of the inflaton can trigger the production of some non-inflaton (iso-curvature) particle \emph{during}
inflation.  Models of this type have attracted considerable interest recently; examples have been studied where particle production occurs via 
parametric resonance \cite{ir,chung,chung2,elgaroy,sasaki,modulated_trapping,brane_brem,trapped,ppcons}, as a result of a phase transition 
\cite{preheatNG,KL,KP,BBS,adams,step_model,gobump,brane_annihilation}, or otherwise \cite{sorbo}.
Such constructions are novel for a variety of reasons:
\begin{enumerate}
  \item The produced iso-inflaton particles may rescatter off the slow-roll condensate and generate a significant contribution
           to the primordial curvature fluctuations through the process of Infra-Red (IR) cascading \cite{ir}.  This provides a new mechanism for 
           generating
           cosmological perturbations that is \emph{qualitatively} different from the standard mechanism \cite{fluctuations}, the curvaton
           \cite{curvaton0,curvaton} or modulated fluctuations \cite{modulated,modulated2}.
  \item Particle production and IR cascading leads to a variety of novel observational signatures, including features in the primordial power spectrum
           and also nongaussianities \cite{ir,ppcons}.  
  \item Particle production arises naturally in a number of microscopically realistic models of inflation, including
           examples from string theory \cite{trapped} 
           and supersymmetric (SUSY) field theory \cite{berrera}.  In particular, inflationary particle production
           is a generic feature of open string inflation models \cite{ppcons}, such as brane/axion monodromy \cite{monodromy1,monodromy2,monodromy3,monodromy4}.
       (See also \cite{amjad}.)
  \item Observable features in the primordial power spectrum, generated by particle 
           production and IR cascading, offer a novel example of the non-decoupling of high scale physics in the Cosmic Microwave Background (CMB)
            \cite{chung,jim}.  In the most interesting examples, the produced 
            particles are
            extremely massive for (almost) the entire history of the universe, however, their effect cannot be integrated out due to the non-adiabatic
            time dependence of the iso-inflaton 
            mode functions during particle production.  In \cite{chung} particle production during large field inflation was
            proposed as a possible probe of Planck-scale physics.
 \item The energetic cost of producing particles during inflation has a dissipative effect on the dynamics of the inflaton.
          Particle production may therefore slow
          the motion of the inflaton, even on a steep potential.  This gives rise to a 
          new inflationary mechanism, called \emph{trapped inflation} \cite{beauty,trapped}, 
          which may circumvent
          some of the fine tuning problems associated with standard slow-roll inflation.  See \cite{trapped} for an explicit string theory realization of 
          trapped inflation and \cite{terminal} for a generalization to higher dimensional moduli spaces and enhanced symmetry loci.
          The idea of using
          dissipative dynamics to slow the motion of the inflaton was exploited also for a very interesting mechanism (which pre-dates trapped inflation) 
          called \emph{warm inflation} \cite{warm,warm2,warm3,berrera}.  See also the variant of natural inflation \cite{natural} that was 
          proposed recently by Anber \& Sorbo \cite{sorbo}.
\end{enumerate}

In this article we study the impact of isolated bursts of inflationary particle production on the observable primordial curvature perturbations.
In order to illustrate the basic physics, we focus on a very simple and general prototype model where the inflaton, $\phi$,
and iso-inflaton, $\chi$, fields interact via the coupling
\begin{equation}
\label{int}
  \mathcal{L}_{\mathrm{int}} = -\frac{g^2}{2} (\phi-\phi_0)^2 \chi^2
\end{equation}
On physical grounds, we expect that our results will generalize in a straightforward way to more complicated models,
such as fermion iso-inflaton fields, gauged interactions and (perhaps) inflationary phase transitions.

Scalar field interactions of the type (\ref{int}) have also been studied recently in connection with non-equilibrium Quantum Field
Theory (QFT) \cite{nonequilibrium}, in particular with applications to the theory of preheating after inflation \cite{KLS,KLS97,FK,MT1,MT2,dmitry}
and also moduli trapping \cite{beauty,terminal} at enhanced symmetry points.  Although our focus is on particle production \emph{during} inflation (as 
opposed to during preheating, after inflation) some of our results nevertheless have implications for preheating, moduli trapping and also non-equilibrium 
QFT more generally.  For example, in \cite{ir} analytical and numerical studies of rescattering and IR cascading during inflation made it possible to 
observe, for the first time, the dynamical approach to the turbulent scaling regime that was discovered in \cite{B1,B2}.

Let us now discuss briefly the physics of the model (\ref{int}).  At the moment when $\phi=\phi_0$ (which we assume occurs during the observable range of $e$-foldings of inflation) the $\chi$ 
particles become instantaneously massless and are produced by quantum effects.  This burst of particle production drains energy from the condensate
$\phi(t)$, temporarily slowing the motion of the inflaton background and violating slow roll.  Shortly after this moment the $\chi$ particles become 
extremely non-relativistic, so that their number density dilutes as $a^{-3}$, and eventually the inflaton settles back onto the slow roll trajectory.

\begin{figure}[htbp]
\bigskip \centerline{\epsfxsize=0.4\textwidth\epsfbox{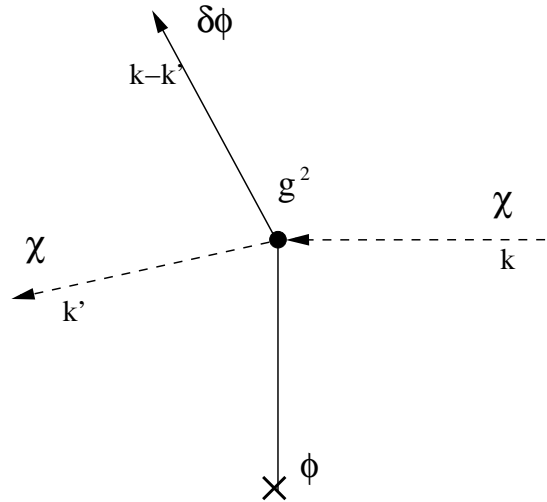}}
\caption{rescattering diagram.
}
\label{Fig:diag}
\end{figure}

The dominant effect of particle production on the observable spectrum of curvature fluctuations arises because the produced, massive $\chi$ particles
can rescatter off the condensate to generate bremsstrahlung radiation of long-wavelength $\delta\phi$ fluctuations via diagrams such as 
Fig.~\ref{Fig:diag}.  Multiple such rescatterings lead to a rapid cascade of power into the IR.  The inflaton modes generated by this IR 
cascading freeze once their wavelength crosses the horizon and lead to a bump-like feature in the primordial power spectrum.   This bump-like
feature is accompanied by a localized, uncorrelated nongaussian feature in the bispectrum \cite{ir}. 

In this paper we extend previous work \cite{ir,ppcons} on the model (\ref{int}).  First, we re-visit the problem of quantifying the magnitude of the produced
nongaussianity.  Using lattice field theory simulations we compute numerically the skewness and kurtosis of the Probability Distribution Function (PDF) of the 
primordial curvature fluctuations.  By comparison to the more familiar local model of nongaussianity, we argue that the bispectrum associated with this mechanism 
may be observable in future missions.  

Next, we provide a detailed analytical theory of the quantum production of $\chi$ particles and the subsequent rescattering off the slow-roll condensate
for the model (\ref{int}).  This new formalism improves significantly upon previous efforts \cite{ir} by consistently incorporating both the expansion of the
universe and also metric perturbations.  We test our approach by comparison to fully nonlinear lattice field theory simulations, finding excellent 
agreement.  We also use our formalism to estimate the shape of the bispectrum.

The outline of this paper is as follows.  In section \ref{sec:overview} we review the key results of \cite{ir,ppcons}, describing heuristically the underlying 
mechanism of IR cascading and the resultant observational signatures.  In section \ref{sec:NG} we characterize the size of the nongaussianity associated
with particle production and IR cascading, relying primarily on lattice field theory simulations.  In section \ref{sec:theory} we provide an analytical theory
of inflationary particle production and IR cascading in the model (\ref{int}), neglecting metric perturbations.  Using this new formalism 
we estimate the shape of the bispectrum in the model (\ref{int}).  In section \ref{sec:perts} we reconsider our analytical approach,
showing how metric perturbations can be consistently incorporated and, further, we demonstrate explicitly that their inclusion does not 
significantly alter the results of section \ref{sec:theory}.  Finally, in section \ref{sec:conclusions}, we conclude.

\section{Overview of the Mechanism}
\label{sec:overview}

In this section we provide a brief overview of the dynamics of particle production and IR cascading in the model (\ref{int}) and we also
summarize the key observational signatures.  This section is largely review of \cite{ir,ppcons}, the reader already familiar with those
works may wish to skip ahead to the next section.

We consider the following model
\begin{eqnarray}
  S &=& \int d^4 x \sqrt{-g} \left[ \frac{M_p^2}{2} R - \frac{1}{2}(\partial \phi)^2 - V(\phi) \right. \nonumber \\
      &&\,\,\,\,\,\,\,\,\,\,\,\,\,   \left. - \frac{1}{2}(\partial \chi)^2 - \frac{g^2}{2}(\phi-\phi_0)^2\chi^2 \right]  \label{L}
\end{eqnarray}
where $R$ is the Ricci curvature constructed from the metric $g_{\mu\nu}$, $\phi$ is the inflaton field and $\chi$ is the iso-inflaton.  
As usual, we assume a flat FRW space-time with scale factor $a(t)$
\begin{equation}
  ds^2 \equiv g_{\mu\nu} dx^{\mu} dx^{\nu} = -dt^2 + a^2(t) d{\bf x}^2
\end{equation}
and employ the reduced Planck mass $M_p \equiv (8\pi G_N)^{-1/2} \cong 2.43\times 10^{18} \mathrm{GeV}$.  
We leave the potential $V(\phi)$ driving inflation unspecified except to assume that it is sufficiently flat in 
the usual sense; that is $\epsilon \ll 1$, $|\eta| \ll 1$ where
\begin{equation}
\label{slow_roll}
  \epsilon \equiv \frac{M_p^2}{2} \left(\frac{V'}{V}\right)^2, \hspace{5mm}
  \eta \equiv M_p^2 \frac{V''}{V}
\end{equation}
are the usual slow roll parameters.  

The coupling $\frac{g^2}{2}(\phi-\phi_0)^2\chi^2$ in (\ref{L}) is introduced to ensure that
the iso-inflaton field can become instantaneously massless at some point $\phi=\phi_0$  along the inflaton trajectory
(which we assume occurs during the observable range of $e$-foldings of inflation).
At this moment $\chi$ particles will be produced by quantum effects.  (In section \ref{sec:theory} we will discuss how particle
production and rescattering are modified by the inclusion of a mass term $\mu^2\chi^2$ for the iso-inflaton.)

\subsection{Quantum Production of $\chi$ Particles}
\label{subsec:prodn}

Let us first consider the homogeneous dynamics of the inflaton field, $\phi(t)$.  Near the point $\phi = \phi_0$ we can generically expand
\begin{equation}
  \phi(t) \cong \phi_0 + v t
\end{equation}
where $v \equiv \dot{\phi}(0)$ and we have arbitrarily set the origin of time so that $t=0$ corresponds to the
moment when $\phi = \phi_0$.  The interaction (\ref{int}) induces an effective (time varying) mass for the $\chi$ 
particles of the form
\begin{equation}
\label{m_approx}
  m_\chi^2 = g^2 (\phi - \phi_0)^2 \cong k_\star^4 t^2
\end{equation}
where we have defined the characteristic scale
\begin{equation}
\label{kstar}
  k_\star = \sqrt{g |v|}
\end{equation}
It is straightforward to verify that the simple expression (\ref{m_approx}) will be a good approximation 
for $(H |t|)^{-1} \lsim \mathcal{O}(\epsilon,\eta)$ which, in most models, will be true for the entire observable $60$ $e$-foldings of 
inflation.

Note that, without needing to specify the background inflationary potential $V(\phi)$, we can write the ratio $k_\star / H$ as
\begin{equation}
\label{ratio}
  \frac{k_\star}{H} = \sqrt{\frac{g}{2\pi \mathcal{P}_\zeta^{1/2}}}
\end{equation}
where $\mathcal{P}_{\zeta}^{1/2} = 5\times 10^{-5}$ is the usual amplitude of the vacuum fluctuations from inflation.  
In this work we assume $k_\star > H$ which is easily satisfied for reasonable values of the coupling $g^2  > 10^{-7}$.
In particular, for $g^2 \sim 0.1$ we have $k_\star / H \sim 30$.

The scenario we have in mind is the following.  Inflation starts at some field value $\phi > \phi_0$ and the inflaton rolls toward
the point $\phi = \phi_0$.  Initially, the iso-inflaton field is extremely massive $m_\chi \gg H$
and hence it stays pinned in the vacuum, $\chi = 0$, and does not contribute to super-horizon curvature fluctuations.  
Eventually, at $t=0$, the inflaton rolls through the point $\phi=\phi_0$ where $m_\chi = 0$
and $\chi$ particles are produced.  To describe this burst of particle production one must solve 
the following equation for the $\chi$-particle mode functions in an expanding universe
\begin{equation}
\label{intro_chi}
  \ddot{\chi}_k + 3 H \dot{\chi}_k + \left[ \frac{k^2}{a^2} + k_\star^4 t^2  \right] \chi_k = 0
\end{equation}
Equations of this type are well-studied in the context of preheating after inflation \cite{KLS97} and moduli trapping \cite{beauty}.
In the regime $k_\star > H$ particle production is fast compared to the expansion time\footnote{In the opposite regime, $k_\star \ll H$, the field $\chi$
will  be light as compared to the Hubble scale for a significant portion of inflation.  In this case it is no longer consistent to treat the background dynamics
as being effectively single-field, hence the scenario has changed considerably.  We will not consider this possibility any further.} and one can solve 
(\ref{intro_chi}) very accurately for the occupation number of the created $\chi$ particles
\begin{equation}
\label{n_k}
  n_k = e^{-\pi k^2 / k_\star^2} 
\end{equation}
Very quickly after the moment $t=0$, within a time $\Delta t \sim k_\star^{-1} \ll H^{-1}$,
these produced $\chi$ particles become non-relativistic ($m_\chi > H$) and their number density starts to dilute as $a^{-3}$.

Following the initial burst of particle production there are two distinct physical effects which take place.  First, the energetic cost of producing the gas of
massive out-of-equilibrium $\chi$ particles drains energy from the inflaton condensate, forcing $\dot{\phi}$ to drop abruptly.  This velocity dip
is the result of the backreaction of the produced $\chi$ fluctuations on homogeneous condensate $\phi(t)$.  The second physical effect is
that the produced massive $\chi$ particles rescatter off the condensate via the diagram Fig.~\ref{Fig:diag} and emit  bremsstrahlung
radiation of light inflaton fluctuations (particles).  Backreaction and rescattering leave distinct imprints in the observable cosmological perturbations.
Let us discuss each separately.

\subsection{Backreaction Effects}
\label{subsec:backreaction}

We first consider the impact of backreaction.  This effect can be studied analytically using the mean field equation
\begin{equation}
\label{mean}
  \ddot{\phi} + 3 H \dot{\phi} + V_{,\phi} + g^2 (\phi - \phi_0) \langle \chi^2 \rangle  = 0
\end{equation}
where the vacuum average is computed following \cite{KLS97,beauty}
\begin{equation}
\label{mean2}
  \langle \chi^2 \rangle \cong \Theta(t) \frac{n_\chi a^{-3}}{ g |\phi - \phi_0|}
\end{equation}
and $n_\chi = \int \frac{d^3 k}{(2\pi)^3} n_k \sim k_\star^3$ is the total number density of produced $\chi$
particles.  The Heaviside function $\Theta(t)$ in (\ref{mean2}) enforces the fact the the backreaction effects become
important only for $t>0$, \emph{after} the $\chi$ particles have been produced.  The factor of $a^{-3}$ in (\ref{mean2})
reflects the usual volume dilution of non-relativistic matter.  

The solutions of (\ref{mean}) display the expected behaviour: the energetic cost of the production of $\chi$ particles at $t=0$
leads to an abrupt dip in the velocity $\dot{\phi}$, momentarily violating the smallness of the slow roll parameter $\ddot{\phi}/(H \dot{\phi})$.  
Within a few $e$-foldings of the moments $t=0$, the produced $\chi$ particles have become
extremely massive and have been diluted away by the inflationary expansion of the universe.  At this time, the inflaton must settle back onto
the slow roll trajectory, $\dot{\phi} \cong -V' / (3 H)$.

Backreaction effects lead to a transient violation of slow roll, and hence we expect an associated ``ringing'' pattern (damped oscillations) in the 
primordial curvature fluctuations, similar to models with a sharp feature in the potential 
\cite{adams,step_model,gobump,star1,star2,chen1,chen2,space_break,jain}.  This effect can be seen by solving the well-known
equation for the curvature perturbation on co-moving hypersurfaces, $\mathcal{R}$, in linear theory:
\begin{equation}
\label{R}
  \mathcal{R}''_k + 2 \frac{z'}{z} \mathcal{R}'_k + k^2 \mathcal{R}_k = 0
\end{equation}
Here the prime denotes derivatives with respect to conformal time $\tau = \int^t a^{-1}(t')dt'$ and $z\equiv a \dot{\phi} / H$.  Note that (\ref{R}) is valid only in the 
absence of entropy perturbations.  However, in our case the $\chi$ field is extremely massive, $m_\chi^2\gg H^2$, for nearly the entire duration of
inflation, hence the direct iso-curvature contribution to $\mathcal{R}$ is negligible.

In \cite{ir} the coupled system (\ref{mean},\ref{R}) was solved numerically and the expected ringing pattern in the power spectrum 
$P_{\mathcal{R}}(k) = \frac{k^3}{2\pi^2}|\mathcal{R}_k|^2$ was obtained. (See also \cite{sasaki}.)  This effect is sub-dominant to the rescattering processes 
described in the next subsection, hence we will not pursue backreaction any further in this work.

\subsection{Rescattering Effects}
\label{subsec:resc}

The second physical effect which takes place after the quantum production of $\chi$ particles in the model (\ref{int}) is rescattering.
This effect was considered for the first time in the context of inflationary particle production in \cite{ir}.  Fig.~\ref{Fig:diag} illustrates the dominant process: 
bremsstrahlung emission of long-wavelength $\delta\phi$ fluctuations from rescattering of the produced $\chi$ particles off the condensate $\phi(t)$.  The time 
scale for such processes is set by the microscopic scale, $k_\star^{-1}$, and is thus very short compared to the expansion time, $H^{-1}$.  Moreover, the 
production of inflaton fluctuations $\delta\phi$ deep in the IR is extremely energetically
inexpensive, since the inflaton is very nearly massless.  The combination of the short time scale for rescattering and the energetic cheapness
of radiating  IR $\delta\phi$ leads to a rapid build-up of power in long wavelength inflaton modes: IR cascading.  This effect leads to a bump-like
feature in the power spectrum of inflaton fluctuations, very different from the ringing pattern associated with backreaction.  The bump-like
feature from rescattering dominates over the ringing pattern from backreaction for all values of parameters.

In \cite{ir} the model (\ref{L}) was studied using lattice field theory simulations, without neglecting any physical processes (that is
to say that full nonlinear structure of the theory, including backreaction and rescattering effects, was accounted for consistently).
However, this same dynamics can be understood analytically by solving the equation for the inflaton fluctuations
$\delta\phi$ in the approximation that all interactions are neglected, except for the diagram Fig.~\ref{Fig:diag}.  The appropriate equation
is
\begin{equation}
\label{inf_eqn}
  \delta\ddot{\phi} + 3 H \delta\dot{\phi} - \frac{\grad^2}{a^2}\delta\phi + V_{,\phi\phi}\delta \phi \cong -g^2\left[\phi(t)-\phi_0\right]\chi^2
\end{equation}
The solution of (\ref{inf_eqn}) may be split into two parts: the solution of the homogeneous equation and the particular solution
which is due to the source term.  Schematically we have
\begin{equation}
\label{hom+par}
  \delta\phi(t,{\bf x}) = \underbrace{\delta\phi_{\mathrm{vac}}(t,{\bf x})}_{\mathrm{homogeneous}} + \underbrace{\delta\phi_{\mathrm{resc}}(t,{\bf x})}_{\mathrm{particular}}
\end{equation}
The former contribution is the homogeneous solution which behaves as $\delta\phi_{\mathrm{vac}} \sim H / (2\pi)$ on large scales and, physically, corresponds
to the usual scale invariant vacuum fluctuations from inflation.  The particular solution, $\delta\phi_{\mathrm{resc}}$, corresponds physically to inflaton
fluctuations which are generated by rescattering.  The abrupt growth of $\chi$ inhomogeneities at $t=0$ sources the particular solution $\delta\phi_{\mathrm{resc}}$,
leading to the production of inflation fluctuations which subsequently cross the horizon and become frozen. 

A detailed analytical theory of equation (\ref{inf_eqn}) will be the subject of sections \ref{sec:theory} and \ref{sec:perts}.  Here we simply point out
that the primordial power spectrum in the model (\ref{int}) may, to good approximation, be described by a simple semi-analytic fitting function \cite{ppcons}
\begin{eqnarray}
  P_{\mathcal{R}}(k) &=& A_s \left(\frac{k}{k_0}\right)^{n_s-1} \nonumber \\
         &+& A_{\mathrm{IR}} \left(\frac{\pi e}{3}\right)^{3/2}\left(\frac{k}{k_{\mathrm{IR}}}\right)^3 e^{-\frac{\pi}{2} \left(\frac{k}{k_{\mathrm{IR}}}\right)^2}
\label{P_fit}
\end{eqnarray}
where the first term corresponds to the usual vacuum fluctuations from inflation (with amplitude $A_s$ and spectral index $n_s$) 
while the second term corresponds to the bump-like feature from particle production and IR cascading.  The amplitude of this feature ($A_{\mathrm{IR}}$) 
depends on $g^2$, while the location ($k_{\mathrm{IR}}$) depends on $\phi_0$.

In \cite{ppcons} the simple fitting function (\ref{P_fit}) was used to place observational constraints on inflationary particle production using a
variety of cosmological data sets.  Current data are consistent with rather large spectral distortions of the type (\ref{P_fit}).
Features as large as $A_{\mathrm{IR}} / A_s \sim 0.1$ are allowed in the case that $k_{\mathrm{IR}}$ falls within the range of scales relevant for CMB experiments.  
A feature of this magnitude corresponds to a realistic coupling $g^2 \sim 0.01$.  Even larger values of $g^2$ are allowed if the feature is localized on smaller scales.
In \cite{forecast} Large Scale Structure forecast constraints were considered for the model (\ref{P_fit}).  It was shown that, for $k_{\mathrm{IR}} \lsim 0.1\, \mathrm{Mpc}^{-1}$,
the constraint on $A_{\mathrm{IR}} / A_s$ will be stengthened to the $0.5\%$ level by Planck or $0.1\%$ including also data from a Square Kilometer Array (SKA).  
With a Cosmic Inflation Probe (CIP) similar constraints could be achieved for $k_{\mathrm{IR}}$ as large as $1\, \mathrm{Mpc}^{-1}$.

\subsection{Nongaussianity from Particle Production}
\label{subsec:NG}

The bump-like feature in $P(k)$, corresponding to the second term in (\ref{P_fit}), must be associated with a nongaussian feature in the 
bispectrum \cite{ir}.  Indeed, it is evident already from inspection of equation (\ref{inf_eqn}) that the inflaton fluctuations generated by rescattering are 
significantly nongaussian; the particular solution of (\ref{inf_eqn}) is bi-linear in the gaussian field $\chi$.  

Nongaussian statistics have attracted a considerable amount of interest recently, owing to their potential as a tool for observationally
discriminating between the plethora of inflationary models in the literature.  Although the simplest single-field slow roll models
are known to produce negligible (primordial) nongaussianity \cite{riotto,maldacena,seerylidsey}, there are a currently a number of alternative models 
which \emph{may} predict an observable signature.  Examples include models with preheating into light fields 
\cite{preheatNG,preheatNG2,preheatNG3,preheatNG4}, nonlocal inflation \cite{NLNG}, the curvaton 
mechanism \cite{curvatonNG}, multi-field models \cite{turnNG}, constructions with a small sound speed \cite{small_sound} 
(such as DBI \cite{DBI} inflation), trapped inflation \cite{trapped}, the gelaton \cite{gelaton},
models with features or rapid oscillations in the inflaton potential \cite{chen1,chen2}, 
non-vacuum initial conditions \cite{small_sound,nonBD1,nonBD2,nonBD3}, warm inflation \cite{warm_NG}, etc.

Nongaussianity is usually characterized in terms of the \emph{bispectrum}, $B_\zeta(k_i)$, which is the 3-point correlation function of the Fourier 
transform of the primordial curvature fluctuation on uniform density hypersurfaces, $\zeta$.  Explicitly, we define
\begin{equation}
\label{B_zeta}
  \langle \zeta_{\bf k_1} \zeta_{\bf k_2} \zeta_{\bf k_3} \rangle = (2\pi)^3 \delta^{(3)}({\bf k_1}+{\bf k_2}+{\bf k_3}) B_\zeta(k_i)
\end{equation} 
where $k_i \equiv |{\bf k_i}|$ and $\zeta_k$ is related to the variable $\mathcal{R}_k$ appearing in (\ref{R}) as 
$\zeta_k \cong -\mathcal{R}_k$ on large scales $k \ll a H$.  The delta function in (\ref{B_zeta}) reflects translational invariance
and ensures that $B_\zeta(k_i)$ depends on three wavenumbers which form a triangle: ${\bf k_1}+{\bf k_2}+{\bf k_3}=0$.  A general 
bispectrum $B_\zeta(k_i)$ may be characterized by specifying its size (amplitude of $B_\zeta$), shape (whether $B_\zeta$ 
peaks on squeezed, equilateral or flattened triangles) and running (the dependence of $B_\zeta$ on the size of the triangle).
The various nongaussian scenarios discussed above may be classified according to the size, shape and running of the bispectrum,
see \cite{ng_rev} for a more detailed review.

The nongaussian signature from IR cascading  is very different from other models, such as the
local, equilateral or enfolded shapes, which have been studied in the literature.  IR cascading only influences modes leaving the horizon near the moment 
$\phi=\phi_0$, when particle production 
occurs, hence we expect the bispectrum to be very far from scale invariant (this is also true for the model considered in \cite{chen1,chen2}).  
The dominant contribution to $B_\zeta(k_i)$ should peak strongly for triangles with a characteristic size $\sim k_{\mathrm{IR}}$, corresponding to the 
location of the bump in the power spectrum (\ref{P_fit}).  
We will estimate the shape of the bispectrum from particle production and IR cascading in more detail in section \ref{sec:theory} and re-visit
this issue also in an upcoming publication \cite{inprog}.

The unusual shape and strong scaling properties of the bispectrum from particle production makes it difficult to compare the 
magnitude of nongaussianity in this model to more familiar bispectra, such as the local shape, which are very close to scale invariant.  We find it
useful to quantify the magnitude of the nongaussianity in the model (\ref{int}) by computing the moments of the Probability Distribution Function (PDF),
$P(\zeta)$, which is the probability that the curvature perturbation has a fluctuations of size $\zeta$.  These moments carry information
about the correlation functions of $\zeta$ integrated over all wavenumbers $k_i$ and therefore provide a useful tool to compare models
with very different shape/running properties \cite{shandera}.  (See also \cite{shellard2} for a related discussion and alternative methodology.)

Let us define the central moments of the PDF as
\begin{equation}
  \langle \zeta^n \rangle = \int \zeta^n P(\zeta) d\zeta
\end{equation}
The $n$-th cummulant $\kappa_n$ is the connected $n$-point function.  For $\langle \zeta \rangle = 0$ the first few non-vanishing cummulants are:
\begin{eqnarray}
  \kappa_2 &=& \langle \zeta^2 \rangle \equiv \sigma_\zeta^2 \\
  \kappa_3 &=& \langle \zeta^3 \rangle \\
  \kappa_4 &=& \langle \zeta^4 \rangle - 3\langle \zeta^2\rangle^2 \\
  \kappa_5 &=& \langle \zeta^5 \rangle - 10 \langle \zeta^3\rangle \langle \zeta^2\rangle
\end{eqnarray}
It is useful to introduce the dimensionless cummulants, defined as
\begin{equation}
\label{kappa_hat}
  \hat{\kappa}_n \equiv \frac{\kappa_n}{\langle\zeta^2\rangle^{n/2}} \equiv \frac{\kappa_n}{\sigma_\zeta^n}
\end{equation}
For a gaussian PDF we have $\hat{\kappa}_n = 0$ for $n\geq 3$, hence these quantify departures from gaussian statistics.  When the nongaussianities
are small, $|\hat{\kappa}_{n\geq 3}| \ll 1$, then the corrections to $P(\zeta)$ are well described by the Edgeworth expansion:
\begin{equation}  
\label{edgeworth}
  P(\zeta) = \frac{1}{\sqrt{2\pi} \sigma_\zeta} e^{-\zeta^2/(2\sigma_\zeta^2)} \left[ 1 + \frac{\hat{\kappa}_3}{3!} H_3\left(\frac{\zeta}{\sigma_\zeta}\right) + \cdots  \right]
\end{equation}
where $H_3(x) = x^3-3x$ is a Hermite polynomial and the $\cdots$ denotes corrections of order $\hat{\kappa}_4$, $\hat{\kappa}_3^2$ and smaller.
See \cite{shandera,structure,perturbative} for more details and \cite{hidalgo} for an alternative derivation.

\subsection{Relation to Other Works}

Before proceeding to study the model (\ref{L}) in detail, it is worth commenting on the relationship between our analysis and previous works.  Trapped
inflation \cite{beauty,trapped} is a very closely related model that uses multiple bursts of particle production, each similar to the event described in 
subsection \ref{subsec:prodn}, in order to slow the motion of the inflaton on a steep potential.  In that case, dissipation which results from the 
backreaction effects discussed in subsection \ref{subsec:backreaction} actually \emph{dominate} over the friction term $3 H \dot{\phi}$ which is due 
to the inflationary expansion of the universe.  In contrast, here we assume that the inflaton potential $V(\phi)$ is sufficiently flat to 
support slow-roll inflation; see equation (\ref{slow_roll}).  In our scenario, dissipative
effects on the homogeneous motion of $\phi(t)$ due to particle production (backreaction effects) are always negligible; see subsection 
\ref{subsec:backreaction}.  Nevertheless, we expect that many of our results and analytical techniques will also be applicable in the case of trapped 
inflation.

Trapped inflation was not the first model to attempt to exploit dissipative effects in order to assist in slowing the motion of the inflaton.  Another 
interesting construction of this type is \emph{warm inflation} \cite{warm,warm2,warm3,berrera}, which employs a gas of particles in thermal equilibrium.  
  Again, our analysis is distinguished from this
mechanism since we assume that the homogeneous dynamics of the inflaton $\phi(t)$ are of the usual slow-roll type.

Our main focus in this work is to determine the observational consequences of isolated burst of particle production on cosmological observables such
as the spectrum and nongaussianity of the primordial fluctluations.  In this sense, the spirit of our investigation is more similar to 
works such as \cite{chung,chung2,elgaroy,sasaki} than to warm inflation or trapped inflation.  However, unlike those papers, we have consistently 
accounted for rescattering effects which provide the dominant contribution to observables in the case at hand; see 
subsection \ref{subsec:resc}.

\section{Nongaussianity of the Probability Distribution Function}
\label{sec:NG}

In order to quantify the magnitude of the nongaussianity generated by particle production, let us now consider the PDF in the model (\ref{int}).  
We proceed numerically, re-visiting the lattice field theory simulations performed in \cite{ir} using the HLattice code \cite{HLattice}.  For illustration, we assume the standard 
chaotic inflation model $V(\phi) = m^2 \phi^2 / 2$ with $m \cong 10^{-6}\sqrt{8\pi} M_p$ and $\phi_0 = 3.2 \sqrt{8\pi} M_p$.  We consider three different 
choices of coupling, $g^2=1,0.1,0.01$.  Our simulations are performed in a $512^3$ box whose co-moving size is initially $\sim 3$ times the horizon 
size $H^{-1}$.  We run our simulations for roughly $3$ $e$-foldings from the moment when $\phi=\phi_0$, which is more than enough to see the feature 
from IR cascading freeze out as an observable, super-horizon density fluctuation.  Our choice of $\phi_0$ ensures that the feature will be frozen-in at 
scales slightly smaller than the current horizon.  Note that our quantitative results do not depend sensitively on the choice of $\phi_0$, nor on the details of 
the background inflationary potential $V(\phi)$; see \cite{ir} for further discussion.

We extract the PDF of $\delta\phi$ from our HLattice simulations by measuring the fraction of the simulation box which contains the fluctuation field $\delta \phi$ at a particular 
value.  Notice that this approach is completely nonperturbative: it does not rely on the validity of the Edgeworth expansion, nor does it assume anything about the size
or ordering of the cummulants.  This procedure implicitly puts a IR cut-off at the box size $L$ and a UV cut-off at the lattice spacing, $\Lambda^{-1}$.  
Since the nongaussian effects in our model are strongly localized in Fourier space, our quantitative results are largely insensitive to $L$ and $\Lambda$.

\begin{figure}[htbp]
\bigskip \centerline{\epsfxsize=0.5\textwidth\epsfbox{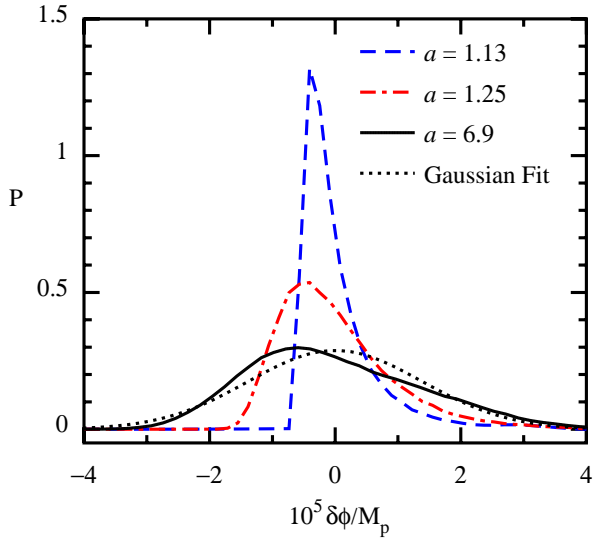}}
\caption{The PDF of the inflaton fluctuations generated by rescattering and IR cascading, at a series of different values 
of the scale factor, $a$.  The dotted black curve shows a Gaussian fit at late times and we have 
normalized the scale factor so that $a=1$ at the moment when particle production occurs.   For illustration, we have chosen $g^2=0.1$ and 
a standard chaotic inflation potential $V(\phi) = m^2\phi^2/2$.}
\label{Fig:onepi}
\end{figure}

In Fig.~\ref{Fig:onepi} we plot our numerical result for the PDF of the inflaton fluctuations generated by rescattering and IR cascading.  In order to make
the physics of inflationary particle production clear, we have subtracted off the contribution coming from the usual vacuum fluctuations of the inflaton.  That
is, the PDF in Fig.~\ref{Fig:onepi} is associated only with the contribution $\delta\phi_{\mathrm{resc}}$ in equation (\ref{hom+par}).

We can understand physically the behaviour of PDF plotted in Fig.~\ref{Fig:onepi}. 
Shortly after the initial burst of particle production the inflaton perturbations $\delta\phi$ are extremely nongaussian, due to the sudden 
appearance of the source term $J \propto \chi^2$ in the equation of motion (\ref{inf_eqn}).  Very quickly, in less than an $e$-folding, nonlinear 
interactions begin to drive the system towards gaussianity.  
A very similar behaviour has been observed in lattice simulations of out-of-equilibrium interacting scalar fields during preheating 
\cite{preheat_pdf1,preheat_pdf2}.  In the case of rescattering during preheating, the system will eventually become gaussian
when the fields thermalize.  However, in our case the universe is still inflating.
As a result, nongaussian inflaton fluctuations generated by rescattering are stretched out by the quasi-de Sitter expansion and must 
freeze once their wavelength crosses the Hubble scale.  Hence, at late times the PDF does not become completely gaussian, but rather
freezes-in with some non-trivial skewness.  Within a few $e$-foldings from the moment of particle production the time evolution of the
PDF has become completely negligible.

\begin{figure}[htbp]
\bigskip \centerline{\epsfxsize=0.5\textwidth\epsfbox{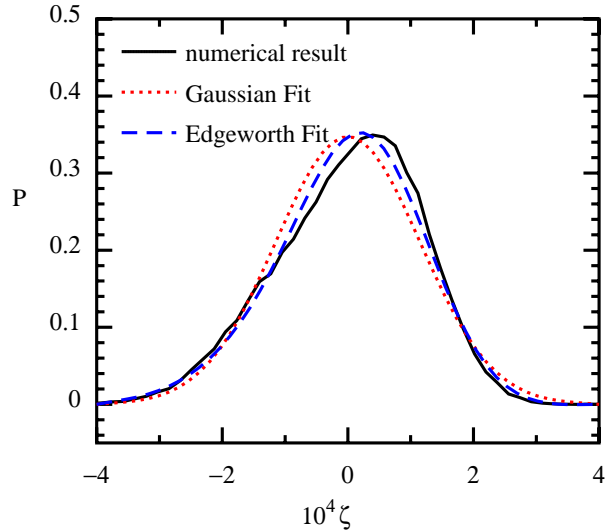}}
\caption{The PDF of the total curvature fluctuation, $\zeta$, at late times (well after all relevant modes have crossed the horizon
and frozen).  The solid black curve is the exact result from our HLattice simulations and the dotted red curve is a gaussian fit.  We have also plotted the 
leading correction to the gaussian result in the Edgeworth expansion, given explicitly by equation (24).  For illustration, we have chosen $g^2=0.1$ and 
a standard chaotic inflation potential $V(\phi) = m^2\phi^2/2$.}
\label{Fig:truePDF}
\end{figure}

In order to characterize the nongaussianity of the observable primordial fluctuations, we would like to construct the PDF for the curvature perturbation $\zeta$, 
including \emph{both} the contributions from the vacuum fluctuations of the inflaton and also from rescattering.
To this end, we construct $\zeta$ using the naive relation $\zeta = -\frac{H}{\dot{\phi}}\delta\phi$ (see section \ref{sec:perts} for justification)
and take into account both contributions to $\delta\phi$ in equation (\ref{hom+par}).  In Fig.~\ref{Fig:truePDF} we plot the full PDF obtained
in this manner, evaluated at very late times, well after all relevant modes have crossed the horizon and become frozen.

Given our numerical results for the PDF of the total observable curvature fluctuation, that is Fig.~\ref{Fig:truePDF}, it is straightforward to compute dimensionless 
cummulants (\ref{kappa_hat}) such as the skewness ($\hat{\kappa}_3$), kurtosis ($\hat{\kappa}_4$) and noltosis ($\hat{\kappa}_5$) for various values of the coupling 
$g^2$.  We have summarized our results in Table \ref{cummulant_table}.  Note that for $g^2=0.01$ both $\hat{\kappa}_4$ and $\hat{\kappa}_5$ are too small to be 
measured accurately from our HLattice simulations.

In order to give some sense of the magnitude of the nongaussianity from particle production we have also computed an 
``equivalent $f_{NL}^{\mathrm{local}}$'' defined by $5 \hat{\kappa}_3 / (18 \sigma_\zeta)$ where the variance is
$\sigma_\zeta \equiv \langle \zeta^2 \rangle^{1/2} \sim 10^{-9/2}$.  For a given $g^2$ this effective $f_{NL}^{\mathrm{local}}$
is the magnitude of $f_{NL}$ which would be necessary to reproduce the skewness $\hat{\kappa}_3$ of the IR cascading PDF using a local ansatz
$\zeta = \zeta_g + \frac{3}{5} f_{NL} \left[ \zeta_g^2 - \langle\zeta_g^2\rangle \right]$.\footnote{Our sign conventions for $f_{NL}$ are 
consistent with WMAP \cite{WMAP7}.  See \cite{shandera} for a discussion of various conventions employed in the literature.}

\begin{table}[htbp]
{\caption{Moments of the Probability Distribution Function}\label{cummulant_table}}
  \begin{center}
  \begin{tabular}{lllll} 
  \hline
  $g^2$ & skewness  & kurtosis & 5-th moment & ``equivalent'' \\
            &  ($\hat{\kappa}_3$) & ($\hat{\kappa}_4$) & ($\hat{\kappa}_5$) &  $f_{NL}^{\mathrm{local}}$ \\
   \hline
  $1$  & $-0.51$ & $0.2$ & $1.2$ & $-4500$ \\
  $0.1$ & $-0.49$ & $-0.1$ & $1.5$ & $-4300$ \\
  $0.01$ & $-0.006$ & $<\mathcal{O}(10^{-3})$ & $<\mathcal{O}(10^{-3})$ & $-53$\\
  \hline
  \end{tabular}
  \end{center}
\end{table}

From Table \ref{cummulant_table} we see that IR cascading during inflation can generate significant nongaussianity.  Even taking
$g^2 = 0.01$ (which is compatible with cosmological data for any choice of $\phi_0$ \cite{ppcons}) we still obtain a skewness $\hat{\kappa}_3=-0.006$,
which is the same value that would be produced by a local model with $f_{NL} \sim -53$.  This ``equivalent'' local nongaussianity
is comparable to current observational bounds, and is well within the expected accuracy of future missions.  This suggests that nongaussian
features from particle production during inflation might be observable for reasonable values of $g^2$.  

The ``equivalent'' $f_{NL}^{\mathrm{local}}$ values presented in Table \ref{cummulant_table} must be interpreted with care.  We have included this
information only to give a heuristic sense of the magnitude of nongaussianity in our model.  It must be stressed that the PDF plotted
in Fig.~\ref{Fig:truePDF} is quite different from the analogous result for local-type nongaussianity.  For example, the value of the kurtosis
(and higher moments) are different, as is the ordering of the cummulants.  Moreover, we should emphasize that observational bounds on $f_{NL}^{\mathrm{local}}$ 
\emph{cannot} be directly applied to our model since the bispectrum in our case is uncorrelated with the vacuum fluctuations and is far from scale invariant.  
A detailed study of the detectability of nongaussianity from particle production will be the subject of a upcoming publication \cite{inprog}.

Depending on the value of $\phi_0$, the model (\ref{int}) may lead to a variety of observable signatures.  As discussed previously, $\phi_0$
controls the location of the feature in the primordial power spectrum (\ref{P_fit}).  Nongaussian effects are also localized near the same characteristic
scale, $k_{\mathrm{IR}}$.  If $k_{\mathrm{IR}}$ corresponds to scales relevant for CMB experiments, then we predict a bump-like feature in 
the primordial power spectrum, $P_{\zeta}(k)$, and an associated feature in the bispectrum $B_{\zeta}(k_i)$ with an unusual shape (that will
be discussed in section \ref{sec:theory}).  A key question is whether the nongaussian feature can be observably large in a regime where the power
spectrum feature is small enough to be compatible with current observations.  Preliminary results are encouraging: for $g^2=0.01$ the power spectrum
is consistent with all observational data \cite{ppcons} while the skewness of the PDF is rather large.  A detailed investigation will require a simple, 
separable template for the bispectrum and will be discussed in a future publication \cite{inprog}.

On the other hand, we could imagine a scenario where the feature from IR cascading shows up on smaller scales, relevant for Large Scale Structure 
(LSS) experiments \cite{shandera,neal,pat,andrew}.  
In this case our scenario could be probed using higher order correlations of LSS probes (such as the galaxy bispectrum) or the
abundance of collapsed objects (or voids).  The latter possibility is interesting since the cluster/void abundance is determined the tails of the PDF
and may be insensitive to the detailed shape of the bispectrum.  Quantitative predictions for observable cluster/void abudances require the PDF of the
evolved density field, smoothed on some relevant scale \cite{inprog}, rather than the PDF of the primordial curvature perturbation (which is plotted in 
Fig.~\ref{Fig:truePDF}).  However, we can nevertheless describe the qualitative signatures which should be expected.  Our model robustly predicts
a negative skewness for both the curvature perturbation, $\zeta$, and the density field, $\delta\rho/\rho$.  Hence, we should expect a \emph{decrease}
in the abundance of the largest collapsed objects and an \emph{increase} in the abundance of the largest voids \cite{void,lss_rev}.  Owing to the localized nature
of the bispectrum feature, we expect that this effect should show up only when the density field is smoothed on a scale close to $k_{\mathrm{IR}}$.

It is worth mentioning that recent weak lensing measurement of the dark matter mass of the high-redshift galaxy cluster XMMUJ2235.3-2557 \cite{cluster} 
have been construed as a possible hint of nongaussian initial conditions \cite{lss2}.  Unfortunately, our model does not produce the correct sign of 
skewness to explain such observations.  

\section{Analytical Formalism}
\label{sec:theory}

In \cite{ir} we studied particle production, rescattering and IR cascading using nonlinear lattice field theory simulations.  In addition to
this numerical approach, a cursory analytical formalism was also presented.  Here we re-visit the analytical analytical theory of particle
production, rescattering and IR cascading in an expanding universe, in order to better understand the results of \cite{ir}  from a physical
perspective.

\subsection{The Prototype Model}

We consider now the theory
\begin{eqnarray}
  S &=& \int d^4 x \sqrt{-g} \left[ \frac{M_p^2}{2} R - \frac{1}{2}(\partial \phi)^2 - V(\phi) \right. \nonumber \\
      &&\,\,\,\,\,\,\,\,\,\,\,\,\,   \left. - \frac{1}{2}(\partial \chi)^2 - \frac{\mu^2}{2}\chi^2 - \frac{g^2}{2}(\phi-\phi_0)^2\chi^2 \right]  \label{L2}
\end{eqnarray}
which differs from our original model (\ref{L}) by the inclusion of a mass term $\Delta \mathcal{L} = -\mu^2\chi^2/2$ for the iso-inflaton.
Such a term is not forbidden by any symmetry and hence one typically expects it to be generated by radiative corrections,
even if the iso-inflaton is classically massless at $\phi=\phi_0$.   The new parameter $\mu$ has the effect of reducing the efficiency of the particle
production effects discussed in subsection \ref{subsec:prodn}; the time-varying mass of the iso-inflaton 
\[
  m_\chi^2 = \mu^2 + g^2 (\phi-\phi_0)^2
\]
does not vanish at $\phi=\phi_0$, but rather reaches a minimum value $\mu^2$, making the adiabaticity condition more difficult to violate.  A concern is 
the possibility that radiative corrections induce a large $\mu$ and suppress the observable effects associated with inflationary particle production.
Indeed, it is well known that fine tuning may be required to keep the mass of any scalar field significantly below the cut-off scale associated with the 
validity of the effective field theory description (\ref{L2}).  Below, we will show that that the suppression of $\chi$-particle production is not significant
provided the following condition is satisfied
\begin{equation}
\label{pp_cond}
  \mu^2 \ll k_\star^2
\end{equation}
where $k_\star \equiv \sqrt{g|v|}$.  Depending on how (\ref{L2}) is embedded within a more complete framework, the constraint (\ref{pp_cond}) may (or may
not) require fine-tuning to satisfy.  Below, we will show that the condition (\ref{pp_cond}) is quite naturally satisfied for a large number of microscopically
realistic constructions.

Our prototype model (\ref{L2}) has been choosen to elucidate the key physics and observational signatures of inflationary particle production in
a simple framework wherein computations are tractable.  We expect, however, that many of our qualitative results will carry over to more complicated 
scenarios.  In particular, one might wish to supplement the action (\ref{L2}) by its supersymmetric (SUSY) completion; see the interesting work 
\cite{berrera} for an explicit example.  Such an embedding has the advantage that the flatness of the inflaton potential $V(\phi)$ may be protected from 
large radiative corrections coming from loops of the $\chi$ field.  Moreover, a SUSY embedding of the model (\ref{L2}) also allows for some control over 
the quantum corrections to the mass scale $\mu$.  
  
For models obtained from string theory or super-gravity (SUGRA), it is natural to have $\mu$ of order the Hubble scale\footnote{In the context of 
SUGRA, the finite energy density driving inflation breaks SUSY and induces soft 
scalar potentials with curvature of order $V''_{\mathrm{soft}} \sim \mu^2 \sim H^2$ \cite{susy_break}.  In the case of string theory,  many scalars are 
conformally coupled to gravity \cite{rapid_roll} through an interaction of the form $\delta\mathcal{L} = - \frac{1}{12} R \chi^2$ where the Ricci scalar is 
$R \sim H^2$ during inflation.  More generally, any non-minimal coupling $\delta\mathcal{L} = -\frac{\xi}{2} R \chi^2$ between gravity and the iso-inflaton 
will induce a contribution of order $H$ to the effective mass of $\chi$, as long as $\xi = \mathcal{O}(1)$.} during inflation 
\cite{false_vac,susy_break,rapid_roll}; hence we expect $\mu^2 \sim H^2$ for such models.
In that case, the constraint  (\ref{pp_cond}) is automatically satisfied, because $k_\star^2 \gg H^2$ whenever particle production is fast as compared to the expansion time (that is, for 
reasonable values of the coupling $g^2 > 10^{-7}$, which we assume throughout this work).  Hence, there exists a very large class of realistic microscopic 
models in which radiative effects will not spoil the observational consequences of inflationary particle production and IR cascading.

Although the condition for the efficiency of particle production - that is equation (\ref{pp_cond}) - can be easily satisfied for models coming from string theory or
SUSY, we prefer to remain agnostic regarding how the prototype action (\ref{L2}) is embedded within a more complete framework.  Throughout our analysis
we will keep the inflaton potential $V(\phi)$ and the iso-inflaton mass parameter $\mu$ more-or-less arbitrary (we assume that the slow roll
conditions are satisfied, and also that $\mu^2 \geq 0$).  This phenomenological approach is not different from the philosophy that is 
employed in the majority of work on inflationary cosmology, since the slow roll conditions (\ref{slow_roll}) may be sensitive to UV physics whose detailed
form is often not specified.  The question of how the model (\ref{L2}), with a given choice of $V(\phi)$ and $\mu$, arises from some complete model of 
particle physics is interesting.  However, this question it is not the main focus of the current investigation.  We refer the reader to \cite{ppcons} for 
several example microscopic embeddings within string theory and also SUSY (see also \cite{trapped}).

Let us now proceed to develop an analytical formalism to study inflationary particle production in the model (\ref{L2}).
The equations of motion that we wish to solve are
\begin{eqnarray}
  -\Box \phi + V'(\phi) + g^2 (\phi-\phi_0) \chi^2 &=& 0 \label{phiKG} \\
  -\Box \chi + \left[\mu^2+ g^2 (\phi-\phi_0)^2\right]\chi &=& 0 \label{chiKG}
\end{eqnarray}
where $\Box = g_{\mu\nu}\nabla^{\mu}\nabla^{\nu}$ is the covariant d'Alembertian.  It will be useful to work with conformal time
$\tau$, related to cosmic time $t$ via $a d\tau = dt$.  In terms of conformal time the metric takes the form
\begin{eqnarray}
  ds^2 &=& -dt^2 + a^2(t) d{\bf x}\cdot d{\bf x} \nonumber \\
           &=& a^2(\tau) \left[ -d\tau^2 + d{\bf x}\cdot d{\bf x}  \right]
\end{eqnarray}
We denote derivatives with respect to cosmic time as $\dot{f} \equiv \partial_t f$ and with respect to conformal time as $f' \equiv \partial_\tau f$.  
The Hubble parameter $H = \dot{a} / a$ has conformal time analogue $\sH = a' / a$.  For an inflationary (quasi-de Sitter) phase 
($H \cong \mathrm{const}$) one has
\begin{equation}
\label{conf_scale}
  a = -\frac{1}{H\tau}\frac{1}{1-\epsilon}, \hspace{3mm} \sH = -\frac{1}{\tau}\frac{1}{1-\epsilon}
\end{equation}
to leading order in the slow roll parameter $\epsilon \ll 1$.

As discussed in section \ref{sec:overview}, the motion of the homogeneous inflaton $\phi(t)$ leads to the production
of a gas of $\chi$ particles at the moment $t=0$ when $\phi=\phi_0$.  The first step in our analytical computation is
to describe this burst of particle production in an expanding universe.  Following the initial burst, both backreaction
and rescattering effects take place.  Our formalism will focus on the latter effect, which has been shown to be much more 
important \cite{ir}.

\subsection{Particle Production in an Expanding Universe}
\label{chi_sec}

The first step in our scenario is the quantum mechanical production of $\chi$-particles due to the motion of $\phi$.
To understand this effect we must solve the equation for the $\chi$ fluctuations in the rolling inflaton background.
Approximating $\phi \cong \phi_0 + v t$ equation (\ref{chiKG}) gives
\begin{equation}
\label{chi_expanding}
  \ddot{\chi} + 3 H \dot{\chi} - \frac{\grad^2}{a^2} \chi + \left[ \mu^2 + k_\star^4 t^2 \right] \chi = 0
\end{equation} 
where $k_\star \equiv \sqrt{g |v|}$.  We remind the reader that $k_\star \gg H$ for reasonable values of the coupling, see equation (\ref{ratio}).

The flat space analogue of equation (\ref{chi_expanding}) is very well understood from studies of broad band parametric resonance during preheating \cite{KLS97}
and also moduli trapping at enhanced symmetry points \cite{beauty}.  One does not expect this treatment to differ significantly in our case
since both the time scale for particle production $\Delta t$ and the characteristic wavelength of the produced fluctuations $\lambda$ are small
compared to the Hubble scale.  Hence, we expect that the occupation number of produced
$\chi$ particles will not differ significantly from the flat-space result, at least on scales $k \gsim H$.
Furthermore, notice that the $\chi$ field is extremely massive for most of inflation.  Indeed, even in the case $\mu^2=0$ we have
\begin{equation}
  \frac{m_\chi^2}{H^2} \cong \frac{k_\star^4 t^2}{H^2}
\end{equation}
Since $k_\star \gg H$ it follows that $m_\chi^2 \gg H^2$,
except in a tiny interval $H |\Delta t| \sim (H / k_\star)^2$ which amounts to roughly $10^{-3}$ $e$-foldings for $g^2 \sim 0.1$.  
Therefore, we do not expect any significant fluctuations of $\chi$ to be produced on super-horizon scales $k \lsim H$.  (Allowing for $\mu^2\not= 0$
only strengthens this conclusion.)

Let us now consider the solutions of equation (\ref{chi_expanding}).  We work with conformal time $\tau$
and write the Fourier transform of the quantum field $\chi$ as
\begin{equation}
\label{chi_fourier}
  \chi(\tau,{\bf x}) = \int \frac{d^3k}{(2\pi)^{3/2}} \frac{\xi^\chi_{\bf k}(\tau)}{a(\tau)} e^{i {\bf k}\cdot {\bf x}}
\end{equation}
Note the explicit factor of $a^{-1}$ in (\ref{chi_fourier}) which is introduced to give $\xi^\chi_{\bf k}$ a canonical kinetic 
term.  The q-number valued Fourier transform $\xi^\chi_{\bf k}(\tau)$ can be written as 
\begin{equation}
\label{chi_mode}
  \xi^\chi_{\bf k}(\tau) = a_{\bf k}\, \chi_k(\tau) + a^\dagger_{-{\bf k}}\, \chi_k^\star(\tau)
\end{equation}
where the annihilation/creation operators satisfy the usual commutation relation
\begin{equation}
\label{commutator}
  \left[a_{\bf k}, a_{\bf k'}^{\dagger}\right] = \delta^{(3)}({\bf k}-{\bf k'})
\end{equation}
and the c-number valued mode functions $\chi_k$ obey the following oscillator-like equation
\begin{equation}
\label{chi_prod}
  \chi_k''(\tau) + \omega_k^2(\tau) \chi_k(\tau) = 0
\end{equation}
The time-dependent frequency is
\begin{eqnarray}
  \omega_k^2(\tau) &=& k^2 + a^2m_\chi^2(\tau) - \frac{a''}{a} \nonumber \\
                              &\cong& k^2 + \frac{1}{\tau^2}\left[ \frac{k_\star^4}{H^2} t^2(\tau) + \left(\frac{\mu}{H}\right)^2 - 2  \right]  \label{omega_chi}
\end{eqnarray}
where 
\begin{equation}
\label{m_chi}
  m_\chi^2(\tau) = \mu^2 + g^2 (\phi-\phi_0)^2 \cong \mu^2 + k_\star^4 t^2(\tau)
\end{equation}
is the time-dependent effective mass of the $\chi$ particles and
\begin{equation}
\label{cosmic_time}
  t(\tau) = \frac{1}{H} \ln\left(\frac{-1}{H \tau}\right)
\end{equation}
is the usual cosmic time variable.  We have arbitrarily set the origin of conformal time so that $\tau=-1/H$ corresponds
to the moment when $\phi=\phi_0$.  

It is useful to define the occupation number $n_k$ of the $\chi$ of particles with momentum ${\bf k}$, defined as the energy
of the mode $\frac{1}{2}|\chi'_k|^2 + \frac{1}{2}\omega_k^2 |\chi_k|^2$ divided by the energy $\omega_k$ of each particle.  Explicitly, we define
\begin{equation}
  n_k = \frac{\omega_k}{2}\left[ \frac{|\chi'_k|^2}{\omega_k^2} + |\chi_k|^2 \right] - \frac{1}{2} \label{n_k_def}
\end{equation}
where the term $-\frac{1}{2}$ comes from extracting the zero-point energy of the linear harmonic oscillator (see \cite{KLS97} for a review).  
Our definition (\ref{n_k_def}) coinicides with the usual notion of particle number in the asymptotic adiabatic regimes ($|t| \gsim k_\star^{-1}$).
During the very brief non-adiabatic period ($|t| \ll k_\star^{-1}$) our result coincides with the usual notion of quasi-particle number, obtained
by instantaneous diagonalization of the Hamiltonian

Let us now try to understand analytically the behaviour of the solutions of (\ref{chi_prod}).
At early times $t \ll -k_\star^{-1}$, the frequency $\omega_k$ varies adiabatically
\begin{equation}
\label{adiabatic_condition}
  \left|\frac{\omega'_k}{\omega_k^2}\right| \ll 1
\end{equation}
In this in-going adiabatic regime the modes $\chi_k$ are not excited and the solution of (\ref{chi_prod}) are well described
by the adiabatic solution $\chi_k(\tau) = f_k(\tau)$ where
\begin{equation}
\label{f_k}
  f_k(\tau) \equiv \frac{1}{\sqrt{2\omega_k(\tau)}} \exp\left[-i \int^\tau d\tau' \omega_k(\tau) \right]
\end{equation}
We have normalized (\ref{f_k}) to be pure positive frequency so that the state of the iso-inflaton field 
at early times corresponds to the adiabatic vacuum with no $\chi$ particles.  (Inserting (\ref{f_k}) into (\ref{n_k_def})
one finds $n_k=0$ for the adiabatic solution, as expected.)

The adiabatic solution (\ref{f_k}) ceases to be a good approximation very close to the moment when $\phi=\phi_0$, 
that is at times $|t| \lsim k_\star^{-1}$.  In this regime the adiabaticity condition (\ref{adiabatic_condition}) is violated for modes
with wave-number $H \lsim k \lsim \sqrt{k_\star^2-\mu^2}$ and $\chi$ particles within this momentum band are produced.  During the
non-adiabatic regime we can still represent the solutions of (\ref{chi_prod}) in terms of the functions $f_k(\tau)$ as
\begin{equation}
\label{chi_bog}
  \chi_k(\tau) = \alpha_k(\tau) f_k(\tau) + \beta_k(\tau) f_k^\star(\tau)
\end{equation}
This expression affords a solution of (\ref{chi_prod}) provided the time-dependent Bogoliubov coefficients obey the following set of coupled equations
\begin{eqnarray}
  \alpha'_k(\tau) &=& \frac{\omega'_k(\tau)}{2\omega_k(\tau)} \exp\left[ +2 i \int^\tau d\tau' \omega_k(\tau') \right] \beta_k(\tau) \label{alpha_dot} \\
  \beta'_k(\tau) &=& \frac{\omega'_k(\tau)}{2\omega_k(\tau)} \exp\left[ -2 i \int^\tau d\tau' \omega_k(\tau') \right] \alpha_k(\tau) \label{beta_dot}
\end{eqnarray}
The Bogoliubov coefficients are normalized as $|\alpha_k|^2 - |\beta_k|^2 = 1$ and the assumption that no $\chi$ particles are present
in the asymptotic past\footnote{This assumption is justified since any initial excitation
of $\chi$ would have been damped out exponentially fast by the expansion of the universe.} fixes the 
initial conditions $\alpha_k = 1$, $\beta_k = 0$ for $t \rightarrow -\infty$.  This is known as the adiabatic initial condition.

From the structure of equations (\ref{alpha_dot},\ref{beta_dot}) it is clear that violations of the condition (\ref{adiabatic_condition}) near $t=0$
leads to a rapid growth in the $|\beta_k|$ coefficient.  The time variation of $\beta_k$ can be interpreted as a corresponding growth in the 
occupation number.  Inserting (\ref{chi_bog}) into (\ref{n_k_def}) we find 
\begin{equation}
  n_k = |\beta_k|^2
\end{equation}

At late times ($t \gsim k_\star^{-1}$) adiabaticity is restored and the growth of $n_k=|\beta_k|^2$ must saturate.  By inspection of equations
(\ref{alpha_dot},\ref{beta_dot}) we can see that the Bogoliubov coefficients must tend to constant values in the out-going adiabatic regime.
Therefore, within less than an $e$-folding from the moment of particle production the solution $\chi_k$ of equation (\ref{chi_prod}) can be represented as a
simple superposition of positive frequency $f_k$ modes and negative frequency $f_k^\star$ modes.  Our goal now is to derive an analytical expression 
for the modes $\chi_k$ which is valid in this out-going adiabatic region.

First, we seek an expression for the Bogoliubov coefficients $\alpha_k$, $\beta_k$ in the out-going adiabatic regime $t \gsim k_\star^{-1}$.
From (\ref{alpha_dot},\ref{beta_dot}) it is clear that the value of the Bogoliubov coefficients at late times can depends only on dynamics
during the interval $|t| \lsim k_\star^{-1}$ where the adiabaticity condition (\ref{adiabatic_condition}) is violated.  This interval
is tiny compared to the expansion time and we are justified in treating $a(\tau)$ as roughly constant during this phase.  Hence, it follows that the 
flat space computation of the Bogoliubov coefficients \cite{beauty,KLS97} must apply, at least for scales $k \gsim H$.  To a very
good approximation we therefore have the well-known result 
\begin{eqnarray}
  \alpha_k &\cong& \sqrt{1+e^{-\pi \mu^2 / k_\star^2} e^{-\pi k^2 / k_\star^2}} \label{alpha_k} \\
  \beta_k &\cong& -i e^{-\pi \mu^2 / (2k_\star^2)}e^{-\pi k^2 / (2k_\star^2)} \label{beta_k}
\end{eqnarray}
in the out-going adiabatic regime.  Equation (\ref{beta_k}) gives the usual expression\footnote{Our result for the Bogoliubov coefficients
is consistent with \cite{beauty}.  In that work $\mu$ was interpreted as an ``impact parameter'' for the motion of the modulus, whereas in our
work we interpret this as a ``bare'' mass term.  This distinction has no impact on the result for the occupation number because, in both cases,
the parameter appears in the same way in the equation of motion for the fluctuations of $\chi$.} for the co-moving occupation number
of particles produced by a single burst of broad-band parametric resonance:
\begin{equation}
\label{n_k_mu}
  n_k = |\beta_k|^2 = e^{-\pi \mu^2 / k_\star^2} e^{-\pi k^2 / k_\star^2}
\end{equation}
Comparing equations (\ref{n_k_mu}) and (\ref{n_k}) we see that the mass parameter $\mu$ for the iso-inflaton has the effect of suppressing the number
density of produced $\chi$ particles by an amount $e^{-\pi \mu^2 / k_\star^2} \leq 1$.  This suppression reflects the reduced phase space of produced particles:
the adiabaticity condition is violated only for modes with $k < \sqrt{k_\star^2 - \mu^2}$.  Notice that the suppression of $\chi$ particle production is 
negligible when $\mu^2 \ll k_\star^2$, precisely the condition (\ref{pp_cond}) that was alluded to earlier.  For the remainder of this work we will assume
that $\mu \lsim k_\star$, since in the opposite regime the observational signatures of particle production effects are exponentially suppressed.

Next, we seek an expression for the adiabatic solution $f_k(\tau)$ in the out-going regine $t \gsim k_\star^{-1}$.  We assume $\mu \lsim k_\star$
and also focus on the interesting region of phase space, $H \lsim k \lsim \sqrt{k_\star^2-\mu^2}$.  In this case, the adiabatic solution (\ref{f_k}) is very well approximated 
by
\begin{equation}
\label{f_k_approx}
  f_k(\tau) \cong \frac{1}{a^{1/2} k_\star \sqrt{2 t(\tau)}} e^{-\frac{i}{2} k_\star^2 t^2(\tau)}
\end{equation}
where $t(\tau)$ is defined by (\ref{cosmic_time}).
It is interesting to note that equation (\ref{f_k_approx}) is identical to the analogous flat-space result \cite{ir}, except for the
factor of $a^{-1/2}$.  Taking into account also the explicit factor of $a^{-1}$ in our definition
of the Fourier transform (\ref{chi_fourier}) we recover  the expected large-scale behaviour for a massive field in de Sitter space, that is $\chi \sim a^{-3/2}$.
This dependence on the scale factor is easy to understand physically, it simply reflects the volume dilution of non-relativistic particles: 
$\rho_\chi \sim m^2_\chi \chi^2 \sim a^{-3}$.  Notice that the parameter $\mu$ does not appear in (\ref{f_k_approx}).  This is so because, for 
the time-varying mass of the $\chi$ field, equation (\ref{m_chi}), is dominated by the interaction term when $k_\star |t| \gsim 1$ and $\mu \lsim k_\star$.

Finally, we arrive at an expression for the out-going adiabatic $\chi$ modes which is accurate
for interesting scales $H \lsim k \lsim \sqrt{k_\star^2-\mu^2}$.  Putting together the results (\ref{f_k_approx}) and (\ref{chi_bog}) along with the well-known
expressions (\ref{alpha_k},\ref{beta_k}) we arrive at
\begin{eqnarray}
  &&\chi_k(\tau) \cong 
  \sqrt{1+e^{-\pi \mu^2/k_\star^2}e^{-\pi k^2 / k_\star^2}} \frac{1}{a^{1/2} k_\star \sqrt{2 t(\tau)}} e^{-\frac{i}{2} k_\star^2 t^2(\tau)} \nonumber \\
   &&- i e^{-\pi \mu^2 / (2k_\star^2)}e^{-\pi k^2 /(2 k_\star^2)}  \frac{1}{a^{1/2} k_\star \sqrt{2 t(\tau)}} e^{+\frac{i}{2} k_\star^2 t^2(\tau)} \label{chi_soln}
\end{eqnarray}
valid for $t \gsim k_\star^{-1}$.
Equation (\ref{chi_soln}) is the main result of this subsection.  We will now justify that this expression is quite sufficient for our purposes.

For modes deep in the UV, $k \gsim k_\star$, our expression (\ref{chi_soln}), is not accurate.\footnote{The expression (\ref{f_k_approx})
for the adiabatic modes $f_k$ is not valid at high momenta where $\omega_k \cong k$.}  
However, such high momentum
particles are not produced, the condition (\ref{adiabatic_condition}) is always satisfied for $k \gg k_\star$.  Note that the absence of particle
production deep in the UV is built into our expression (\ref{chi_soln}): as $k\rightarrow \infty$ this function tends to the vacuum solution 
$\chi_k \rightarrow f_k$.

Our expression (\ref{chi_soln}) is also not
valid deep in the IR, for modes $k < H$.  To justify this neglect requires somewhat more care.  Notice that, even very far from the 
point $\phi=\phi_0$ long wavelength modes $k \ll H$ should not be thought of as particle-like.  The large-scale mode functions are not oscillatory
but rather damp exponentially fast as $\chi \sim a^{-3/2}$.  Hence, even if we started with some super-horizon fluctuations of $\chi$ at the beginning of
inflation, these would be suppressed by an exponentially small factor before the time when particle production occurs.  Any super-horizon
fluctuation generated near $t=0$ would need to be exponentially huge to overcome this damping.  However, resonant particle production 
during inflation does
\emph{not} lead to exponential growth of mode functions.\footnote{In this regard our scenario is very different from preheating
at the end of inflation.  In the latter case the inflaton passes \emph{many} times through the massless point $m_\chi = 0$ and there are, 
correspondingly, many bursts of particle production.  After many oscillations of the inflaton field, the $\chi$ particle occupation numbers
build up to become exponentially large and, averaged over many oscillations of the background, the $\chi$ mode functions grow exponentially.
However, in our case case there is only a \emph{single} burst of particle production at $t=0$.  The resulting occupation number (\ref{n_k}) is always
less than unity and the solutions of (\ref{chi_prod}) never display exponential growth.}

To verify explicitly that there is no significant effect for super-horizon fluctuations let us consider solving equation (\ref{chi_expanding}) 
neglecting gradient terms.  The equation we wish to solve, then, is
\begin{equation}
\label{no_grad}
\partial_t^2(a^{3/2} \chi) + \left[ k_\star^4 t^2 - \frac{9}{4}H^2 \right] (a^{3/2}\chi) = 0
\end{equation} 
(For simplicity we take $\mu=0$ and $\epsilon=0$ for this paragraph, however, this has no effect on our results.)  The solution of this equation
may be written in terms of parabolic cylinder functions $D_{\nu}(z)$ as
\begin{eqnarray}
  \chi(t,{\bf x}) &\sim& \frac{1}{a^{3/2}} \left(\,\,\,\,  C_1 \, D_{ - \frac{1}{2} + \frac{9H^2}{8k_\star^2}i }\left[(1+i)k_\star t\right] \right. \nonumber \\
 && \,\,\,\,\,\,\,\,\,\,\,\,\,       \left.            + \,\,\,   C_2 \, D_{ - \frac{1}{2} - \frac{9H^2}{8k_\star^2}i }\left[(-1+i)k_\star t\right]   \,\,    \right) \label{cylinder}
\end{eqnarray}
For our purposes the precise values of the coefficients $C_1$, $C_2$ are not important.  Rather, it suffices to note that for $k_\star|t| \gsim 1$
the function (\ref{cylinder}) behaves as 
\begin{equation}
\label{chi_large_damp_thing}
  \chi(t,{\bf x}) \sim |t|^{-1/2} e^{-3 H t / 2} \times \left[\mathrm{oscillatory}\right]
\end{equation}
This explicit large-scale asymptotics confirms our previous claims that the super-horizon fluctuations of $\chi$ damp to zero
exponentially fast, as $a^{-3/2} \sim e^{-3Ht/2}$.  As discussed previously, this damping is easy to understand in terms of the volume
dilution of non-relativistic particles.  (If we had included the parameter $\mu^2$
in equation (\ref{no_grad}) then our conclusions would only be strengthened, since this parameter has the effect of making the iso-inflaton even 
more massive.)
We can also understand the power-law damping that appears in (\ref{chi_large_damp_thing}) from
a physical perspective.  The properly normalized modes behave as $a^{3/2} \chi \sim \omega_k^{-1/2}$ while on large scales we have
$\omega_k \sim |m_\chi| \sim |t|$.  Hence, the late-time damping factor $t^{-1/2}$ which appears in (\ref{chi_large_damp_thing}) reflects the 
fact that the $\chi$ particles become ever more massive as $\phi$ rolls away from the point $\phi_0$.  

Finally, it is straightforward to see that the function (\ref{cylinder}) does not display
any exponential growth near $t=0$.  Hence, we conclude that there is no significant generation of super-horizon $\chi$ fluctuations
due to particle production.\footnote{This is strictly true only in the linearized theory.  It is possible that $\chi$ particles are generated
by nonlinear effects such as rescattering.  However, even such second order $\chi$ fluctuations will be extremely massive
compared to the Hubble scale and must therefore suffer exponential damping $a^{-3/2}$ on large scales.}

In this subsection we have seen that the quantum production of $\chi$ particles in an expanding universe proceeds very much as it does
in flat space.  This is reasonable since particle production occurs on a time scale short compared to the expansion time and involves
modes which are inside the horizon at the time of production.  

\subsection{Inflaton Fluctuations}

In section \ref{chi_sec} we studied the quantum production of $\chi$ particles which occurs when $\phi$ rolls past the
massless point $\phi=\phi_0$.  Subsequently, there are two distinct physical processes which take place: backreaction and rescattering.
As we discussed, the former effect has a negligible impact of the observable spectrum of cosmological
perturbations and may be neglected.

In this subsection we study the rescattering of produced $\chi$ particle off the inflaton condensate.  
The dominant process to consider is the diagram illustrated in Fig.~\ref{Fig:diag}, corresponding to 
bremsstrahlung emission of $\delta\phi$ fluctuations (particles) in the background of the external field.  (There is also
a sub-dominant process of the type $\chi\chi\rightarrow\delta\phi\delta\phi$ which is phase space suppressed.)  Taking into
account only the rescattering diagram illustrated in Fig.~\ref{Fig:diag} is equivalent to solving the following equation for the q-number inflaton
fluctuation
\begin{equation}
\label{delta_phi}
  \left[\partial_t^2 + 3 H \partial_t - \frac{\grad^2}{a^2} + m^2 \right] \delta \phi = -g^2 \left[\phi(t)-\phi_0\right] \chi^2
\end{equation}
where we have introduced the notation $m^2 \equiv V_{,\phi\phi}$ for the inflaton effective mass. (Note that we are not assuming a background
potential of the form $m^2 \phi^2 / 2$, only that $V_{,\phi\phi}\not= 0$ in the vicinity of the point $\phi=\phi_0$.)

Equation (\ref{delta_phi}) may be derived by noting that (\ref{L}) gives an interaction of the form $g^2(\phi-\phi_0) \delta \phi\chi^2$ between
the inflaton and iso-inflaton, in the background of the external field $\phi(t)$.  Equivalently, one may construct this equation by a straightforward
iterative solution of (\ref{phiKG}).

We work in conformal time and define the q-number Fourier
transform $\xi^\phi_{\bf k}(\tau)$ of the inflaton fluctuation analogously to (\ref{chi_fourier}):
\begin{equation}
\label{phi_fourier}
  \delta\phi(\tau,{\bf x}) = \int \frac{d^3 k}{(2\pi)^{3/2}}\frac{\xi_{\bf k}^\phi(\tau)}{a(\tau)} e^{i {\bf k}\cdot {\bf x}}
\end{equation}
(To avoid potential
confusion we again draw the attention of the reader to the explicit factor $a^{-1}$ in our convention for the Fourier 
transform.)  The equation of motion (\ref{delta_phi}) now takes the form
\begin{eqnarray}
\label{phi_mode}
&&  \left[\partial_\tau^2 + k^2 + a^2 m^2 - \frac{a''}{a} \right] \xi_{\bf k}^\phi(\tau) \nonumber \\
&&= -g k_\star^2 a(\tau) t(\tau) \int \frac{d^3k'}{(2\pi)^{3/2}} \xi^\chi_{\bf k'}\xi^\chi_{\bf k-k'}(\tau)
\end{eqnarray}
The solution of (\ref{phi_mode}) consists of two parts: the solution of the homogeneous equation and the particular solution which is due
to the source.  The former corresponds, physically, to the usual vacuum fluctuations from inflation.  On the other hand, the particular solution
corresponds physically to the secondary inflaton modes which are generated by rescattering.

\subsection{Homogeneous Solution and Green Function}

We consider first the homogeneous solution of (\ref{phi_mode}).  Since the homogeneous solution is a gaussian field, we may
expand the q-number Fourier transform in terms of annihilation/creation operators $b_{\bf k}$, $b_{\bf k}^\dagger$ and c-number mode
functions $\phi_k(\tau)$ as
\begin{equation}
\label{phi_annihilation}
  \xi^{\phi}_{\bf k}(\tau) = b_{\bf k}\, \phi_k(\tau) + b^\dagger_{-{\bf k}} \,\phi_k^\star(\tau)
\end{equation}
Here the inflaton annihilation/creation operators $b_{\bf k}$, $b_{\bf k}^\dagger$ obey 
\begin{equation}
  \left[b_{\bf k}, b_{\bf k'}^{\dagger}\right] = \delta^{(3)}({\bf k} - {\bf k'})
\end{equation}
and commute with the annihilation/creation operators of the $\chi$-field:
\begin{equation}
  \left[a_{\bf k}, b_{\bf k'} \right] = \left[ a_{\bf k}, b_{\bf k'}^{\dagger} \right] = 0
\end{equation}

Using (\ref{conf_scale}) and (\ref{slow_roll}) it is straightforward to see that the homogeneous inflaton mode functions obey the following equation
\begin{equation}
\label{simple_inf_eqn}
  \partial_\tau^2 \phi_k + \left[ k^2 - \frac{1}{\tau^2}\left( \nu^2 - \frac{1}{4}  \right)  \right] \phi_k = 0
\end{equation}
where we have defined
\begin{equation}
\label{nu1}
  \nu \cong \frac{3}{2} - \eta + \epsilon
\end{equation}
The properly normalized mode function solutions are well known and may be written 
in terms of the Hankel function of the first kind as
\begin{equation}
\label{phi_k}
  \phi_k(\tau) = \frac{\sqrt{\pi}}{2} \sqrt{-\tau} H_{\nu}^{(1)}(-k\tau)
\end{equation}
This solution corresponds to the usual quantum vacuum fluctuations of the inflaton field during inflation.

In passing, let us compute the power spectrum of the quantum vacuum fluctuations from inflation.  Using the solutions (\ref{phi_k})
we have
\begin{equation}
\label{P_phi_vac}
  P_\phi^{\mathrm{vac}}(k) = \frac{k^3}{2\pi^2}\left|\frac{\phi_k}{a} \right|^2 \cong \frac{H^2}{(2\pi)^2} \left(\frac{k}{a H}\right)^{n_s-1}
\end{equation}
on large scales $k \ll aH$.  The explicit factor of $a^{-2}$ in (\ref{P_phi_vac}) appears to cancel the $a^{-1}$ in our definition of the
Fourier transform (\ref{phi_fourier}).   The spectral index is
\begin{equation}
\label{n_s_no_metric}
  n_s - 1 = 3 - 2\nu \cong 2\eta - 2\epsilon
\end{equation}
using (\ref{nu1}).

Given the solution (\ref{phi_k}) of the homogeneous equation, it is now trivial to construct the retarded Green function for equation 
(\ref{phi_mode}).  This may be written in terms of the free theory mode functions (\ref{phi_k}) as
\begin{eqnarray}
  G_k(\tau-\tau') &=& i \Theta(\tau-\tau') \left[ \,\,\,\phi_k(\tau) \phi_k^\star(\tau') 
                                                 -  \phi^\star_k(\tau) \phi_k(\tau')\,\,\, \right] \nonumber \\
  &=& \frac{i \pi}{4} \Theta(\tau-\tau') \sqrt{\tau\tau'}\left[\,\,\, H_{\nu}^{(1)}(-k\tau) H_{\nu}^{(1)}(-k\tau')^{\star} \right. \nonumber \\
  && \,\,\,\,\,\,\,\,\,\,\,\,\,\,\, \left. - \,\,\,  H_{\nu}^{(1)}(-k\tau)^{\star} H_{\nu}^{(1)}(-k\tau') \,\,\, \right] \label{green}
\end{eqnarray}

\subsection{Particular Solution: Rescattering Effects}

We now consider the particular solution of (\ref{phi_mode}).  This is readily constructed using the Green function (\ref{green})
as
\begin{eqnarray}
  && \xi_{\bf k}^\phi(\tau) =  \\
  && \,\,\,\,\,\,\, -\frac{g k_\star^2}{(2\pi)^{3/2}} \int d\tau' d^3k' G_k(\tau-\tau') \,
  a(\tau') t(\tau') \,\xi^\chi_{\bf k'}\xi^\chi_{\bf k-k'}(\tau') \nonumber \label{xi_k_particular}
\end{eqnarray}
Notice that this particular solution is statistically independent of the homogeneous solution (\ref{phi_annihilation}).  
In other words, the particular solution can be expanded in terms of the annihilation/creation operators $a_{\bf k}, a_{\bf k}^\dagger$ 
associated with the $\chi$ field, whereas the homogeneous solution is written in terms of the annihilation/creation operators $b_{\bf k}, b_{\bf k}^\dagger$ associated
with the inflaton vacuum fluctuations.  These two sets of operators commute with one another.

We will ultimately be interested in computing the $n$-point correlation functions of the particular solution (\ref{xi_k_particular}).  
For example, carefully carrying out the Wick contractions, the connected contribution to the 2-point function is
\begin{eqnarray}
&& \langle \xi^\phi_{\bf k_1} \xi^\phi_{\bf k_2}(\tau) \rangle = \frac{2 g^2 k_\star^4}{(2\pi)^{3}} \delta^{(3)}({\bf k_1}+{\bf k_2})
   \nonumber \\
&& \times \int d\tau' d\tau'' a(\tau')a(\tau'')t(\tau')t(\tau'')  G_{k_1}(\tau-\tau')  G_{k_2}(\tau-\tau'') \nonumber \\
&& \,\,\times \int d^3k' \chi_{k_1-k'}(\tau')\chi_{k_1-k'}^\star(\tau'')\chi_{k'}(\tau')\chi_{k'}^\star(\tau'') \label{2pt}
\end{eqnarray}
The power spectrum of $\delta\phi$ fluctuations generated by rescattering is then defined in terms of the 2-point function
in the usual manner
\begin{equation}
  \langle \xi^\phi_{\bf k}(t) \xi^\phi_{\bf k'}(\tau) \rangle \equiv
  \delta^{(3)}( {\bf k} + {\bf k'} ) \frac{2\pi^2}{k^3} a(\tau)^2 P_\phi^{\mathrm{resc}} \label{pwr_def}
\end{equation}
(The explicit factor of $a^2$ in the definition (\ref{pwr_def}) appears to cancel the factor of $a^{-1}$ in our convention
for Fourier transforms (\ref{phi_fourier}).)

The total power spectrum is simply the sum of the contribution from the vacuum fluctuations (\ref{P_phi_vac}) and the contribution from rescattering
(\ref{pwr_def}):
\begin{equation}
\label{Psum}
  P_\phi(k) = P_\phi^{\mathrm{vac}}(k) + P_\phi^{\mathrm{resc}}(k)
\end{equation}
There are no cross-terms, owing to the fact $a_{\bf k}$ and $b_{\bf k}$ commute.

\subsection{Renormalization}

We now wish to evaluate the 2-point correlator (\ref{2pt}).  
In principle, this is straightforward: first substitute the result (\ref{chi_soln}) 
for the $\chi_k$ modes and the result (\ref{green}) for the Green function into (\ref{2pt}), next evaluate the integrals.
However, there is a subtlety. 
The resulting power spectrum is formally infinite.  Moreover, the 2-point correlation function (\ref{2pt}) receives contributions 
from two distinct effects.  There is a contribution from particle production, which we are interested in.  However, there is also 
a contribution coming from quantum vacuum fluctuations of the $\chi$ field interacting non-linearly with the inflaton.  The latter
contribution would be present even in the absence of particle production, when $\alpha_k = 1$, $\beta_k=0$.

In order to isolate the effects of particle production on the inflaton fluctuations, we would like to subtract off the contribution to
the 2-point correlation function (\ref{2pt}) which is coming from the quantum vacuum fluctuations of $\chi$.  This subtraction
also has the effect of rendering the power spectrum (\ref{pwr_def}) finite, since it extracts the usual UV divergent contribution 
associated with the Minkowski-space vacuum fluctuations.

As a step towards renormalizing the 2-point correlation function of inflaton fluctuations from rescattering (\ref{2pt}), let us first
consider the simpler problem of renormalizing the 2-point function of the gaussian field $\chi$.
We defined the renormalized 2-point function in momentum space as follows:
\begin{eqnarray}
  \langle \xi^\chi_{k_1}(t_1) \xi^\chi_{k_2}(t_2) \rangle_{\mathrm{ren}} &=& 
  \langle \xi^\chi_{k_1}(t_1) \xi^\chi_{k_2}(t_2) \rangle \nonumber \\
  &&- \langle \xi^\chi_{k_1}(t_1) \xi^\chi_{k_2}(t_2) \rangle_{\mathrm{in}} \label{ren_rule}
\end{eqnarray}
In (\ref{ren_rule}) the quantity $\langle \xi^\chi_{k_1}(t_1) \xi^\chi_{k_2}(t_2) \rangle_{\mathrm{in}}$ is the contribution which would be present 
even in the absence of particle production, computed by simply taking the solution (\ref{chi_bog}) with $\alpha_k=1$, $\beta_k=0$.  Explicitly, we
have
\begin{equation}
  \langle \xi^\chi_{k_1}(t_1) \xi^\chi_{k_2}(t_2) \rangle_{\mathrm{in}} = \delta^{(3)}({\bf k_1} + {\bf k_2}) f_{k_1}(t_1) f_{k_2}^\star(t_2)
\end{equation}
where $f_k$ are the adiabatic solutions (\ref{f_k}).

To see the impact of this subtraction, let us consider the renormalized variance for the iso-inflaton field, $\langle\chi^2\rangle$.  Employing the 
prescription (\ref{ren_rule}) we have
\begin{eqnarray}
  \langle \chi^2(\tau,{\bf x}) \rangle_{\mathrm{ren}} &=& 
    \int \frac{d^3 k}{(2\pi)^3 a^2(\tau)}\left[  |\chi_k(\tau)|^2 - \frac{1}{2\omega_k(\tau)}   \right] \nonumber \\
    &=& \langle \chi^2(\tau,{\bf x}) \rangle - \delta_M \label{ren_var}
\end{eqnarray}
where $\delta_M$ is the contribution from the Coleman-Weinberg potential.  This proves that our prescription reproduces the scheme advocated
in \cite{beauty}.  The renormalized variance (\ref{ren_var}) is finite and may be computed explicitly using our solutions (\ref{chi_soln}).  
We find
\begin{equation}
\label{ren_var_explicit}
  \langle\chi^2(t,{\bf x}) \rangle_{\mathrm{ren}} \cong \frac{n_\chi a^{-3}}{g |\phi-\phi_0|}
\end{equation}
where
\begin{equation}
  n_\chi \equiv \int \frac{d^3k}{(2\pi)^{3}} n_k \sim e^{-\pi \mu^2 / k_\star^2} k_\star^3
\end{equation}
is the total co-moving number density of produced $\chi$ particles.  The result (\ref{ren_var_explicit}) was employed in \cite{sasaki} to quantify the
effect of backreaction on the inflaton condensate in the mean field treatment (\ref{mean}).  Hence, the renormalization scheme (\ref{ren_rule}) was implicit
in that calculation also.

At the level of the 2-point function, our renormalization scheme is tantamount to assuming that Coleman-Weinberg corrections are already absorbed into 
the definition of the inflaton potential, $V(\phi)$.  In general, such corrections might steepen $V(\phi)$ and spoil slow-roll inflation.  Here, we assume 
that this problem has already been dealt with, either by fine-tuning the bare inflaton potential or else by including extended SUSY (which can minimize 
dangerous corrections).  See also \cite{beauty} for a related discussion.  Note, also, that our renormalization procedure is equivalent to the quasi-particle
normal ordering scheme described in \cite{russian_text}.

Having established a scheme for remormalizing the 2-point function of the gaussian field $\chi$, it is now straightforward to 
consider higher order correlation functions.  We simply re-write the 4-point function as a product of 2-point functions 
using Wick's theorem.  Next, each Wick contraction is renormalized as (\ref{ren_rule}).  Applying this prescription to (\ref{2pt}) 
amounts to
\begin{eqnarray}
  \langle \xi^\phi_{k_1}(\tau)\xi^\phi_{k_2}(\tau) \rangle_{\mathrm{ren}} 
  &=& \frac{2 g^2 k_\star^4}{(2\pi)^3} \delta^{(3)}({\bf k_1}+{\bf k_2}) \nonumber \\
  &&\hspace{-28mm}\times \int d\tau'd\tau''t(\tau') t(\tau'') a(\tau')a(\tau'') G_{k_1}(\tau-\tau')  G_{k_2}(\tau-\tau'') \nonumber \\
  &&\hspace{-28mm} \times \int d^3k' \left[ \chi_{k_1-k'}(\tau')\chi_{k_1-k'}^\star(\tau'') - f_{k_1-k'}(\tau')f_{k_1-k'}^\star(\tau'')\right] 
              \nonumber \\
  &&\hspace{-17mm} \times \left[\chi_{k'}(\tau')\chi_{k'}^\star(\tau'')- f_{k'}(t')f_{k'}^\star(\tau'')\right]  \label{pwr_phi_ren}
\end{eqnarray}
where $f_k(t)$ are the adiabatic solutions defined in (\ref{f_k}).

\subsection{Power Spectrum}

We are now in a position to compute the renormalized power spectrum of inflation fluctuations generated by rescattering, $P_\phi^{\mathrm{resc}}(k)$.
We renormalize the 2-point correlator of the inflaton fluctuations generated by rescatter according to (\ref{pwr_phi_ren}) and extract the power spectrum
by comparison to (\ref{pwr_def}).  We have relegated the technical details to Appendix A and here we simply state the final result
\begin{widetext}
\begin{eqnarray}
  P_\phi^{\mathrm{resc}}(k) &=& \frac{g^2 k^3 k_\star}{16\pi^5} \left[ \,\,\,\,\,\,\,\, \frac{e^{-2\pi \mu^2 / k_\star^2}e^{-\pi k^2 / (2k_\star^2)}}{2\sqrt{2}}\left( I_2(k,\tau)^2 + |I_1(k,\tau)|^2   \right)   \right. \nonumber \\
 &+& \left[ e^{-\pi\mu^2/k_\star^2}e^{-\pi k^2/(4k_\star^2)} + \frac{e^{-2\pi\mu^2/k_\star^2}}{2\sqrt{2}}\,e^{-3\pi k^2/(8k_\star^2)} \right]\left( I_2(k,\tau)^2 - \mathrm{Re}\left[I_1(k,\tau) \right] \right)   \nonumber \\
  &+& \left.  \left[ \frac{8\sqrt{2}}{3\sqrt{3}} \,e^{-3\pi\mu^2/(2k_\star^2)} e^{-\pi k^2 / (3k_\star^2)} + \frac{4\sqrt{2}}{5\sqrt{5}} \,e^{-5\pi\mu^2/(2k_\star^2)}e^{-3 \pi k^2/(5k_\star^2)} \right]
                \mathrm{Im}\left[ I_1(k,\tau)I_2(k,\tau) \right]\,\,\,\,\,\,\,\, \right] \label{full_pwr_result}
\end{eqnarray}
\end{widetext}
where the functions $I_1$, $I_2$ are the curved space generalization of the characteristic integrals defined in \cite{ir}.  Explicitly we have
\begin{eqnarray}
  I_1(k,\tau) &=& \frac{1}{a(\tau)}\int d\tau' G_k(\tau-\tau') e^{i k_\star^2 t^2(\tau')} \label{I1_text}\\
  I_2(k,\tau) &=& \frac{1}{a(\tau)}\int d\tau' G_k(\tau-\tau') \label{I2_text}
\end{eqnarray}
The characteristic integral $I_2$ can be evaluated analytically, however, the resulting expression is not particularly enlightening.  Evaluation
of the integral $I_1$ requires numerical methods.  More details in Appendix A.  Equation (\ref{full_pwr_result}) is the main result of this section. 

\begin{figure}[htbp]
\bigskip \centerline{\epsfxsize=0.35\textwidth\epsfbox{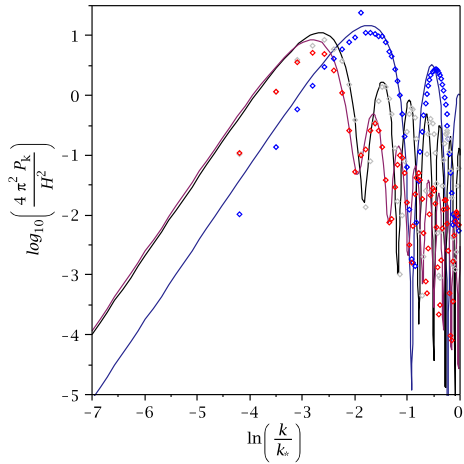}}
\caption{The power spectrum of inflaton modes induced by rescattering.
(normalized to the usual vacuum fluctuations) as a function of $\ln(k/k_\star)$, 
plotted for three representative time steps in the late-time evolution.  
For each time step we plot the analytical result (the solid line) and the data points obtained 
using lattice field theory simulations (diamonds).  The agreement between these two independent results
is evident.  For illustration, we have set $\mu^2=0$.}
\label{Fig:pwr_late}
\end{figure}

To test our analytical formalism, let us compare the result (\ref{full_pwr_result}) with the out-put of fully nonlinear HLattice simulations.
In Fig.~\ref{Fig:pwr_late} we plot our results for $P^{\mathrm{resc}}_\phi(k)$ as a function of $k$, for several time steps in the evolution.  
We have normalized  $P^{\mathrm{resc}}_\phi(k)$ to the amplitude of the usual vacuum fluctuations from inflation, 
$P^{\mathrm{vac}}_\phi(k) \sim H^2/(2\pi)^2$.  This figure illustrates the final stages of IR cascading; we see the peak of the bump-like
feature slide to $k \sim e^{-3} k_\star$, at which point the associated mode functions $\delta\phi_k$
have crossed the horizon and become frozen.  At later times in the evolution the peak of the feature
and also the IR tail ($\sim k^3$) remain fixed.  Modes associated with the UV end of the spectrum
are still inside the horizon and continue to evolve as $\delta\phi_k \sim a^{-1}$, which explains the
damping of the $k>e^{-2} k_{\star}$ part of the spectrum.  At late times, the shape of the feature that is frozen outside the horizon can be
very well approximated by the semi-analytic fitting function (\ref{P_fit}).

The agreement between our analytical formalism and the exact numerical results is quite evident from Fig.~\ref{Fig:pwr_late} and provides
a highly nontrivial check on our calculation.

\subsection{The Bispectrum}

So far, we have shown how to compute analytically the power spectrum generated by particle production, rescattering and IR 
cascading in the model (\ref{L}).  We found that IR cascading leads to a bump-like contribution to the primordial power
spectrum of the inflaton fluctuations.  However, this same dynamics must also have a nontrivial impact on nongaussian statistics,
such as the bispectrum.  Indeed, it is already evident from our previous analysis that the inflaton fluctuations generated by rescattering
may be significantly nongaussian.  From the expression (\ref{xi_k_particular}) we see that the particular solution (due to rescattering) is bi-linear is the 
gaussian field $\chi$.

We define the bispectrum of the inflaton field fluctuations in terms of the three point correlation function as
\begin{equation}
  \langle \xi_{\bf k_1}^\phi \xi_{\bf k_2}^\phi \xi_{\bf k_3}^\phi (\tau) \rangle  = (2\pi)^3 a^3(\tau) \delta({\bf k_1} + {\bf k_2} + {\bf k_3})
  B_\phi(k_i)
\label{B_phi}
\end{equation}
The factor $a^3$ appears in (\ref{B_phi}) to cancel the explicit factors of $a^{-1}$ in our convention (\ref{phi_fourier}) for the Fourier transform.
It is well-known that the nongaussianity associated with the usual quantum vacuum fluctuations of the inflaton is negligible 
\cite{riotto,maldacena,seerylidsey}, therefore, when evaluating the bispectrum (\ref{B_phi}) we consider only the particular solution 
(\ref{xi_k_particular}) which is due to rescattering.  Carefully
carrying out the Wick contractions, we find the following result for the renormalized 3-point function
\begin{widetext}
\begin{eqnarray}
  &&  \langle \xi_{\bf k_1}^\phi \xi_{\bf k_2}^\phi \xi_{\bf k_3}^\phi(\tau) \rangle_{\mathrm{ren}} = \frac{4 g^3 k_\star^6}{(2\pi)^{9/2}} \, \delta({\bf k_1} + {\bf k_2} + {\bf k_3}) \,
		\prod_{i=1}^{3} \int d\tau_i t(\tau_i)a(\tau_i) G_{k_i}(\tau-\tau_i) \nonumber \\
  && \,\,\,\,\, \times \int d^3p \left[ \chi_{k_1-p}(\tau_1)\chi_{k_1-p}^\star(\tau_2) - f_{k_1-p}(\tau_1)f_{k_1-p}^\star(\tau_2)   \right] 
			\left[   \chi_{k_3+p}(\tau_2)\chi_{k_3+p}^\star(\tau_3) - f_{k_3+p}(\tau_2)f_{k_3+p}^\star(\tau_3)    \right] \nonumber \\
  && \,\,\,\,\,\,\,\,\,\,\,\,\,\,\,\,\,\,\,\,\,\, \times \left[   \chi_{p}(\tau_1)\chi_{p}^\star(\tau_3) - f_{p}(\tau_1)f_{p}^\star(\tau_3)    \right] \nonumber \\ 
  && \,\,\,\,\,\,\,\,\,\,\,\,  + \,\,\,\,\, (k_2 \leftrightarrow k_3)\label{bispectrum_soln}
\end{eqnarray}
\end{widetext}
where the modes $\chi_k$ are defined by (\ref{chi_bog}) and $f_k$ are the adiabatic solutions (\ref{f_k}).  On the last line of (\ref{bispectrum_soln})
we have labeled schematically terms which are identical to the preceding three lines, only with $k_2$ and $k_3$ interchanged.  One may verify
that this expression is symmetric under interchange of the momenta $k_i$ by changing dummy variables of integration.

\subsection{Estimating the Shape of the Bispectrum}

It is straightforward (but tedious) to plug the expressions (\ref{f_k_approx}) and (\ref{chi_soln}) into (\ref{bispectrum_soln}) and evaluate
the integrals.  The resulting expression is extremely cumbersome and not particularly enlightening.  We are interested here in extracting some
information about the shape of the bispectrum $B_\phi(k_i)$.  For this purpose, it suffices to work in the flat-space limit, $H\rightarrow 0$.  This will
give a reasonable qualitative picture of the full result since the entire process of IR cascading occurs over a time scale somewhat shorter than
the expansion time.  In \cite{ir} this same approximation was employed to study the power spectrum from IR cascading and was found to reproduce
the $H\not= 0$ results to good accuracy.  

A detailed calculation of $B_\phi(k_i)$ has been relegated to appendix B.  Here we simply provide a representative contribution, in order to give
a rough sense of the qualitative behaviour:
\begin{equation}
\label{B_schematic}
  B_\phi(k_i) \sim C \prod_{i=1}^3 e^{-\pi k_i^2 / (3k_\star^2)} \left[\frac{1-\cos\left(\sqrt{k_i^2 + m^2} \, t\right)}{(k_i^2 + m^2)}\right]
\end{equation}
for some constant $C$.
This expression captures some of the qualitative features of the full result, in particular the dynamical cascading of nongaussianity into the 
IR to generate a localized bispectrum feature.  It should be stressed that (\ref{B_schematic}) is a heuristic estimate and \emph{not} a fitting
function nor a systematic approximation to the full result.  Hence, equation (\ref{B_schematic}) should not be used to make quantitative 
predictions of any kind.

As anticipated, our expression for $B_\phi(k_i)$ peaks only over when all wavenumbers are close to the characteristic scale corresponding to the 
location of the bump in (\ref{P_fit}).  Therefore, particle production and IR cascading leads to a localized nongaussian feature in the bispectrum, rather 
than the nearly scale-invariant signatures that are usually considered.  We will discuss the phenomenology of this new type of nongaussianity in a
forthcoming publication \cite{inprog}.

Now we would like to attempt to characterize the shape of the nongaussianity from particle production and IR cascading.
To this end we define a ``shape function'' $S(k_i)$ as follows
\begin{equation}
\label{S}
  S(k_i) = N^{-1} (k_1 k_2 k_3)^2 B_\phi(k_i)
\end{equation}
where $N$ is a normalization factor which will not concern 
us.\footnote{As we argued in section \ref{sec:NG}, the size of the nongaussianity in this model is most naturally quantified by 
evaluating the cummulants.  Here we are interested \emph{only} in discussing the shape of this novel type of nongaussianity.}
The function $S(k_i)$ has the advantage that the strong $k^6$ running of the bispectrum is extracted.  Hence, any residual 
scaling behaviour displayed by $S(k_i)$ must be a result of nonlinear interactions; see also \cite{chen1,chen2}. 

\begin{figure}[tbp]
\begin{center}
\includegraphics[width=1.6in]{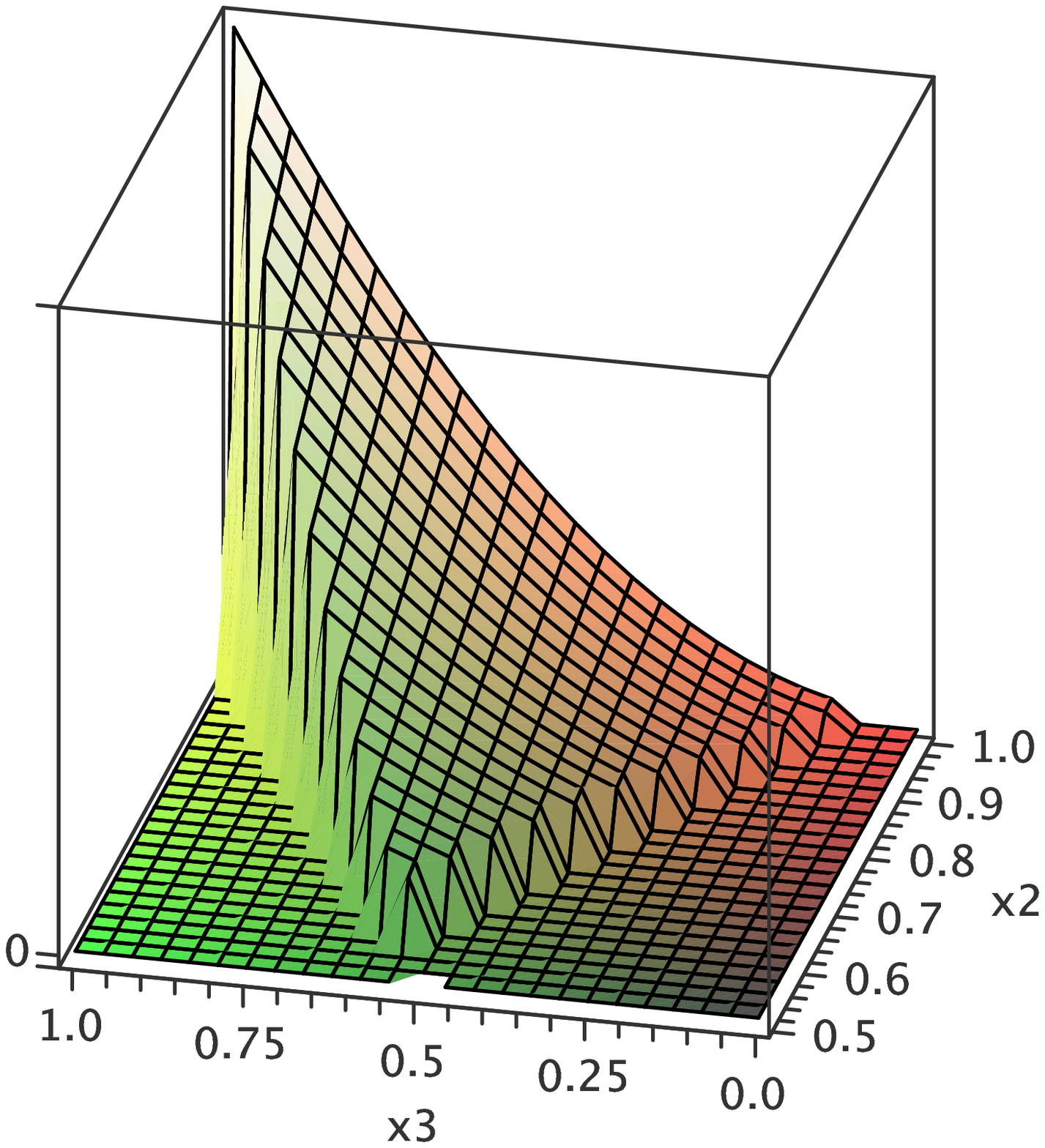}
\includegraphics[width=1.7in]{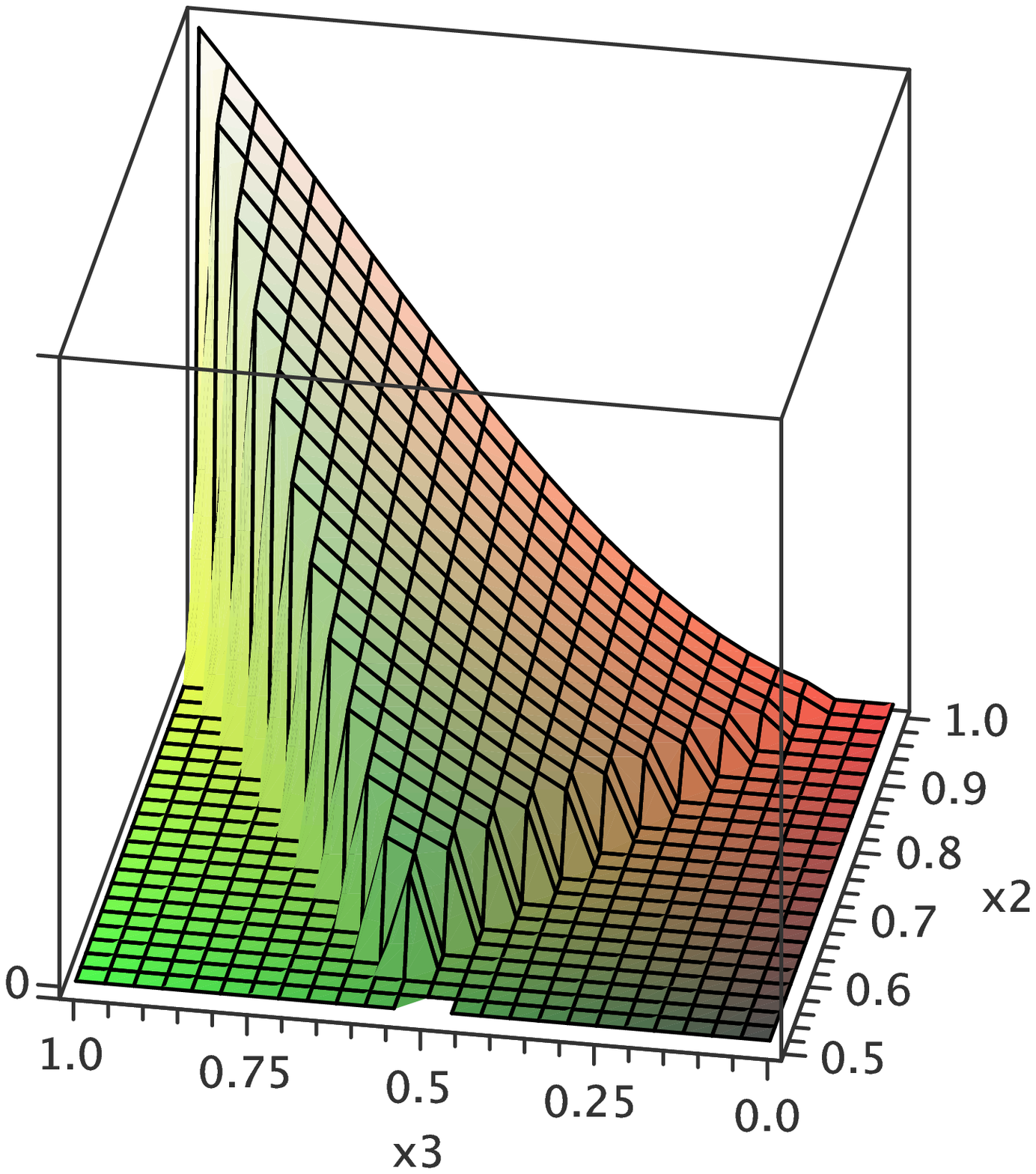}
\includegraphics[width=1.6in]{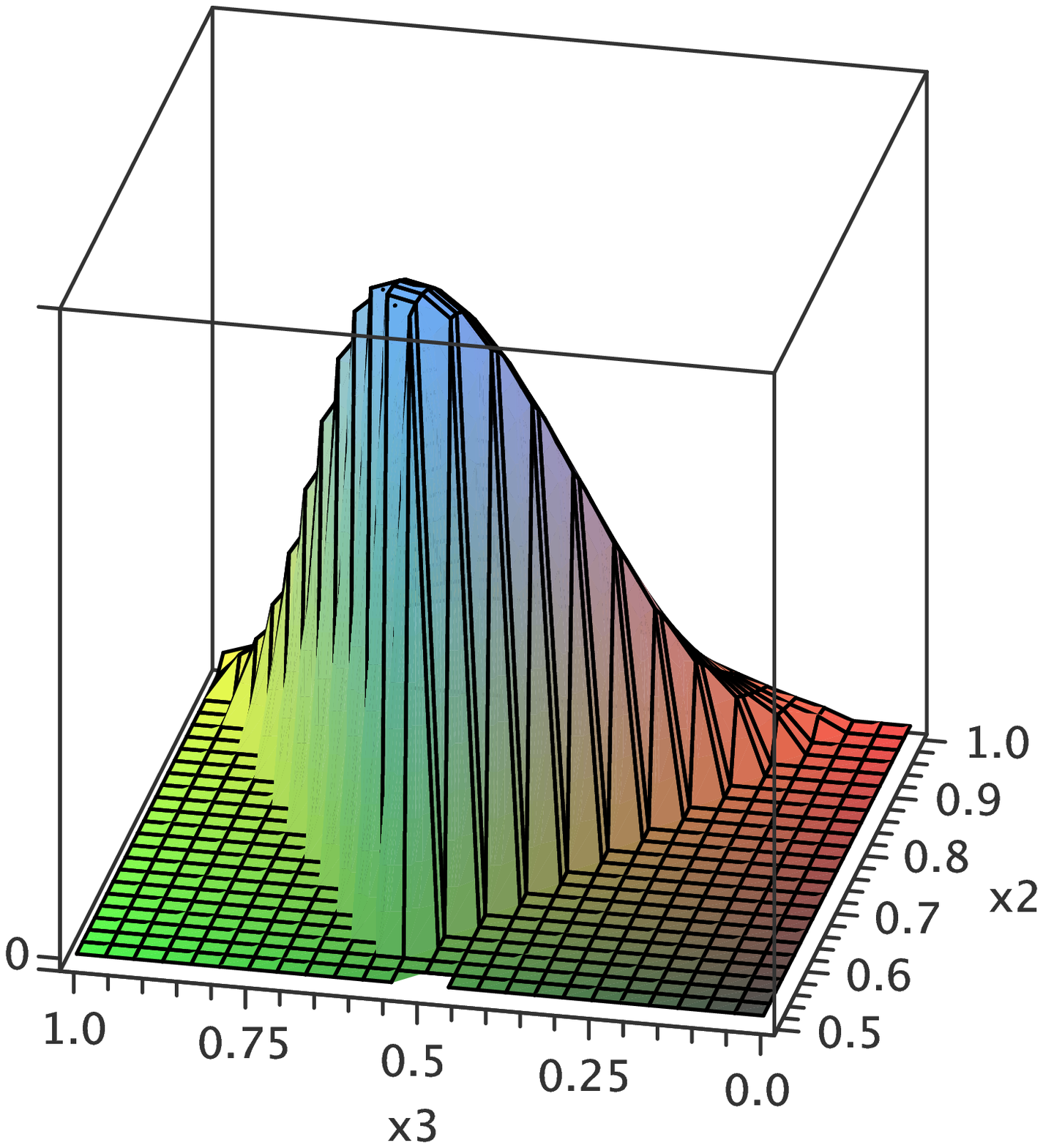}
\includegraphics[width=1.6in]{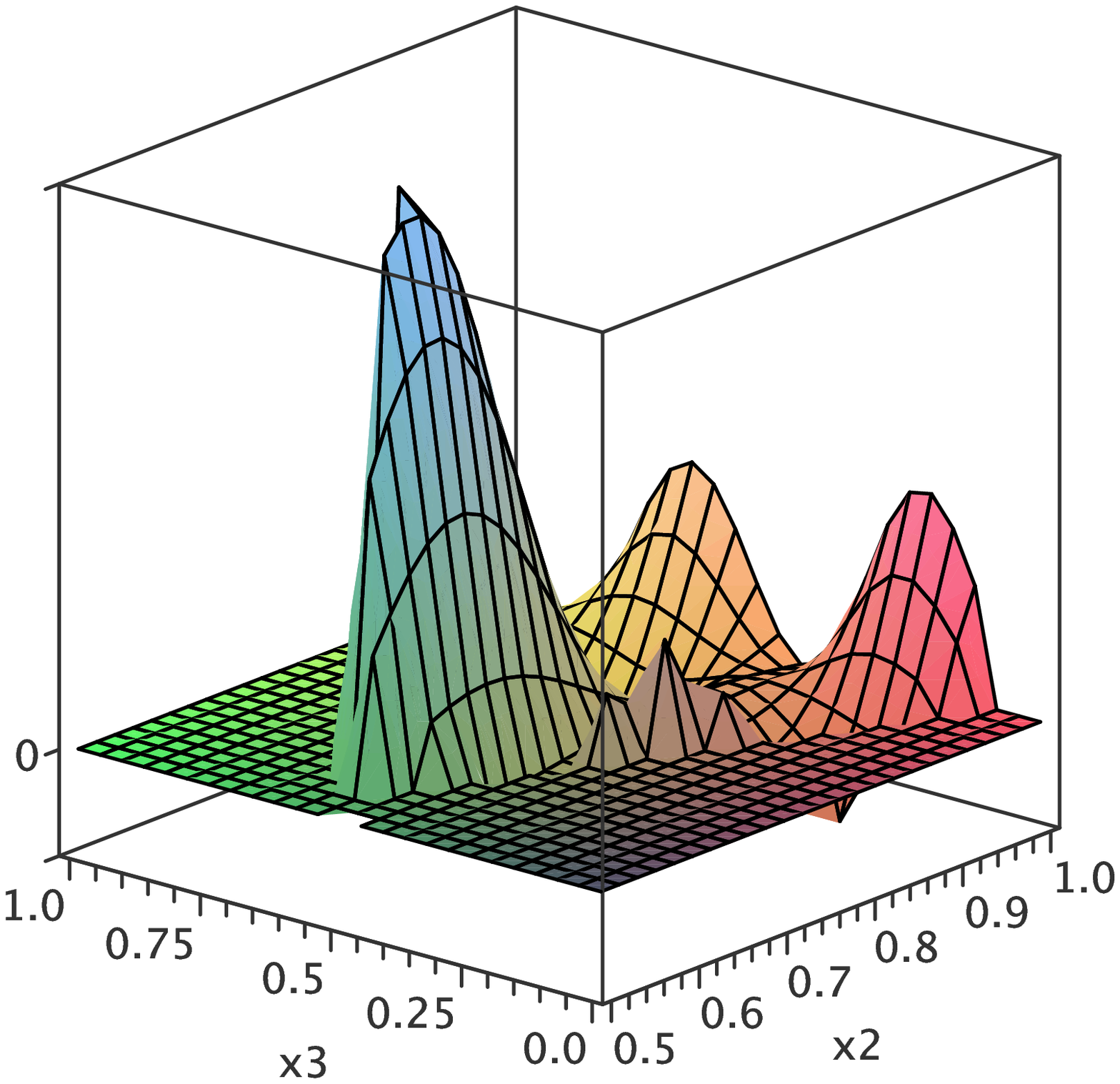}
\caption{The shape function $S(k,k x_2, k x_3)$, defined by (\ref{S}), as a function of the dimensionless quantities $x_2,x_3$ which parametrize
the shape of the triangle.  The upper left panel corresponds to $k = e^{-1}k_{\mathrm{bump}}$, the upper right panel is $k=k_{\mathrm{bump}}$,
the lower left panel is $k=e^{+1} k_{\mathrm{bump}}$ and the lower right panel is $k=e^{+2}k_{\mathrm{bump}}$.
In the IR 
($k \leq k_{\mathrm{bump}}$) 
the shape of the bispectrum is similar to the equilateral shape, however, there is also some support on 
flattened triangles near 
$k \sim e^{+1} k_{\mathrm{bump}}$.  
At larger values of $k$ the shape is unlike any other template proposed 
in the literature.  For illustration we have chosen $\mu^2=0$.
}\label{Fig:bispectrum_shape}
\end{center}
\end{figure}

Symmetry of the bispectrum under permutations of momenta implies that we can focus only on the region $k_1 \geq k_2 \geq k_3$, to avoid
counting the same configuration twice.  Moreover, the triangle inequality implies that $1-\frac{k_2}{k_1} \leq \frac{k_3}{k_1}$.  Therefore
we can completely specify the shape of the bispectrum for a given size of triangle $k$ by plotting $S(k,kx_2,kx_3)$ in the region
$x_3 \leq x_2 \leq 1$ and $1-x_2 \leq x_3$.  (See also \cite{babich}.)  Because our bispectrum is very far from scale-invariant, it follows that this shape function
is sensitive to the choice of $k$.  Therefore, in Fig.~\ref{Fig:bispectrum_shape} we choose several representative choices: 
$\ln (k / k_{\mathrm{bump}}) = -1,0,1,2$.  


We see that a rich array of shape are possible: for $k \lsim k_{\mathrm{bump}}$ the bispectrum is qualitatively similar to the equilateral model,
however, at slightly larger $k$ there is considerable support on flattened triangles also.  
Note that for $k \gsim 7.4\, k_{\mathrm{bump}}$ the shape of the bispectrum is extremely unusual and is not easily comparable to any shape
that has been proposed in previous literature.

\section{Cosmological Perturbation Theory}
\label{sec:perts}

In section \ref{sec:theory} we developed an analytical theory of particle production and IR cascading during inflation which is in very good
agreement with nonlinear lattice field theory simulations.  However, this formalism suffers from a neglect of metric perturbations and, consequently, we
were unable to rigorously discuss the gauge invariant curvature perturbation $\zeta$. Hence, the reader may be concerned about gauge ambiguities in 
our results.  In this 
section we address such concerns, showing that metric perturbations may be incorporated in a straightforward manner and that their consistent inclusion does not change our results in any significant
way.  We will do so by showing explicitly that, with appropriate choice of gauge, equations (\ref{delta_phi}) and (\ref{chi_expanding}) for the fluctuations of the inflaton
and iso-inflaton still hold, to first approximation.  We will also go beyond our previous analysis by explicitly showing that in this same gauge the spectrum of the curvature fluctuations, $P_\zeta$,
is trivially related to the spectrum of inflaton fluctuations, $P_\phi$ (and similarly for the bispectrum).  

To render the analysis tractable we would like to take full advantage of the results derived in the last section.  To do so, we employ the Seery
et al.\ formalism for working directly with the field equations \cite{seery} and make considerable use of results derived by
Malik in \cite{malik1,malik2}. (Note that our notations differ somewhat from those employed by Malik.
The reader is therefore urged to take care in comparing our formulae.)

We expand the inflaton and iso-inflaton fields up to second order in perturbation theory as
\begin{eqnarray}
  \phi(\tau,{\bf x}) &=& \phi(\tau) + \delta_1\phi(\tau,{\bf x}) + \frac{1}{2} \delta_2\phi(\tau,{\bf x}) \\
  \chi(\tau,{\bf x}) &=& \delta_1\chi(\tau,{\bf x}) + \frac{1}{2} \delta_2\chi(\tau,{\bf x})
\end{eqnarray}
The perturbations are defined to average to zero $\langle \delta_n\phi\rangle =\langle\delta_n\chi\rangle = 0$ so that 
$\langle \phi(t,{\bf x}) \rangle = \phi(t)$ and $\langle \chi(t,{\bf x}) \rangle = 0$.  (The condition $\langle\chi\rangle = 0$ is 
ensured by the fact that $m_\chi \gg H$ for nearly the entire duration of inflation.)

We employ the flat slicing and threading throughout this section.  With this gauge choice the perturbed metric takes the form
\begin{eqnarray}
  g_{00} &=& -a^2 (1 + 2\psi_1 + \psi_2) \\
  g_{0i} &=& a^2 \partial_i \left[ B_1 + \frac{1}{2} B_2  \right] \\
  g_{ij} &=& a^2 \delta_{ij}
\end{eqnarray}
so that spatial hyper-surfaces are flat.  Note also that in this gauge the field perturbations $\delta_n\phi$, $\delta_n\chi$
coincide with the Sasaki-Mukhanov variables \cite{SMvariable} at  both first and second order.


\subsection{Gaussian Perturbations}

In \cite{malik1} Malik has derived closed-form evolution equations for the field perturbations $\delta_n \phi$, $\delta_n \chi$ at both first ($n=1$) and 
second ($n=2$) order in perturbation theory.  Let us first study the gaussian perturbations.  The closed-form Klein-Gordon equation for $\delta_1\phi$ derived in \cite{malik1}
can be written as
\begin{eqnarray}
&&  \delta_1 \phi'' + 2\sH \delta_1 \phi' - \grad^2 \delta_1\phi + \left[ a^2 m^2  - 3 \left(\frac{\phi'}{M_p}\right)^2   \right]\delta_1 \phi  \nonumber \\
&&\,\,\,\,= 0  \label{d1phi}
\end{eqnarray}
Following our previous analysis we expand
the first-order perturbation in terms of annihilation/creation operators as
\begin{equation}
  \delta_1 \phi(t,{\bf x}) = \int \frac{d^3k}{(2\pi)^{3/2}} \left[ b_{\bf k} \frac{\delta_1\phi_k(\tau)}{a(\tau)} e^{i {\bf k}\cdot {\bf x}} + \mathrm{h.c.}  \right]
\end{equation}
where $\mathrm{h.c.}$ denotes the Hermitian conjugate of the preceding term and we draw the attention of the reader to the 
the explicit factor of $a^{-1}$ in our definition of the Fourier transform.  Working to leading order in slow roll parameters we have
\begin{equation}
\label{d1phi_k}
  \delta_1\phi_k'' + \left[k^2 + \frac{1}{\tau^2}\left( -2 + 3\eta - 9\epsilon  \right)\right]\delta_1\phi_k = 0
\end{equation}
This equation coincides exactly with (\ref{simple_inf_eqn}) and the properly normalized solutions again take the form (\ref{phi_k}).
The only difference is that the order of the Hankel function, $\nu$, is now given by
\begin{equation}
\label{nu2}
  \nu \cong \frac{3}{2} - \eta + 3 \epsilon
\end{equation}
rather  than by equation (\ref{nu1}).  The power spectrum of the gaussian fluctuations is, again, given by (\ref{P_phi_vac}).  The correction
to the order of the Hankel function $\nu$ translates into a correction to the spectral index: instead of (\ref{n_s_no_metric}) we now have
\begin{equation}
  n_s - 1 = 2\eta - 6\epsilon
\end{equation}
which is precisely the standard result \cite{riotto_rev}.

Thus, as far as the quantum vacuum fluctuations of the inflaton are concerned, the only impact of consistently including metric perturbations
is an $\mathcal{O}(\epsilon)$ correction to the spectral index $n_s$.

Let us now turn our attention to the first order fluctuations of the iso-inflaton.  The closed-form Klein-Gordon equation for $\delta_1\chi$ derived 
in \cite{malik1} can be written as
\begin{equation}
  \delta_1 \chi'' + 2\sH \delta_1 \chi' - \grad^2 \delta_1\chi + a^2\left[ \mu^2 +  k_\star^4 t^2(\tau) \right]\delta_1 \chi = 0
\end{equation}
This coincides \emph{exactly} with equation (\ref{chi_expanding}), which we have already solved.  The fact that linear perturbations of $\chi$
do not couple to the metric fluctuations follows from the condition $\langle \chi \rangle  = 0$.

\subsection{Nongaussian Perturbations}

Now let us consider now the second order perturbation equations. The closed-form Klein-Gordon equation for $\delta_2\phi$ derived in \cite{malik1}
can be written as
\begin{eqnarray}
&&  \delta_2 \phi'' + 2\sH \delta_2 \phi' - \grad^2 \delta_2\phi + \left[ a^2 m^2  - 3 \left(\frac{\phi'}{M_p}\right)^2   \right]\delta_2 \phi \nonumber \\
&&\,\,\,\, = J(\tau,{\bf x}) \label{d2phi}
\end{eqnarray}
As usual, the left-hand-side is identical to the first order equation (\ref{d1phi}) while the source term $J$ is constructed from a bi-linear 
combination of the first order quantities $\delta_1\phi$ and $\delta_1\chi$.  In order to solve equation (\ref{d2phi}) we require explicit
expressions for the Green function $G_k$ and the source term $J$.  The Green function is trivial for the case at hand; it is still given
by our previous result (\ref{nu1}), provided one takes into account the fact that the order of the Hankel functions $\nu$ is now given by (\ref{nu2}), 
rather than (\ref{nu1}).  In other words, the Green function for the nongaussian perturbations (\ref{d2phi}) differs from the result obtained neglecting 
metric perturbations only by $\mathcal{O}(\epsilon)$ corrections.

Next, we would like to consider the source term, $J$, appearing in (\ref{d2phi}).  Schematically, we can split the source into contributions bi-linear in 
the gaussian inflaton fluctuation $\delta_1\phi$ and contributions bi-linear in the iso-inflaton $\delta_1\chi$:
\begin{equation}
\label{Jsplit}
  J = J_{\phi} + J_{\chi}
\end{equation}
The contribution $J_\phi$ would be present even in the absence of the iso-inflaton.  These correspond, physically, to the usual nongaussian corrections
to the inflaton vacuum fluctuations coming from self-interactions.  This contribution to the source is well-studied in the literature and is known to contribute
negligibly to the bispectrum \cite{seery}.  Thus, in what follows, we will ignore $J_\phi$.

On the other hand, the contribution $J_\chi$ appearing in (\ref{Jsplit}) depends only on the iso-inflaton fluctuations $\delta_1\chi$.  This contribution can be understood, physically,
as generating nongaussian inflaton fluctuations $\delta_2\phi$ by rescattering of the produced $\chi$ particles off the condensate.  Hence, the contribution $J_\chi$ may source
large nongaussianity and is most interesting for us.  It is straightforward to compute $J_\chi$ explicitly for our model using the general results of \cite{malik1}.  We find
\begin{eqnarray}
  J_\chi &=& -2 a^2 g^2 (\phi-\phi_0) (\delta_1\chi)^2 \nonumber \\
     && \pm \frac{\sqrt{2\epsilon}}{M_p} \left[ \,\,\,\,   -a^2 \left( \mu^2 + g^2 (\phi-\phi_0)^2 \right) (\delta_1\chi)^2 \right. \nonumber \\
      && \,\,\,\, \left. - \frac{1}{2} (\grad\delta_1 \chi)^2   - \frac{1}{2} (\delta_1\chi')^2         \right. \nonumber \\
     && \,\,\,\, + \nabla^{-2}\left(   \partial_i(\delta_1\chi)\grad^2\partial^i(\delta_1\chi)  + \grad^2(\delta_1\chi)\grad^2(\delta_1\chi)   \right. \nonumber \\ 
     &&\,\,\,\, \left. \left. + \delta_1\chi' \grad^2 \delta_1\chi + (\grad\delta_1\chi')^2 \,\,\,\,  \right) \,\,\,\, \right] \label{J} 
\end{eqnarray}
where the upper sign is for $\phi' > 0$, the lower sign is for $\phi' < 0$.  
Notice that the contributions to $J_\chi$ on the fourth and fifth line of (\ref{J}) contain the inverse spatial Laplacian $\nabla^{-2}$ and are thus nonlocal.  
These terms all contain at least as many gradients as inverse gradients and hence the large
scale limit is well-defined.  In \cite{vernizzi} it was argued that these terms nearly always contribute negligibly to the curvature perturbation on large 
scales.

Let us now examine the structure of the iso-inflaton source $J_\chi$, equation (\ref{J}).  The first line of (\ref{J}) goes like $a^2g^2(\phi-\phi_0)(\delta_1\chi)^2$.  
This coincides exactly with the source term in equation (\ref{delta_phi}) which was already studied in section \ref{sec:theory}.  On the other hand, the terms on 
the second, third, fourth and fifth lines of (\ref{J}) are new.  These represent corrections to IR cascading which result from the consistent inclusion of metric perturbations.
We will now argue that these ``extra'' terms are negligible as compared to the first line.  
If we denote the energy density in gaussian iso-inflaton fluctuations as $\rho_\chi \sim m_\chi^2 (\delta_1\chi)^2$ then, by inspection, we see that
the first line of (\ref{J}) is parametrically of order $\rho_\chi / |\phi-\phi_0|$ while the remaining terms are or order $\sqrt{\epsilon} \rho_\chi / M_p$.
Hence, we expect the first term to dominate for the field values $\phi \cong \phi_0$ which are relevant for IR cascading.  This suggests that the dominant
contribution to $J_\chi$ is the term which we have already taken into account in section \ref{sec:theory}.

Let us now make this argument more quantitative.  We assume that $\mu^2 \lsim k_\star^2$, since otherwise particle production effects are exponentially
suppressed.  Inspection reveals that the only ``new'' contribution to (\ref{J}) which has any chance of competing with the ``old'' term $a^2g^2(\phi-\phi_0)(\delta_1\chi)^2$ 
is the one proportional to $\sqrt{\epsilon} a^2 g^2 (\phi-\phi_0)^2 (\delta_1\chi)^2 / M_p$ (on the second line).  This new correction has the
possibility of becoming significant because it grows after particle production, as $\phi$ rolls away from $\phi_0$.  This growth, which reflects
the fact that the energy density in the $\chi$ particles increases as they become more massive, cannot persist indefinitely.  Within a few
$e$-foldings of particle production the iso-inflaton source term must behave as $J_\chi \sim a^{-3}$, corresponding to the volume dilution of non-relativistic
particles.  Hence, in order to justify the analysis of section \ref{sec:theory} we must check that the term 
\begin{equation}
  J_{\mathrm{new}} \sim \frac{\sqrt{\epsilon}}{M_p} a^2 g^2 (\phi-\phi_0)^2 (\delta_1\chi)^2
\end{equation}
does not dominate over the term which we have already considered
\begin{equation}
  J_{\mathrm{old}} \sim a^2 g^2 (\phi-\phi_0) (\delta_1\chi)^2
\end{equation}
during the relevant time $H \Delta t = \mathcal{O}(1)$ after particle production.  It is straightforward to show that
\begin{equation}
  \frac{J_{\mathrm{old}}}{J_{\mathrm{new}}} \sim \frac{M_p}{\sqrt{\epsilon}} \frac{1}{\phi-\phi_0}
  \sim \frac{M_p H}{\dot{\phi}\sqrt{\epsilon}}\,\frac{1}{N} \sim \frac{1}{\epsilon}\, \frac{1}{N}
\end{equation}
where $N=Ht$ is the number of $e$-foldings elapsed from particle production to the time when IR cascading has completed.  Hence, 
$N = \mathcal{O}(1)$ and we conclude that the second, third, fourth and fifth lines of (\ref{J}) are (at least) slow roll suppressed as compared 
to the first line.  

In summary, we have shown that consistent inclusion of metric perturbations yields corrections to the inflaton fluctuations $\delta\phi$
which fall into two classes:
\begin{enumerate}
  \item  Slow-roll suppressed corrections to the inflaton vacuum fluctuations $\delta_1\phi$ (these amount to changing the definition of $\nu$
            in the solution (\ref{phi_k})).  These corrections have two physical effects.  First, they yield an $\mathcal{O}(\epsilon)$ correction to the
            spectral index. Second, they modify the propagator $G_k$ by an $\mathcal{O}(\epsilon)$ correction. 
  \item Corrections to the source $J$ for the nongaussian inflaton perturbation $\delta_2\phi$.  These corrections
           are the second, third and fourth lines of (\ref{J})) which, as we have seen, are slow roll suppressed.
\end{enumerate}
It should be clear that neither of these corrections alters our previous analysis in any significant way.

\subsection{Correlators}

So far, we have shown that a consistent inclusion of metric perturbations does not significantly alter our previous results for the field perturbations.
Specifically, $\delta_1\chi$ is identical to our previous solution of equation (\ref{chi_expanding}) for the iso-inflaton, while $\delta_1\phi$ coincides 
with the homogeneous solution of equation (\ref{delta_phi}), up to slow-roll corrections.  At second order in perturbation theory, we have seen that
\[
  \delta_2 \phi = \int d^4 x' G(x-x') J_\chi(x') + \mathcal{O}\left[(\delta_1\phi)^2\right]
\]
To leading order in slow roll $J_\chi \cong -2 a^2 g^2 (\phi-\phi_0)(\delta_1\chi)^2$ and the first term coincides with our previous result for the particular 
solution of  equation (\ref{delta_phi}).  The terms of order $(\delta_1\phi)^2$ represent nongaussian corrections to the vacuum fluctuations from inflation 
(coming from self-interactions of $\delta\phi$ and the nonlinearity of gravity). These would be present even in the absence of particle production, and are known 
to have a negligible impact on the spectrum and bispectrum \cite{seery}.

We are ultimately interested in the connected $n$-point correlation functions of $\delta\phi$.  For example, the 2-point function $\langle (\delta\phi)^2 \rangle$ get a contribution
of the form $\langle (\delta_1\phi)^2 \rangle$ which gives the usual nearly scale invariant large-scale power spectrum from inflation.  The cross term 
$\langle\delta_1\phi \delta_2\phi \rangle$ is of order $\langle (\delta_1\phi)^4 \rangle$ and represents a negligible ``loop'' correction to the scale-invariant
spectrum from inflation.  (The cross term does not involve the iso-inflaton since $\delta_1\phi$ and $\delta_1\chi$ are statistically independent.)
Finally, there is a contribution $\langle (\delta_2\phi)^2 \rangle$ which involves terms of order $\langle \chi^4 \rangle$ coming from rescattering and terms of order
$\langle (\delta_1\phi)^4 \rangle$ which represent (more) loop corrections to the scale-invariant spectrum from inflation.  Thus, we can schematically write
\[
  P_\phi(k) = P_\phi^{\mathrm{vac}}(k) \left[1 + (\mathrm{loops})\right] + P_\phi^{\mathrm{resc}}(k)
\]
Here $P_\phi^{\mathrm{vac}} \sim k^{n_s-1}$ is the usual nearly scale invariant spectrum from inflation and $P_\phi^{\mathrm{resc}}$ is the bump-like contribution
from rescattering and IR cascading which we have studied in the previous section.  The ``loop'' corrections to $P^{\mathrm{vac}}(k)$ have been studied in detail
in the literature (see, for example, \cite{loop_corrections,loops1,loops2,loops3}) and are known to be negligible in most models.

We can also make a similar schematic decomposition of the bispectrum by considering the structure of the 3-point correlator $\langle (\delta\phi)^3 \rangle$.  Following
our previous line of reasoning, it is clear that the dominant contribution comes from rescattering and is of order $\langle \chi^6 \rangle$.  The terms involving 
$\langle(\delta_1\phi)^3\rangle$, on the other hand, represent the usual nongaussianity generated during single field slow roll inflation and are known to be small \cite{seery}.

\subsection{The Curvature Perturbation}

Ultimately one wishes to compute not the field perturbations $\delta_n\phi$, $\delta_n\chi$, but rather the gauge invariant curvature fluctuation, $\zeta$.  We expand this in perturbation
theory in the usual manner
\begin{equation}
  \zeta  = \zeta_1 + \frac{1}{2}\zeta_2
\end{equation}
In \cite{malik2} Malik has derived expressions for the large scale curvature perturbation 
in terms of the Sasaki-Mukhanov variables at both first and second order in perturbation theory.  
We remind the reader that in the flat slicing (which we employ) the Sasaki-Mukhanov variable for each field simply coincides with the field
perturbation (\emph{i.e.}~- $Q_\phi = \delta \phi$ and $Q_\chi = \delta\chi$). 

At first order in perturbation theory the iso-inflaton does not contribute to the curvature perturbation (since $\langle\chi\rangle = 0$) and we have
\begin{equation} 
\label{zeta1}
  \zeta_1 = -\frac{\sH}{\phi'}\delta_1\phi
\end{equation}

At second order in perturbation theory the expression for the curvature perturbation is more involved.  Using the results of \cite{malik2} and working to 
leading order in slow roll parameters we find\footnote{We have dropped a spurious additive $2\zeta_1^2$ which stems from using the Malik
and Wands \cite{M&W} definition of the curvature perturbation, rather than the definition employed by Lyth and Rodriguez \cite{L&R} and also by
Maldacena \cite{maldacena}.  (See also \cite{preheatNG}.)}
\begin{eqnarray}
  \zeta_2 &\cong& -\frac{\sH}{\phi'}\left[ \delta_2\phi - \frac{\delta_2\phi'}{3\sH}  \right] \label{zeta2} \\
          &+& \frac{1}{3(\phi')^2}\left[ (\delta_1\chi')^2 + a^2\left( \mu^2 + g^2 v^2 t^2(\tau)\right)(\delta_1\chi)^2   \right] \nonumber \\
          &+& \frac{1}{3(\phi')^2}\left[ (\delta_1\phi')^2 + a^2 m^2 (\delta_1\phi)^2   \right] \nonumber 
\end{eqnarray}
Let us discuss the various contributions to this equation.  The third line contributes to the nongaussianity of the vacuum fluctuations 
during inflation.  These terms are known to be negligible \cite{riotto,maldacena,seerylidsey,seery} and, indeed, one may explicitly verify that
(\ref{zeta2}) would predict $f_{NL} \sim \mathcal{O}(\epsilon,\eta)$ in the absence of particle production.

Next, we consider the
second line of (\ref{zeta2}).  This represents the direct contribution of the gaussian fluctuations $\delta_1\chi$ to the curvature perturbation.  
This contribution is tiny since the $\chi$ particles are extremely massive for nearly the entire duration of inflation and hence $\delta_1\chi \sim a^{-3/2}$
(see also \cite{trapped} for a related discussion).  The smallness of this contribution to $\zeta$ can be understood physically by noting that the 
super-horizon iso-curvature fluctuations in our model are negligible.

Finally, let us consider the contribution on the first line of (\ref{zeta2}).  This contribution is the most interesting.  
To make contact with observations we must compute the curvature perturbation at late times and on large scales.
In section \ref{sec:theory} we have already shown that $\delta_n\phi$ is constant on large scales and at late times for both $n=1$ and $n=2$.
This is the expected result: the curvature fluctuations are frozen far outside the horizon  and in the absence of entropy perturbations 
\cite{Sconserved}.\footnote{Note that, in some cases, the curvature fluctuations may evolve significantly after horizon exit \cite{liddle,sasaki2}.
(See also \cite{jain2}.)
This is a concern in models where there are significant violations of slow-roll.  In \cite{ir} we have already shown that the transient
violation of slow roll has a negligible effect on the curvature fluctuations in our model; see also \cite{ng_rev}.  Hence, the result 
$\zeta_n \sim \delta_n\phi \sim \mathrm{const}$ far outside
the horizon is consistent with previous studies.}  Hence $\delta_2\phi'$ is completely negligible and the first term on the first line of (\ref{zeta2}) must 
dominate over the second term.
We conclude that ,at late times and on large scales, the second order curvature perturbation is very well approximated by
\begin{equation}
\label{zeta2_approx}
  \zeta_2 \cong -\frac{\sH}{\phi'}\delta_2\phi + \cdots
\end{equation}

In summary, we have shown that the power spectrum of curvature fluctuations from inflation in the model (\ref{L}) is trivially related to the power spectrum
of inflaton fluctuations
\begin{equation}
\label{P_zeta_prop}
  P_\zeta(k) \cong \frac{H^2}{\dot{\phi}^2} P_\phi(k) = \frac{1}{2\epsilon M_p^2} P_\phi(k)
\end{equation}
at both first and second order in cosmological perturbation theory.  This relation is valid at late times and for scales far outside the horizon.  The curvature spectrum
(\ref{P_zeta_prop}) may be written as
\begin{equation}
\label{P_zeta_decomp}
  P_\zeta(k) = P_\zeta^{\mathrm{vac}}(k) \left[ 1 + (\mathrm{loops}) \right] + P_\zeta^{\mathrm{resc}}(k)
\end{equation}
The power spectrum of the inflaton vacuum fluctuations agrees with the usual result obtained in linear theory \cite{riotto_rev} 
\begin{equation}
  P_{\zeta}^{\mathrm{vac}}(k) \cong \frac{H^2}{8\pi^2 \epsilon M_p^2}\left(\frac{k}{aH}\right)^{2\eta-6\epsilon}
\end{equation}
In (\ref{P_zeta_decomp}) we have schematically labeled the corrections arising from the third line of (\ref{zeta2}) and the source $J_\phi$ as ``loop''.
These are nongaussian corrections to the inflaton vacuum fluctuations arising from self-interactions of the inflaton and also the nonlinearity of gravity.  Such corrections are negligible.
The most interesting contribution to the power spectrum (\ref{P_zeta_decomp}) is due to rescattering, $P_\zeta^{\mathrm{resc}}(k)$.  This quantity is proportional to our previous result (\ref{full_pwr_result}).

In passing, notice that the bispectrum $B_\phi$ (defined by (\ref{B_phi})) of inflaton fluctuations will differ from the bispectrum $B$ of the curvature 
fluctuations (defined by (\ref{B_zeta})) only by a simple re-scaling:
\begin{equation}
  B(k_i) \cong -\left(\frac{H}{\dot{\phi}}\right)^3 B_\phi(k_i) = -\frac{1}{(2\epsilon)^{3/2} M_p^3} B_\phi(k_i)
\end{equation}
The dominant contribution to $B_\phi$ comes from rescattering effects and scales as 
$\langle\delta_2\phi^3 \rangle \sim \langle \delta_1\chi^6 \rangle$.  

The analysis of this section justifies our neglect of metric fluctuations in section \ref{sec:theory}.

\section{Conclusions}
\label{sec:conclusions}

In the context of a realistic microscopic frame-work, we might generically expect the inflaton to couple to a large
number of fields whose energy density does not play any important role in driving inflation.  Such couplings can lead to isolated bursts
of particle production during inflation.  The associated observational signatures provide a rare opportunity to learn about how $\phi$ 
couples to other species, as opposed to the self-coupling information which is encoded in $V(\phi)$.  In this paper we have considered a simple
example of this effect which is dynamically rich and derivable from realistic particle physics models, such as string theory.  

Inflationary particle production leads to features in the primordial curvature fluctuations via the mechanism of IR cascading.  This 
process is interesting in its own right: it is qualitatively different from other mechanisms in the literature (in that we do not rely on 
the quantum vacuum fluctuations of some light iso-curvature fields) and the underlying dynamics are relevant for preheating, moduli trapping
and non-equilibrium QFT more generally.  Moreover, particle production and IR cascading lead to a variety of novel observable signatures,
including localized features in both the spectrum and bispectrum of the cosmological fluctuations.  

In this paper we have extended previous work \cite{ir,ppcons} on inflationary particle production in two directions.
Firstly, we have developed an analytical theory of particle production and IR cascading during inflation, which is in excellent agreement with 
lattice field theory simulations.  This formalism helps to clarify the underlying physics of the mechanism, and provides a crucial cross-check 
on our numerical methods.

Our second main result has been a more detailed investigation of the nongaussian signature associated with particle production and IR cascading.
The bispectrum in this model is rather unusual: it peaks only for triangles with a size comparable to some characteristic scale.  We have argued that the 
magnitude of this type of nongaussianity is best characterized by studying the moments of the PDF.  For realistic values of the coupling,
the skewness of the PDF is quite large.  For example, with $g^2 \sim 0.01$ the power spectrum for our model is compatible with all observational data \cite{ppcons}
while the skewness of the PDF is equivalent to what would be produced in a local model with $f_{NL}^{\mathrm{equiv}} \sim -53$.  This value is somewhat larger than current observational bounds,
suggesting that nongaussianity from inflationary particle production may be observable in future missions.  However, we stress that the nongaussian signature in our model is quite
different from what would be expected for a local model with $\zeta = \zeta_g + \frac{3}{5} f_{NL}^{\mathrm{equiv}} \left[\zeta_g^2 - \langle\zeta_g^2\rangle\right]$.  In particular, the 
higher order cummulants (such as the kurtosis) are different, as are the shape and running of the bispectrum.  

Note that, if it were to be detected, the nongaussian signature from IR cascading must be correlated with an observable feature in the power spectrum and also with signatures in polarization.  
Hence, it should be possible to robustly rule out the possibility that massive iso-curvature particles were produced at some point during the observable range of $e$-foldings of inflation.

The nongaussian signature predicted by inflationary particle production is rather complicated as compared to the local or equilateral models.
However, the underlying field theory description of our model is extremely simple and rather generic from the low-energy perspective.  In order to obtain 
large nongaussianity it was not necessary to fine-tune the inflaton trajectory or appeal to re-summation of an infinite series of high 
dimension operators.  Indeed, the only ``tuning'' which is required for our signal to be observable is the requirement that $\phi=\phi_0$ during the 
observable range of $e$-foldings.  We believe that this type of nongaussianity is very natural and merits further investigation
from the observational perspective.  

There are a variety of directions for future studies.  From the theoretical perspective, it would be interesting to explicitly generalize our results to
more complicated models with particle production during inflation (such as SUSY models, higher spin iso-inflatons and phase transitions).  There are
also a wide range of interesting phenomenological possibilities.  Varying the location of the feature we can have a variety of possible signatures for
the CMB and LSS.  We expect that IR cascading will also have implications for the spectrum of gravity waves from inflation and also primordial black
holes.  We could imagine superposing multiple bursts of particle production to obtain an even richer variety of signatures.  It would be interesting to construct
a simple, separable estimator for the bispectrum from IR cascading which can be confronted with observational data in order to obtain explicit constraints
on the underlying model parameters.  We leave these possibilities for future investigation.


\section*{Acknowledgments}
This work is dedicated to the memory of L.~Kofman, who played a key role in initiating this line of research.
Thanks to T.~Battefeld, I.~Huston, K.~Malik, M.~Sasaki, D.~Seery and S.~Shandera for helpful discussions, comments and correspondence.
I am especially grateful to Z.~Huang for numerous discussions, insightful suggestions and extensive help with the lattice field
theory simulations.

\renewcommand{\theequation}{A-\arabic{equation}}
\setcounter{equation}{0}
\section*{APPENDIX A: Detailed Computation of $P(k)$}
\label{appA}

\renewcommand{\thesubsection}{A.\arabic{subsection}}
\setcounter{subsection}{0}

In this appendix we discuss some of the technical details associated with the computation of the renormalized power spectrum (\ref{full_pwr_result}).
First, notice that using (\ref{f_k_approx}) and (\ref{chi_soln}) we can write the quantity appearing in each renormalized Wick contraction as
\begin{eqnarray}
&& \chi_k(\tau) \chi_k^\star(\tau') - f_k(\tau) f_k^\star(\tau') \cong \nonumber \\
&& \frac{1}{k_\star^2}\frac{1}{\sqrt{a(\tau) a(\tau')}} \frac{1}{\sqrt{t(\tau)t(\tau')}} 
\left[ n_k \cos\left( \frac{k_\star^2 t^2(\tau)}{2} - \frac{k_\star^2 t^2(\tau')}{2} \right) \right. \nonumber \\
&&\left. \,\,\,\,\,\,\,    + \sqrt{n_k} \sqrt{1+n_k} \sin\left( \frac{k_\star^2 t^2(\tau)}{2} - \frac{k_\star^2 t^2(\tau')}{2} \right)  \right] \label{ren_wick_result}
\end{eqnarray}
where the occupation number $n_k$ is defined by (\ref{n_k_mu}).
Plugging (\ref{ren_wick_result}) into (\ref{full_pwr_result}) we find
\begin{widetext}
\begin{eqnarray}
 && \hspace{-5mm} P_\phi(k) = \frac{g^2 k^3}{8 \pi^5} \left[ \,\,\,\,\,\,  \int d^3k' n_{k-k'} n_{k'}\times \int d\tau' d\tau'' \frac{G_{k}(\tau-\tau')}{a(\tau)} \frac{G_k(\tau-\tau'')}{a(\tau)}
                                                                       \cos^2\left[  \frac{k_\star^2 t^2(\tau')}{2} - \frac{k_\star^2 t^2(\tau'')}{2} \right]      \right. \nonumber \\
&+& \int d^3k' \sqrt{ n_{k-k'} n_{k'} }\sqrt{ 1 + n_{k-k'}  } \sqrt{ 1 + n_{k'}}\times
        \int d\tau' d\tau'' \frac{G_{k}(\tau-\tau')}{a(\tau)} \frac{G_k(\tau-\tau'')}{a(\tau)}
                                                                       \sin^2\left[  \frac{k_\star^2 t^2(\tau')}{2} + \frac{k_\star^2 t^2(\tau'')}{2} \right] \nonumber \\
&+& \int d^3k' \left( n_{k-k'}\sqrt{n_{k'}}\sqrt{1+n_{k'}} + n_{k'}\sqrt{n_{k-k'}}\sqrt{1+n_{k-k'}} \right) \nonumber \\
&& \,\,\,\,\,\,\,\,\,\,\,\,   \left.   \times \int d\tau' d\tau'' \frac{G_{k}(\tau-\tau')}{a(\tau)} \frac{G_k(\tau-\tau'')}{a(\tau)}
                             \cos\left[  \frac{k_\star^2 t^2(\tau')}{2} - \frac{k_\star^2 t^2(\tau'')}{2} \right]  
                             \sin\left[  \frac{k_\star^2 t^2(\tau')}{2} + \frac{k_\star^2 t^2(\tau'')}{2} \right]  \,\,\,\,\,\,\,\,\,\,\right] \label{pwr_step}
\end{eqnarray}
\end{widetext}
Notice that the time and phase space integrations in (\ref{pwr_step}) decouple.  This is the key simplification which makes an analytical
evaluation of this expression tractable.  Let us consider these integrations separately.

\subsection{Time Integrals}

All of the integrals over conformal time that appear in (\ref{pwr_step}) can be expressed in terms of two characteristic integrals which
we call $I_1$ and $I_2$.  Explicitly, these are defined as
\begin{eqnarray}
  I_1(k,\tau) &=& \frac{1}{a(\tau)}\int d\tau' G_k(\tau-\tau') e^{i k_\star^2 t^2(\tau')} \\
  I_2(k,\tau) &=& \frac{1}{a(\tau)}\int d\tau' G_k(\tau-\tau') 
\end{eqnarray}
The second characteristic integral, $I_2$, can be evaluated analytically.  However, the resulting expression is not particularly enlightening.  
Evaluation of $I_1$, on the other hand, requires numerical methods.  

Let us now show how the various integrals appearing in (\ref{pwr_step}) may be re-written in terms of $I_1$, $I_2$.  First, consider the first line
of (\ref{pwr_step}) where the following integral appears:
\begin{eqnarray}
  && \int d\tau' d\tau'' \frac{G_{k}(\tau-\tau')}{a(\tau)} \frac{G_k(\tau-\tau'')}{a(\tau)} \nonumber \\
  && \,\,\,\,\,\,\,\,\,\,  \cos^2\left[  \frac{k_\star^2 t^2(\tau')}{2} - \frac{k_\star^2 t^2(\tau'')}{2} \right] \nonumber \\
  &=& \frac{|I_1(k,\tau)|^2}{2} + \frac{I_2(k,\tau)^2}{2} \label{tint1}
\end{eqnarray}
Next, consider the second line of (\ref{pwr_step}) where the following integral appears:
\begin{eqnarray}
  && \int d\tau' d\tau'' \frac{G_{k}(\tau-\tau')}{a(\tau)} \frac{G_k(\tau-\tau'')}{a(\tau)} \nonumber \\
  && \,\,\,\,\,\,\,\,\,\,  \sin^2\left[  \frac{k_\star^2 t^2(\tau')}{2} + \frac{k_\star^2 t^2(\tau'')}{2} \right] \nonumber \\
  &=& -\frac{\mathrm{Re}\left[I_1(k,\tau)^2\right]}{2} + \frac{I_2(k,\tau)^2}{2} \label{tint2}
\end{eqnarray}
Finally, consider the fourth line of (\ref{pwr_step}) where the following integral appears:
\begin{eqnarray}
  && \int d\tau' d\tau'' \frac{G_{k}(\tau-\tau')}{a(\tau)} \frac{G_k(\tau-\tau'')}{a(\tau)} \nonumber \\
  && \,\,\,\,\, \cos\left[  \frac{k_\star^2 t^2(\tau')}{2} - \frac{k_\star^2 t^2(\tau'')}{2} \right]\sin\left[  \frac{k_\star^2 t^2(\tau')}{2} + \frac{k_\star^2 t^2(\tau'')}{2} \right] \nonumber \\
  &=& \mathrm{Im}\left[I_1(k,\tau)I_2(k,\tau)\right] \label{tint3}
\end{eqnarray}
In the expressions (\ref{tint2}) and (\ref{tint3}) the notations $\mathrm{Re}$ and $\mathrm{Im}$ denote the real and imaginary parts, respectively.

\subsection{Phase Space Integrals}

As a warm-up to the subsequent calculation consider the following integral:
\begin{eqnarray}
   && \int d^3k' n_{k-k'}^a n_{k'}^b \nonumber \\
  && = \int d^3k' \exp\left[-a\pi |{\bf k} - {\bf k'}|^2 / k_\star^2 \right]\exp\left[-b\pi |{\bf k'}|^2 / k_\star^2 \right]  \nonumber \\
   && =  \frac{k_\star^3}{(a+b)^{3/2}} \exp\left[-\pi (a+b) \frac{\mu^2}{k_\star^2}\right]\nonumber \\
  && \,\,\,\,\,\,\,\,\,\,\,\,\times \exp\left[-\frac{ab}{a+b}\frac{\pi k^2}{k_\star^2}\right] \label{phase_identity}
\end{eqnarray}
This formula is valid when $a$, $b$ are positive real numbers. Notice that this expression is symmetric under interchange of $a$ and $b$.

The phase space integral in the first line of (\ref{pwr_step}) is computed by a trivial application of the identity (\ref{phase_identity}):
\begin{equation}
 \int d^3k' n_{k-k'}n_{k'} = \frac{k_\star^3}{2\sqrt{2}} e^{-2\pi\mu^2 / k_\star^2} e^{-\pi k^2 / (2 k_\star^2)} \label{kint1}
\end{equation}
However, the remaining phase space integrals appearing in (\ref{pwr_step}) cannot be obtained exactly in closed form because they contain terms like 
$\sqrt{1+n_{k'}}$ where the gaussian factors appear under the square root.  In order to deal with such expressions, we note because $n_k \ll 1$
over most of the domain of integration, it is reasonable to replace $\sqrt{1+n_{k'}} \cong 1 + n_{k'}/2$.  
Let us now proceed in this manner.  The phase space integral on the second line of (\ref{pwr_step}) is:
\begin{eqnarray}
 && \int d^3k' \sqrt{n_{k-k'}n_{k'}}\sqrt{1+n_{k-k'}}\sqrt{1+n_{k'}} \nonumber \\
 && \cong \int d^3k'\left[  n_{k-k'}^{1/2}n_{k'}^{1/2}  + \frac{1}{2} n_{k-k'}^{3/2}n_{k'}^{1/2} + \frac{1}{2} n_{k-k'}^{1/2}n_{k'}^{3/2}  \right] \nonumber \\
 && = k_\star^3 \left[ e^{-\pi \mu^2 / k_\star^2 }\exp\left(-\frac{\pi k^2}{4k_\star^2}\right) + \right. \nonumber \\
  && \,\,\,\,\,\,\,\,\,\,\,\,\,\,\,\,\, \left. \frac{e^{-2\pi\mu^2/k_\star^2}}{2\sqrt{2}}\exp\left(-\frac{3 \pi k^2}{8k_\star^2}\right) \right] \label{kint2}
\end{eqnarray}
Finally, consider the phase space integral on the third line of (\ref{pwr_step}):
\begin{eqnarray}
 && \int d^3k' \left[ n_{k-k'} \sqrt{n_{k'}}\sqrt{1+n_{k'}}  +  n_{k'} \sqrt{n_{k-k'}}\sqrt{1+n_{k-k'}}  \right] \nonumber \\
 && \cong \int d^3k' \left[  n_{k-k'}n_{k'}^{1/2} + n_{k'}n_{k-k'}^{1/2}  \right. \nonumber \\
 && \,\,\,\,\,\,\,\, \hspace{15mm}\left. + \frac{1}{2} n_{k-k'}n_{k'}^{3/2}  + \frac{1}{2} n_{k'}n_{k-k'}^{3/2}       \right] \nonumber \\
 && = k_\star^3 \left[\frac{4\sqrt{2}}{3\sqrt{3}} e^{-3\pi\mu^2/(2k_\star^2)}\exp\left(-\frac{\pi k^2}{3k_\star^2}\right) 
+ \right. \nonumber \\ 
&& \,\,\,\,\,\,\,\,\,\,\,\,\,\,\,\, \left. \frac{2\sqrt{2}}{5\sqrt{5}} e^{-5\pi\mu^2/(2k_\star^2)}\exp\left(-\frac{3 \pi k^2}{5k_\star^2}\right) \right] \label{kint3}
\end{eqnarray}
We have verified the formulae (\ref{kint2},\ref{kint3}) numerically.  In both cases that the numerical results agree with these 
semi-analytical expressions up to the percent level.

We can now, finally, insert the results (\ref{tint1},\ref{tint2},\ref{tint3}) and (\ref{kint1},\ref{kint2},\ref{kint3}) into the expression (\ref{pwr_step}).
Doing so, we arrive at our main analytical result, which is equation (\ref{full_pwr_result}).

\renewcommand{\theequation}{B-\arabic{equation}}
\setcounter{equation}{0}
\section*{APPENDIX B: Evaluation of the Bispectrum}
\label{appB}

\renewcommand{\thesubsection}{B.\arabic{subsection}}
\setcounter{subsection}{0}

In this appendix we discuss some of the technical details associated with the explicit evaluation of 
the renormalized bispectrum (\ref{full_pwr_result}).  To render the analysis tractable we will work in the
flat-space limit $H \rightarrow 0$, which is sensible since the process of IR cascading takes only a single
$e$-folding and, moreover, this approximation was shown to yield sensible results for the power spectrum
in \cite{ir}.  Neglecting the expansion of the universe we have $t=\tau$ and the retarded Green function (\ref{green}) becomes
\begin{equation}
\label{green_flat}
  G_k(t-t') = \frac{\Theta(t-t')}{\Omega_k} \sin\left[\Omega_k(t-t')\right]
\end{equation}
with $\Theta(x)$ the Heaviside function and $\Omega_k\equiv \sqrt{k^2 + m^2}$.  In this limit the characteristic integrals $I_1(k,t)$, 
$I_2(k,t)$ defined by (\ref{I1_text}) and (\ref{I2_text}) can be computed analytically.  For $I_1(k,t)$ we find
\begin{eqnarray}
  I_1(k,t) &=& \frac{\sqrt{\pi}}{2k_\star \Omega_k} e^{i\Omega_k t - i\Omega_k^2 / (4k_\star^2) - i\pi / 4} F(k,t) \label{I1}\\
  F(k,t) &=& \frac{1}{2}\left[ \left(1+e^{-2i\Omega_k t}\right)\mathrm{erf}\left(\frac{e^{-i\pi/4}}{2}\frac{\Omega_k}{k_\star}\right) \right. \label{F} \\
  &&\,\,\,\,\, -  \mathrm{erf}\left(\frac{e^{-i\pi/4}}{2}\left(\frac{\Omega_k}{k_\star} - 2k_\star t\right)\right)  \nonumber \\
  &&\,\,\,\,\, \left. -  e^{-2i\Omega_k t}\,\mathrm{erf}\left(\frac{e^{-i\pi/4}}{2}\left(\frac{\Omega_k}{k_\star} + 2k_\star t\right)\right) \right] \nonumber
\end{eqnarray}
while, for $I_2(k,t)$, we have
\begin{equation}
\label{I2}
  I_2(k,t) = \frac{1}{\Omega_k^2}\left[1-\cos(\Omega_k t)\right]
\end{equation}
(Note that our definition of $I_1$, $I_2$ differs from \cite{ir} by a factor of $\Omega_k^{-1}$.)  Finally, the renormalized Wick contraction 
(\ref{ren_wick_result}) also simplifies in the limit $H\rightarrow 0$:
\begin{eqnarray}
&& \chi_k(t) \chi_k^\star(t') - f_k(t) f_k^\star(t') \cong \nonumber \\
&& \frac{1}{k_\star^2} \frac{1}{\sqrt{t t' }} 
\left[ n_k \cos\left( \frac{k_\star^2 t^2}{2} - \frac{k_\star^2 (t')^2}{2} \right) \right. \nonumber \\
&&\left. \,\,\,\,\,\,\,    + \sqrt{n_k} \sqrt{1+n_k} \sin\left( \frac{k_\star^2 t^2}{2} - \frac{k_\star^2 (t')^2}{2} \right)  \right] \label{ren_wick_result_flat}
\end{eqnarray}
where the occupation number $n_k$ is defined by (\ref{n_k}).  

Inserting (\ref{green_flat}) and (\ref{ren_wick_result_flat}) into (\ref{bispectrum_soln}) we find the following expression for the renormalized 3-point 
correlation function:
\begin{widetext}
\begin{eqnarray}
&& \langle \xi^\phi_{\bf k_1} \xi^\phi_{\bf k_2} \xi^\phi_{\bf k_3}(t) \rangle = \frac{4 g^3}{(2\pi)^{9/2}} \delta^{(3)}({\bf k_1}+{\bf k_2}+{\bf k_3})
\prod_{j=1}^3 \frac{dt_j}{\Omega_{k_j}} \sin\left[\Omega_{k_j}(t-t_j)\right] \nonumber \\
&& \times\int d^3 p \left[  \,\,\,\,\,  \left[ n_{k_1-p} \cos\left( \frac{(k_\star t_1)^2}{2} - \frac{(k_\star t_2)^2}{2} \right) +
                                                   n_{k_1-p}^{1/2}\sqrt{1+n_{k_1-p}} \sin\left( \frac{(k_\star t_1)^2}{2} + \frac{(k_\star t_2)^2}{2} \right)    \right]  \right. \nonumber \\
&& \,\,\,\,\,\,\,\,\,\,\,\,\,\,\, \,\,\,\,\,\,\,\,\,\,\times\left[ n_{k_3+p} \cos\left( \frac{(k_\star t_2)^2}{2} - \frac{(k_\star t_3)^2}{2} \right) +
                                                   n_{k_3+p}^{1/2}\sqrt{1+n_{k_3+p}} \sin\left( \frac{(k_\star t_2)^2}{2} + \frac{(k_\star t_3)^2}{2} \right)     \right] \nonumber \\
&& \left.\,\,\,\,\, \,\,\,\,\,\,\,\,\,\,\,\,\,\,\,\,\,\,\,\,\times  
\left[ n_{p} \cos\left( \frac{(k_\star t_1)^2}{2} - \frac{(k_\star t_3)^2}{2} \right) +
                      n_{p}^{1/2}\sqrt{1+n_{p}} \sin\left( \frac{(k_\star t_1)^2}{2} + \frac{(k_\star t_3)^2}{2} \right)     \right]\,\,\,\,\,\right] \label{explicit_3_pt}
\end{eqnarray}
\end{widetext}
It only remains to expand out the expression (\ref{explicit_3_pt}) and evaluate the various integral which arise.  As in the case of the 2-point
function, the phase space and time integrals decouple, making an analytical evaluation tractable.  Let us consider the various integrals which arise
separately.

\subsection{Phase Space Integrals}

First, let us introduce a notation for the fundamental phase space integral which arises
\begin{eqnarray}
  && \hspace{-7mm}K_{a,b,c} = \int d^3 p \,\,n_{k_1-p}^a n_{k_3+p}^b n_p^c \nonumber \\
                              && \hspace{-7mm} = \frac{k_\star^3}{(a+b+c)^{3/2}} \exp\left[-\pi (a+b+c)\frac{\mu^2}{k_\star^2}\right] \nonumber \\
  && \hspace{-7mm}\,\,\,\,\,\,\,\,\,\,\times\exp\left[ -\pi \frac{\left( ac k_1^2 + bc k_3^2 + ab k_2^2  \right)}{k_\star^2(a+b+c)}  \right]
\label{K_abc_def}
\end{eqnarray}
where we have used the fact that ${\bf k_1} + {\bf k_2} + {\bf k_3} = 0$.  To evaluate integrals containing radicals such as $\sqrt{1+n_p}$ we use
the same trick as was employed for (\ref{kint2}).  That is, we approximate:
\begin{eqnarray}
 && \int d^3p\,\, n_{k_1-p} n_{k_3+p} n_{p}^{1/2} \sqrt{1+n_p} \nonumber \\ 
 &&\cong  \int d^3p \,\,n_{k_1-p} n_{k_3+p} n_{p}^{1/2} + \frac{1}{2}\int d^3 p\,\, n_{k_1-p} n_{k_3+p} n_{p}^{3/2}  \nonumber \\
 &&= K_{1,1,1/2} + \frac{1}{2}\,K_{1,1,3/2}
\end{eqnarray}
and similarly for the other combinations which arise in the expansion of (\ref{explicit_3_pt}).  We have checked numerically that this gives a
good approximation to the exact result.

\subsection{Time Integrals}

The evaluation of the time integrals appearing in (\ref{explicit_3_pt}) is a straightforward generalization of the results presented in appendix A.
Let us introduce some notations for the various combinations of the fundamental integrals $I_1$ and $I_2$ that will appear in the final result:
\begin{widetext}
\begin{eqnarray}
  A &\equiv& \prod_j \int \frac{dt_j}{\Omega_{k_j}} \sin(\Omega_{k_j}(t-t_j)) \times\cos\left[ \frac{(k_\star t_1)^2}{2}-\frac{(k_\star t_2)^2}{2} \right]
                                                                                                                      \cos\left[ \frac{(k_\star t_2)^2}{2}-\frac{(k_\star t_3)^2}{2} \right]
                                                                                                                      \cos\left[ \frac{(k_\star t_1)^2}{2}-\frac{(k_\star t_3)^2}{2} \right]    \nonumber \\
   &=& \frac{1}{4} \left[ I_2(k_1,t)I_2(k_2,t)I_2(k_3,t) + I_2(k_1,t)\mathrm{Re}\left[ I_1(k_2,t)I_1(k_3,t) \right] + (2\hspace{2mm}\mathrm{permutations}) \right]
  \label{A} \\
B &\equiv& \prod_j \int \frac{dt_j}{\Omega_{k_j}} \sin(\Omega_{k_j}(t-t_j)) \times\sin\left[ \frac{(k_\star t_1)^2}{2}+\frac{(k_\star t_2)^2}{2} \right]
                                                                                                                      \sin\left[ \frac{(k_\star t_2)^2}{2}+\frac{(k_\star t_3)^2}{2} \right]
                                                                                                                      \sin\left[ \frac{(k_\star t_1)^2}{2}+\frac{(k_\star t_3)^2}{2} \right]    \nonumber \\
   &=& \frac{1}{4} \left[ -\mathrm{Im}\left[I_1(k_1,t)I_1(k_2,t)I_1(k_3,t)\right] + I_2(k_1,t)I_2(k_2,t)\mathrm{Im}\left[ I_1(k_3,t) \right] + (2\hspace{2mm}\mathrm{permutations}) \right]
\label{B} \\
C_1 &\equiv& \prod_j \int \frac{dt_j}{\Omega_{k_j}} \sin(\Omega_{k_j}(t-t_j)) \times\cos\left[ \frac{(k_\star t_1)^2}{2}-\frac{(k_\star t_2)^2}{2} \right]
                                                                                                                      \cos\left[ \frac{(k_\star t_2)^2}{2}-\frac{(k_\star t_3)^2}{2} \right]
                                                                                                                      \sin\left[ \frac{(k_\star t_1)^2}{2}+\frac{(k_\star t_3)^2}{2} \right]    \nonumber \\
   &=& \frac{1}{4} \left[ \mathrm{Im}\left[ I_1(k_1,t)I_1^\star(k_2,t)I_1(k_3,t)\right] + I_2(k_1,t)I_2(k_2,t)\mathrm{Im}\left[ I_1(k_3,t) \right] + (2\hspace{2mm}\mathrm{permutations}) \right] 
\label{C1} \\
C_2 &\equiv& \prod_j \int \frac{dt_j}{\Omega_{k_j}} \sin(\Omega_{k_j}(t-t_j)) \times\cos\left[ \frac{(k_\star t_1)^2}{2}-\frac{(k_\star t_2)^2}{2} \right]
                                                                                                                      \sin\left[ \frac{(k_\star t_2)^2}{2}+\frac{(k_\star t_3)^2}{2} \right]
                                                                                                                      \cos\left[ \frac{(k_\star t_1)^2}{2}-\frac{(k_\star t_3)^2}{2} \right]    \nonumber \\
   &=& \frac{1}{4} \left[ \mathrm{Im}\left[ I_1^\star(k_1,t)I_1(k_2,t)I_1(k_3,t)\right] + I_2(k_1,t)I_2(k_2,t)\mathrm{Im}\left[ I_1(k_3,t) \right] + (2\hspace{2mm}\mathrm{permutations}) \right] 
\label{C2} \\
C_3 &\equiv& \prod_j \int \frac{dt_j}{\Omega_{k_j}} \sin(\Omega_{k_j}(t-t_j)) \times\sin\left[ \frac{(k_\star t_1)^2}{2}+\frac{(k_\star t_2)^2}{2} \right]
                                                                                                                      \cos\left[ \frac{(k_\star t_2)^2}{2}-\frac{(k_\star t_3)^2}{2} \right]
                                                                                                                      \cos\left[ \frac{(k_\star t_1)^2}{2}-\frac{(k_\star t_3)^2}{2} \right]    \nonumber \\
   &=& \frac{1}{4} \left[ \mathrm{Im}\left[ I_1(k_1,t)I_1(k_2,t)I_1^\star(k_3,t)\right] + I_2(k_1,t)I_2(k_2,t)\mathrm{Im}\left[ I_1(k_3,t) \right] + (2\hspace{2mm}\mathrm{permutations}) \right] 
\label{C3} \\
D_1 &\equiv& \prod_j \int \frac{dt_j}{\Omega_{k_j}} \sin(\Omega_{k_j}(t-t_j)) \times\cos\left[ \frac{(k_\star t_1)^2}{2}-\frac{(k_\star t_2)^2}{2} \right]
                                                                                                                      \sin\left[ \frac{(k_\star t_2)^2}{2}+\frac{(k_\star t_3)^2}{2} \right]
                                                                                                                      \sin\left[ \frac{(k_\star t_1)^2}{2}+\frac{(k_\star t_3)^2}{2} \right]    \nonumber \\
   &=& \frac{1}{4} \left[ I_2(k_1,t)I_2(k_2,t)I_2(k_3,t) + I_2(k_3,t)\mathrm{Re}\left[ I_1(k_1,t)I_1^\star(k_2,t) \right] \right. \nonumber \\
    &&  \,\, \left.    - I_2(k_2)\mathrm{Re}\left[I_1(k_1,t)I_1(k_3,t)\right] - I_2(k_1,t)\mathrm{Re}\left[I_1(k_2,t)I_1(k_3,t)\right] \right] 
\label{D1} \\
D_2 &\equiv& \prod_j \int \frac{dt_j}{\Omega_{k_j}} \sin(\Omega_{k_j}(t-t_j)) \times\sin\left[ \frac{(k_\star t_1)^2}{2}+\frac{(k_\star t_2)^2}{2} \right]
                                                                                                                      \cos\left[ \frac{(k_\star t_2)^2}{2}-\frac{(k_\star t_3)^2}{2} \right]
                                                                                                                      \sin\left[ \frac{(k_\star t_1)^2}{2}+\frac{(k_\star t_3)^2}{2} \right]    \nonumber \\
   &=& \frac{1}{4} \left[ I_2(k_1,t)I_2(k_2,t)I_2(k_3,t) + I_2(k_1,t)\mathrm{Re}\left[ I_1(k_2,t)I_1^\star(k_3,t) \right] \right. \nonumber \\
    &&  \,\, \left.    - I_2(k_3)\mathrm{Re}\left[I_1(k_1,t)I_1(k_2,t)\right] - I_2(k_2,t)\mathrm{Re}\left[I_1(k_1,t)I_1(k_3,t)\right] \right] 
\label{D2}\\
D_3 &\equiv& \prod_j \int \frac{dt_j}{\Omega_{k_j}} \sin(\Omega_{k_j}(t-t_j)) \times\sin\left[ \frac{(k_\star t_1)^2}{2}+\frac{(k_\star t_2)^2}{2} \right]
                                                                                                                      \sin\left[ \frac{(k_\star t_2)^2}{2}+\frac{(k_\star t_3)^2}{2} \right]
                                                                                                                      \cos\left[ \frac{(k_\star t_1)^2}{2}-\frac{(k_\star t_3)^2}{2} \right]    \nonumber \\
   &=& \frac{1}{4} \left[ I_2(k_1,t)I_2(k_2,t)I_2(k_3,t) + I_2(k_2,t)\mathrm{Re}\left[ I_1(k_1,t)I_1^\star(k_3,t) \right] \right. \nonumber \\
    &&  \,\, \left.    - I_2(k_3)\mathrm{Re}\left[I_1(k_1,t)I_1(k_2,t)\right] - I_2(k_1,t)\mathrm{Re}\left[I_1(k_2,t)I_1(k_3,t)\right] \right] \label{D3}
\end{eqnarray}
\end{widetext}

\subsection{The Full Bispectrum}

We are now finally in a position to write out an explicit expression for the renormalized 3-point function of the inflaton fluctuations generated by IR
cascading.  That expression is given below, in terms of the various that were defined explicitly in equations (\ref{K_abc_def}) and (\ref{A})-(\ref{D3}).
As promised, the explicit result for the $3$-point correlation function is cumbersome and not entirely enlightening.
\begin{widetext}
\begin{eqnarray}
&& \langle \xi^\phi_{\bf k_1} \xi^\phi_{\bf k_2} \xi^\phi_{\bf k_3}(t) \rangle = \frac{4 g^3}{(2\pi)^{9/2}} \delta^{(3)}({\bf k_1}+{\bf k_2}+{\bf k_3}) \nonumber \\
&& \left[ \,\,\,\,\, K_{1,1,1} A +  \left[ K_{1/2,1/2,1/2} + \frac{1}{2}\left( K_{3/2,1/2,1/2} + K_{1/2,3/2,1/2} + K_{1/2,1/2,3/2}  \right) \right] B             \right. \nonumber \\
&& \,\,\,\,\, + \left[ K_{1,1,1/2} + \frac{1}{2} K_{1,1,3/2}  \right] C_1 + \left[K_{1,1/2,1} + \frac{1}{2}K_{1,3/2,1}\right] C_2 
     + \left[K_{1/2,1,1} +\frac{1}{2}K_{3/2,1,1}\right]C_3 \nonumber \\
&&  \,\,\,\,\, + \left[ K_{1,1/2,1/2} + \frac{1}{2} \left( K_{1,3/2,1/2} + K_{1,1/2,3/2}  \right) \right] D_1 
    + \left[K_{1/2,1,1/2} + \frac{1}{2}\left( K_{3/2,1,1/2}+K_{1/2,1,3/2}\right) \right] D_2 \nonumber \\
&& \left.    + \left[K_{1/2,1/2,1} +\frac{1}{2}\left(K_{3/2,1/2,1}+K_{1/2,3/2,1}\right)\right]D_3 + (k_2 \leftrightarrow k_3) \,\,\,\,\, \right] \label{full_3_pt_app}
\end{eqnarray}
\end{widetext}


\begin{thebibliography}{99}

\bibitem{ir}

  N.~Barnaby, Z.~Huang, L.~Kofman and D.~Pogosyan,
  ``Cosmological Fluctuations from Infra-Red Cascading During Inflation,''
  Phys.\ Rev.\  D {\bf 80}, 043501 (2009)
  [arXiv:0902.0615 [hep-th]].

\bibitem{ppcons}

 N.~Barnaby and Z.~Huang,
  ``Particle Production During Inflation: Observational Constraints and
  Signatures,''
  arXiv:0909.0751 [astro-ph.CO].

\bibitem{trapped}

 D.~Green, B.~Horn, L.~Senatore and E.~Silverstein,
  ``Trapped Inflation,''
  Phys.\ Rev.\  D {\bf 80}, 063533 (2009)
  [arXiv:0902.1006 [hep-th]].

\bibitem{chung}

  D.~J.~H.~Chung, E.~W.~Kolb, A.~Riotto and I.~I.~Tkachev,
  ``Probing Planckian physics: Resonant production of particles during
  inflation and features in the primordial power spectrum,''
  Phys.\ Rev.\  D {\bf 62}, 043508 (2000)
  [arXiv:hep-ph/9910437].

\bibitem{chung2}

 G.~J.~Mathews, D.~J.~H.~Chung, K.~Ichiki, T.~Kajino and M.~Orito,
  ``Constraints on resonant particle production during inflation from the
  matter and CMB power spectra,''
  Phys.\ Rev.\  D {\bf 70}, 083505 (2004)
  [arXiv:astro-ph/0406046].

\bibitem{elgaroy}

  O.~Elgaroy, S.~Hannestad and T.~Haugboelle,
  ``Observational constraints on particle production during inflation,''
  JCAP {\bf 0309}, 008 (2003)
  [arXiv:astro-ph/0306229].

\bibitem{sasaki}

  A.~E.~Romano and M.~Sasaki,
  ``Effects of particle production during inflation,''
  Phys.\ Rev.\  D {\bf 78}, 103522 (2008)
  [arXiv:0809.5142 [gr-qc]].

\bibitem{modulated_trapping}

  D.~Langlois and L.~Sorbo,
  ``Primordial perturbations and non-Gaussianities from modulated trapping,''
  arXiv:0906.1813 [astro-ph.CO].

\bibitem{brane_brem}

 P.~Brax and E.~Cluzel,
  ``Brane Bremsstrahlung in DBI Inflation,''
  arXiv:0912.0806 [hep-th].

\bibitem{preheatNG}

  N.~Barnaby and J.~M.~Cline,
  ``Nongaussian and nonscale-invariant perturbations from tachyonic  preheating
  in hybrid inflation,''
  Phys.\ Rev.\  D {\bf 73}, 106012 (2006)
  [arXiv:astro-ph/0601481].

  N.~Barnaby and J.~M.~Cline,
  ``Nongaussianity from Tachyonic Preheating in Hybrid Inflation,''
  Phys.\ Rev.\  D {\bf 75}, 086004 (2007)
  [arXiv:astro-ph/0611750].


\bibitem{KL}

  L.~A.~Kofman and A.~D.~Linde,
 ``Generation of Density Perturbations in the Inflationary Cosmology,''
  Nucl.\ Phys.\  B {\bf 282}, 555 (1987).

\bibitem{KP}

  L.~A.~Kofman and D.~Y.~Pogosian,
 ``NONFLAT PERTURBATIONS IN INFLATIONARY COSMOLOGY,''
  Phys.\ Lett.\  B {\bf 214}, 508 (1988).

\bibitem{BBS}

 D.~S.~Salopek, J.~R.~Bond and J.~M.~Bardeen,
  ``Designing Density Fluctuation Spectra in Inflation,''
  Phys.\ Rev.\  D {\bf 40}, 1753 (1989).

\bibitem{adams}

  J.~A.~Adams, B.~Cresswell and R.~Easther,
  ``Inflationary perturbations from a potential with a step,''
  Phys.\ Rev.\  D {\bf 64}, 123514 (2001)
  [arXiv:astro-ph/0102236].

\bibitem{step_model}

  P.~Hunt and S.~Sarkar,
  ``Multiple inflation and the WMAP 'glitches',''
  Phys.\ Rev.\  D {\bf 70}, 103518 (2004)
  [arXiv:astro-ph/0408138].

  P.~Hunt and S.~Sarkar,
  ``Multiple inflation and the WMAP 'glitches' II. Data analysis and
  cosmological parameter extraction,''
  Phys.\ Rev.\  D {\bf 76}, 123504 (2007)
  [arXiv:0706.2443 [astro-ph]].

\bibitem{gobump}

  M.~J.~Mortonson, C.~Dvorkin, H.~V.~Peiris and W.~Hu,
  ``Things that go bump in the CMB polarization: features from inflation versus
  reionization,''
  arXiv:0903.4920 [astro-ph.CO].

\bibitem{brane_annihilation}


D.~Battefeld, T.~Battefeld, H.~Firouzjahi and N.~Khosravi,
``Brane Annihilations during Inflation,''
arXiv:1004.1417 [hep-th].

\bibitem{sorbo}

  M.~M.~Anber and L.~Sorbo,
  ``Naturally inflating on steep potentials through electromagnetic
  dissipation,''
  arXiv:0908.4089 [hep-th].


\bibitem{fluctuations}

 V.~F.~Mukhanov and G.~V.~Chibisov,
``Quantum Fluctuation And 'Nonsingular' Universe,''
JETP Lett.\  {\bf 33}, 532 (1981)
[Pisma Zh.\ Eksp.\ Teor.\ Fiz.\  {\bf 33}, 549 (1981)]; 

S.~W.~Hawking,
``The Development Of Irregularities In A Single Bubble Inflationary Universe,''
Phys.\ Lett.\ B {\bf 115}, 295 (1982); 

A.~A.~Starobinsky,
``Dynamics Of Phase Transition In The New Inflationary Universe Scenario And Generation Of Perturbations,''
Phys.\ Lett.\ B {\bf 117}, 175 (1982); 
A.~H.~Guth and S.~Y.~Pi,
``Fluctuations In The New Inflationary Universe,''
Phys.\ Rev.\ Lett.\  {\bf 49}, 1110 (1982); J.~M.~Bardeen, P.~J.~Steinhardt and

M.~S.~Turner,
``Spontaneous Creation Of Almost Scale - Free Density Perturbations In An Inflationary Universe,''
Phys.\ Rev.\ D {\bf 28}, 679 (1983).


\bibitem{curvaton0}

  K.~Enqvist and M.~S.~Sloth,
  ``Adiabatic CMB perturbations in pre big bang string cosmology,''
  Nucl.\ Phys.\  B {\bf 626}, 395 (2002)
  [arXiv:hep-ph/0109214].

\bibitem{curvaton}

  D.~H.~Lyth and D.~Wands,
  ``Generating the curvature perturbation without an inflaton,''
  Phys.\ Lett.\  B {\bf 524}, 5 (2002)
  [arXiv:hep-ph/0110002].

\bibitem{modulated}

  L.~Kofman,
  ``Probing string theory with modulated cosmological fluctuations,''
  arXiv:astro-ph/0303614.

  F.~Bernardeau, L.~Kofman and J.~P.~Uzan,
  ``Modulated fluctuations from hybrid inflation,''
  Phys.\ Rev.\  D {\bf 70}, 083004 (2004)
  [arXiv:astro-ph/0403315].

\bibitem{modulated2}

   G.~Dvali, A.~Gruzinov and M.~Zaldarriaga,
  ``A new mechanism for generating density perturbations from inflation,''
  Phys.\ Rev.\  D {\bf 69}, 023505 (2004)
  [arXiv:astro-ph/0303591].

  G.~Dvali, A.~Gruzinov and M.~Zaldarriaga,
  ``Cosmological perturbations from inhomogeneous reheating, freezeout, and
  mass domination,''
  Phys.\ Rev.\  D {\bf 69}, 083505 (2004)
  [arXiv:astro-ph/0305548].


\bibitem{warm}
  A.~Berera,
  ``Warm Inflation,''
  Phys.\ Rev.\ Lett.\  {\bf 75}, 3218 (1995)
  [arXiv:astro-ph/9509049].

\bibitem{warm2}

  A.~Berera, M.~Gleiser and R.~O.~Ramos,
  ``Strong dissipative behavior in quantum field theory,''
  Phys.\ Rev.\  D {\bf 58}, 123508 (1998)
  [arXiv:hep-ph/9803394].

\bibitem{warm3}

  A.~Berera,
  ``Warm inflation at arbitrary adiabaticity: A model, an existence proof  for
  inflationary dynamics in quantum field theory,''
  Nucl.\ Phys.\  B {\bf 585}, 666 (2000)
  [arXiv:hep-ph/9904409].

\bibitem{berrera}

  A.~Berera and T.~W.~Kephart,
  ``Ubiquitous inflaton in string-inspired models,''
  Phys.\ Rev.\ Lett.\  {\bf 83}, 1084 (1999)
  [arXiv:hep-ph/9904410].

\bibitem{monodromy1}

 E.~Silverstein and A.~Westphal,
  ``Monodromy in the CMB: Gravity Waves and String Inflation,''
  Phys.\ Rev.\  D {\bf 78}, 106003 (2008)
  [arXiv:0803.3085 [hep-th]].

\bibitem{monodromy2}

 L.~McAllister, E.~Silverstein and A.~Westphal,
  ``Gravity Waves and Linear Inflation from Axion Monodromy,''
  arXiv:0808.0706 [hep-th].

\bibitem{monodromy3}

  R.~Flauger, L.~McAllister, E.~Pajer, A.~Westphal and G.~Xu,
  ``Oscillations in the CMB from Axion Monodromy Inflation,''
  arXiv:0907.2916 [hep-th].

\bibitem{monodromy4}

  S.~Hannestad, T.~Haugbolle, P.~R.~Jarnhus and M.~S.~Sloth,
  ``Non-Gaussianity from Axion Monodromy Inflation,''
  JCAP {\bf 1006}, 001 (2010)
  [arXiv:0912.3527 [hep-ph]].

\bibitem{amjad}

 A.~Ashoorioon and A.~Krause,
 ``Power spectrum and signatures for cascade inflation,''
 arXiv:hep-th/0607001.

\bibitem{beauty}

  L.~Kofman, A.~D.~Linde, X.~Liu, A.~Maloney, L.~McAllister and E.~Silverstein,
  ``Beauty is attractive: Moduli trapping at enhanced symmetry points,''
  JHEP {\bf 0405}, 030 (2004)
  [arXiv:hep-th/0403001].

\bibitem{terminal}


D.~Battefeld and T.~Battefeld,
``A Terminal Velocity on the Landscape: Particle Production near Extra Species Loci in Higher Dimensions,''
arXiv:1004.3551 [hep-th].

\bibitem{natural}

 K.~Freese, J.~A.~Frieman and A.~V.~Olinto,
  ``Natural inflation with pseudo - Nambu-Goldstone bosons,''
  Phys.\ Rev.\ Lett.\  {\bf 65}, 3233 (1990).

  F.~C.~Adams, J.~R.~Bond, K.~Freese, J.~A.~Frieman and A.~V.~Olinto,
  ``Natural Inflation: Particle Physics Models, Power Law Spectra For Large
  Scale Structure, And Constraints From Cobe,''
  Phys.\ Rev.\  D {\bf 47}, 426 (1993)
  [arXiv:hep-ph/9207245].

\bibitem{jim}

  C.~P.~Burgess, J.~M.~Cline, F.~Lemieux and R.~Holman,
  ``Decoupling, trans-Planckia and inflation,''
  arXiv:astro-ph/0306236.

  C.~P.~Burgess, J.~M.~Cline and R.~Holman,
  ``Effective field theories and inflation,''
  JCAP {\bf 0310}, 004 (2003)
  [arXiv:hep-th/0306079].

  C.~P.~Burgess, J.~M.~Cline, F.~Lemieux and R.~Holman,
  ``Are inflationary predictions sensitive to very high energy physics?,''
  JHEP {\bf 0302}, 048 (2003)
  [arXiv:hep-th/0210233].

\bibitem{nonequilibrium}

 J.~Berges and J.~Serreau,
  ``Progress in nonequilibrium quantum field theory,''
  arXiv:hep-ph/0302210.

  J.~Berges and J.~Serreau,
  ``Progress in nonequilibrium quantum field theory. II,''
  arXiv:hep-ph/0410330.

  J.~Berges and S.~Borsanyi,
  ``Progress in nonequilibrium quantum field theory. III,''
  Nucl.\ Phys.\  A {\bf 785}, 58 (2007)
  [arXiv:hep-ph/0610015].


\bibitem{false_vac}

  E.~J.~Copeland, A.~R.~Liddle, D.~H.~Lyth, E.~D.~Stewart and D.~Wands,
  ``False vacuum inflation with Einstein gravity,''
  Phys.\ Rev.\  D {\bf 49}, 6410 (1994)
  [arXiv:astro-ph/9401011].

\bibitem{susy_break}

  M.~Dine, L.~Randall and S.~D.~Thomas,
  ``Supersymmetry breaking in the early universe,''
  Phys.\ Rev.\ Lett.\  {\bf 75}, 398 (1995)
  [arXiv:hep-ph/9503303].

\bibitem{rapid_roll}

  L.~Kofman and S.~Mukohyama,
  ``Rapid roll Inflation with Conformal Coupling,''
  Phys.\ Rev.\  D {\bf 77}, 043519 (2008)
  [arXiv:0709.1952 [hep-th]].

\bibitem{KLS}

 L.~Kofman, A.~D.~Linde and A.~A.~Starobinsky,
  ``Reheating after inflation,''
  Phys.\ Rev.\ Lett.\  {\bf 73}, 3195 (1994)
  [arXiv:hep-th/9405187].

\bibitem{KLS97}

  L.~Kofman, A.~D.~Linde and A.~A.~Starobinsky,
 ``Towards the theory of reheating after inflation,''
  Phys.\ Rev.\  D {\bf 56}, 3258 (1997)
  [arXiv:hep-ph/9704452].

\bibitem{FK}
  
G.~N.~Felder and L.~Kofman,
  ``The development of equilibrium after preheating,''
  Phys.\ Rev.\  D {\bf 63}, 103503 (2001)
  [arXiv:hep-ph/0011160].

\bibitem{MT1}

  R.~Micha and I.~I.~Tkachev,
  ``Relativistic turbulence: A long way from preheating to equilibrium,''
  Phys.\ Rev.\ Lett.\  {\bf 90}, 121301 (2003)
  [arXiv:hep-ph/0210202].

\bibitem{MT2}

  R.~Micha and I.~I.~Tkachev,
  ``Turbulent thermalization,''
  Phys.\ Rev.\  D {\bf 70}, 043538 (2004)
  [arXiv:hep-ph/0403101].

\bibitem{dmitry}

  D.~I.~Podolsky, G.~N.~Felder, L.~Kofman and M.~Peloso,
  ``Equation of state and beginning of thermalization after preheating,''
  Phys.\ Rev.\  D {\bf 73}, 023501 (2006)
  [arXiv:hep-ph/0507096].

\bibitem{B1}

  J.~Berges, A.~Rothkopf and J.~Schmidt,
  ``Non-thermal fixed points: effective weak-coupling for strongly correlated
  systems far from equilibrium,''
  Phys.\ Rev.\ Lett.\  {\bf 101}, 041603 (2008)
  [arXiv:0803.0131 [hep-ph]].

\bibitem{B2}

  J.~Berges and G.~Hoffmeister,
  ``Nonthermal fixed points and the functional renormalization group,''
  arXiv:0809.5208 [hep-th].

\bibitem{star1}

A.~A.~Starobinsky, JETP Lett. 55, 489 (1992).

\bibitem{star2}

 M.~Joy, V.~Sahni and A.~A.~Starobinsky,
  ``A New Universal Local Feature in the Inflationary Perturbation Spectrum,''
  Phys.\ Rev.\  D {\bf 77}, 023514 (2008)
  [arXiv:0711.1585 [astro-ph]].

\bibitem{chen1}

  X.~Chen, R.~Easther and E.~A.~Lim,
  ``Large non-Gaussianities in single field inflation,''
  JCAP {\bf 0706}, 023 (2007)
  [arXiv:astro-ph/0611645].

\bibitem{chen2}

  X.~Chen, R.~Easther and E.~A.~Lim,
  ``Generation and Characterization of Large Non-Gaussianities in Single Field
  Inflation,''
  JCAP {\bf 0804}, 010 (2008)
  [arXiv:0801.3295 [astro-ph]].

\bibitem{space_break}

  R.~Lerner and J.~McDonald,
  ``Space-Dependent Step Features: Transient Breakdown of Slow-roll,
  Homogeneity and Isotropy During Inflation,''
  Phys.\ Rev.\  D {\bf 79}, 023511 (2009)
  [arXiv:0811.1933 [astro-ph]].

\bibitem{jain}

  R.~K.~Jain, P.~Chingangbam, J.~O.~Gong, L.~Sriramkumar and T.~Souradeep,
  ``Double inflation and the low CMB multipoles,''
  JCAP {\bf 0901}, 009 (2009)
  [arXiv:0809.3915 [astro-ph]].

  R.~K.~Jain, P.~Chingangbam, L.~Sriramkumar and T.~Souradeep,
  ``The tensor-to-scalar ratio in punctuated inflation,''
  Phys.\ Rev.\  D {\bf 82}, 023509 (2010)
  [arXiv:0904.2518 [astro-ph.CO]].

  D.~K.~Hazra, M.~Aich, R.~K.~Jain, L.~Sriramkumar and T.~Souradeep,
  ``Primordial features due to a step in the inflaton potential,''
  arXiv:1005.2175 [astro-ph.CO].

\bibitem{forecast}

  T.~Chantavat, C.~Gordon and J.~Silk,
  ``Large Scale Structure Forecast Constraints on Particle Production During Inflation,''
  arXiv:1009.5858 [astro-ph.CO].

\bibitem{riotto}

  V.~Acquaviva, N.~Bartolo, S.~Matarrese and A.~Riotto,
  ``Second-order cosmological perturbations from inflation,''
  Nucl.\ Phys.\  B {\bf 667}, 119 (2003)
  [arXiv:astro-ph/0209156].

\bibitem{maldacena}

  J.~M.~Maldacena,
  ``Non-Gaussian features of primordial fluctuations in single field
  inflationary models,''
  JHEP {\bf 0305}, 013 (2003)
  [arXiv:astro-ph/0210603].

\bibitem{seerylidsey}

  D.~Seery and J.~E.~Lidsey,
  ``Primordial non-gaussianities in single field inflation,''
  JCAP {\bf 0506}, 003 (2005)
  [arXiv:astro-ph/0503692].

\bibitem{preheatNG2}

  J.~R.~Bond, A.~V.~Frolov, Z.~Huang and L.~Kofman,
  ``Non-Gaussian Spikes from Chaotic Billiards in Inflation Preheating,''
  Phys.\ Rev.\ Lett.\  {\bf 103}, 071301 (2009)
  [arXiv:0903.3407 [astro-ph.CO]].

\bibitem{preheatNG3}

  C.~T.~Byrnes,
  ``Constraints on generating the primordial curvature perturbation and
  non-Gaussianity from instant preheating,''
  JCAP {\bf 0901}, 011 (2009)
  [arXiv:0810.3913 [astro-ph]].

\bibitem{preheatNG4}

  D.~Mulryne, D.~Seery and D.~Wesley,
  ``Non-Gaussianity constrains hybrid inflation,''
  arXiv:0911.3550 [astro-ph.CO].

\bibitem{NLNG}

  N.~Barnaby, T.~Biswas and J.~M.~Cline,
  ``p-adic inflation,''
  JHEP {\bf 0704}, 056 (2007)
  [arXiv:hep-th/0612230].

  N.~Barnaby and J.~M.~Cline,
  ``Large Nongaussianity from Nonlocal Inflation,''
  JCAP {\bf 0707}, 017 (2007)
  [arXiv:0704.3426 [hep-th]].

  N.~Barnaby and J.~M.~Cline,
  ``Predictions for Nongaussianity from Nonlocal Inflation,''
  JCAP {\bf 0806}, 030 (2008)
  [arXiv:0802.3218 [hep-th]].

\bibitem{curvatonNG}

  N.~Bartolo, S.~Matarrese and A.~Riotto,
  ``On non-Gaussianity in the curvaton scenario,''
  Phys.\ Rev.\  D {\bf 69}, 043503 (2004)
  [arXiv:hep-ph/0309033].

  K.~Enqvist and S.~Nurmi,
  ``Non-gaussianity in curvaton models with nearly quadratic potential,''
  JCAP {\bf 0510}, 013 (2005)
  [arXiv:astro-ph/0508573].

  K.~A.~Malik and D.~H.~Lyth,
  ``A numerical study of non-gaussianity in the curvaton scenario,''
  JCAP {\bf 0609}, 008 (2006)
  [arXiv:astro-ph/0604387].

  M.~Sasaki, J.~Valiviita and D.~Wands,
  ``Non-gaussianity of the primordial perturbation in the curvaton model,''
  Phys.\ Rev.\  D {\bf 74}, 103003 (2006)
  [arXiv:astro-ph/0607627].

\bibitem{turnNG}

  G.~I.~Rigopoulos, E.~P.~S.~Shellard and B.~J.~W.~van Tent,
  ``Large non-Gaussianity in multiple-field inflation,''
  Phys.\ Rev.\  D {\bf 73}, 083522 (2006)
  [arXiv:astro-ph/0506704].

  F.~Vernizzi and D.~Wands,
  ``Non-Gaussianities in two-field inflation,''
  JCAP {\bf 0605}, 019 (2006)
  [arXiv:astro-ph/0603799].

 C.~T.~Byrnes, K.~Y.~Choi and L.~M.~H.~Hall,
  ``Conditions for large non-Gaussianity in two-field slow-roll inflation,''
  JCAP {\bf 0810}, 008 (2008)
  [arXiv:0807.1101 [astro-ph]].

  C.~T.~Byrnes and G.~Tasinato,
  ``Non-Gaussianity beyond slow roll in multi-field inflation,''
  JCAP {\bf 0908}, 016 (2009)
  [arXiv:0906.0767 [astro-ph.CO]].

  X.~Chen and Y.~Wang,
  ``Quasi-Single Field Inflation and Non-Gaussianities,''
  arXiv:0911.3380 [hep-th].


\bibitem{small_sound}

  X.~Chen, M.~x.~Huang, S.~Kachru and G.~Shiu,
  ``Observational signatures and non-Gaussianities of general single field
  inflation,''
  JCAP {\bf 0701}, 002 (2007)
  [arXiv:hep-th/0605045].

\bibitem{DBI}

  E.~Silverstein and D.~Tong,
  ``Scalar Speed Limits and Cosmology: Acceleration from D-cceleration,''
  Phys.\ Rev.\  D {\bf 70}, 103505 (2004)
  [arXiv:hep-th/0310221].

 M.~Alishahiha, E.~Silverstein and D.~Tong,
  ``DBI in the sky,''
  Phys.\ Rev.\  D {\bf 70}, 123505 (2004)
  [arXiv:hep-th/0404084].

\bibitem{gelaton}

  A.~J.~Tolley and M.~Wyman,
  ``The Gelaton Scenario: Equilateral non-Gaussianity from multi-field
  dynamics,''
  arXiv:0910.1853 [hep-th].


\bibitem{nonBD1}

  R.~Holman and A.~J.~Tolley,
  ``Enhanced Non-Gaussianity from Excited Initial States,''
  JCAP {\bf 0805}, 001 (2008)
  [arXiv:0710.1302 [hep-th]].

\bibitem{nonBD2}

 P.~D.~Meerburg, J.~P.~van der Schaar and P.~S.~Corasaniti,
  ``Signatures of Initial State Modifications on Bispectrum Statistics,''
  JCAP {\bf 0905}, 018 (2009)
  [arXiv:0901.4044 [hep-th]].

\bibitem{nonBD3}

  P.~D.~Meerburg, J.~P.~van der Schaar and M.~G.~Jackson,
  ``Bispectrum signatures of a modified vacuum in single field inflation with a
  small speed of sound,''
  arXiv:0910.4986 [hep-th].

\bibitem{warm_NG}

  S.~Gupta, A.~Berera, A.~F.~Heavens and S.~Matarrese,
  ``Non-Gaussian signatures in the cosmic background radiation from warm
  inflation,''
  Phys.\ Rev.\  D {\bf 66}, 043510 (2002)
  [arXiv:astro-ph/0205152].


\bibitem{ng_rev}

  N.~Barnaby, ``Nongaussianity from Particle Production During Inflation,'' invited review for Advances in Astronomy.

\bibitem{inprog}

  N.~Barnaby, ``Nongaussian Signatures of Inflationary Particle Production,'' \emph{work in progress}.

\bibitem{HLattice}

  Z.~Huang, ``HLattice: a New Code for Simulating Cosmological Scalar Fields,'' \emph{work in progress}.

\bibitem{preheat_pdf1}

  G.~N.~Felder and L.~Kofman,
  ``The development of equilibrium after preheating,''
  Phys.\ Rev.\  D {\bf 63}, 103503 (2001)
  [arXiv:hep-ph/0011160].

\bibitem{preheat_pdf2}

 G.~N.~Felder and O.~Navros,
  ``Inflaton fragmentation after lambda phi**4 inflation,''
  JCAP {\bf 0702}, 014 (2007)
  [arXiv:hep-ph/0701128].

\bibitem{shandera}

  M.~LoVerde, A.~Miller, S.~Shandera and L.~Verde,
  ``Effects of Scale-Dependent Non-Gaussianity on Cosmological Structures,''
  JCAP {\bf 0804}, 014 (2008)
  [arXiv:0711.4126 [astro-ph]].

\bibitem{shellard2}

J.~R.~Fergusson, M.~Liguori and E.~P.~S.~Shellard,
``The CMB Bispectrum,''
arXiv:1006.1642 [astro-ph.CO].

\bibitem{structure}

 S.~Shandera,
  ``The structure of correlation functions in single field inflaiton,''
  Phys.\ Rev.\  D {\bf 79}, 123518 (2009)
  [arXiv:0812.0818 [astro-ph]].

\bibitem{perturbative}

  L.~Leblond and S.~Shandera,
  ``Simple Bounds from the Perturbative Regime of Inflation,''
  JCAP {\bf 0808}, 007 (2008)
  [arXiv:0802.2290 [hep-th]].

\bibitem{hidalgo}

  D.~Seery and J.~C.~Hidalgo,
  ``Non-Gaussian corrections to the probability distribution of the  curvature
  perturbation from inflation,''
  JCAP {\bf 0607}, 008 (2006)
  [arXiv:astro-ph/0604579].

\bibitem{WMAP7}

  E.~Komatsu {\it et al.},
  ``Seven-Year Wilkinson Microwave Anisotropy Probe (WMAP) Observations:
  Cosmological Interpretation,''
  arXiv:1001.4538 [astro-ph.CO].

\bibitem{neal}

  N.~Dalal, O.~Dore, D.~Huterer and A.~Shirokov,
  ``The imprints of primordial non-gaussianities on large-scale structure:
  scale dependent bias and abundance of virialized objects,''
  Phys.\ Rev.\  D {\bf 77}, 123514 (2008)
  [arXiv:0710.4560 [astro-ph]].

\bibitem{pat}

  P.~McDonald,
  ``Primordial non-Gaussianity: large-scale structure signature in the
  perturbative bias model,''
  Phys.\ Rev.\  D {\bf 78}, 123519 (2008)
  [arXiv:0806.1061 [astro-ph]].

\bibitem{andrew}

  N.~Afshordi and A.~J.~Tolley,
  ``Primordial non-gaussianity, statistics of collapsed objects, and the
  Integrated Sachs-Wolfe effect,''
  Phys.\ Rev.\  D {\bf 78}, 123507 (2008)
  [arXiv:0806.1046 [astro-ph]].

\bibitem{seljak}

  A.~Slosar, C.~Hirata, U.~Seljak, S.~Ho and N.~Padmanabhan,
  ``Constraints on local primordial non-Gaussianity from large scale
  structure,''
  JCAP {\bf 0808}, 031 (2008)
  [arXiv:0805.3580 [astro-ph]].

\bibitem{void}

  M.~Kamionkowski, L.~Verde and R.~Jimenez,
  ``The Void Abundance with Non-Gaussian Primordial Perturbations,''
  JCAP {\bf 0901}, 010 (2009)
  [arXiv:0809.0506 [astro-ph]].

\bibitem{lss_rev}

  V.~Desjacques and U.~Seljak,
  ``Primordial non-Gaussianity from the large scale structure,''
  Class.\ Quant.\ Grav.\  {\bf 27}, 124011 (2010)
  [arXiv:1003.5020 [astro-ph.CO]].

\bibitem{cluster}

  M.~J.~Jee {\it et al.},
  ``Hubble Space Telescope Weak-lensing Study of the Galaxy Cluster XMMU
  J2235.3-2557 at z=1.4: A Surprisingly Massive Galaxy Cluster when the
  Universe is One-third of its Current Age,''
  Astrophys.\ J.\  {\bf 704}, 672 (2009)
  [arXiv:0908.3897 [astro-ph.CO]].

\bibitem{lss2}

  R.~Jimenez and L.~Verde,
  ``Implications for Primordial Non-Gaussianity ($f_{NL}$) from weak lensing masses
  of high-z galaxy clusters,''
  Phys.\ Rev.\  D {\bf 80}, 127302 (2009)
  [arXiv:0909.0403 [astro-ph.CO]].

\bibitem{russian_text}

A.~A.~Grid, S.~G.~Mamayev and V.~M.~Mostepaneko, ``Vacuum Quantum Effects in Strong Fields,'' 
Friedmann Laboratory Publishing, St.~Petersburg (1994).


\bibitem{seery}

  D.~Seery, K.~A.~Malik and D.~H.~Lyth,
  ``Non-gaussianity of inflationary field perturbations from the field
  equation,''
  JCAP {\bf 0803}, 014 (2008)
  [arXiv:0802.0588 [astro-ph]].


\bibitem{babich}

  D.~Babich, P.~Creminelli and M.~Zaldarriaga,
  ``The shape of non-Gaussianities,''
  JCAP {\bf 0408}, 009 (2004)
  [arXiv:astro-ph/0405356].

\bibitem{malik1}

  K.~A.~Malik,
  ``A not so short note on the Klein-Gordon equation at second order,''
  JCAP {\bf 0703}, 004 (2007)
  [arXiv:astro-ph/0610864].

\bibitem{malik2}

  K.~A.~Malik,
  ``Gauge-invariant perturbations at second order: Multiple scalar fields  on
  large scales,''
  JCAP {\bf 0511}, 005 (2005)
  [arXiv:astro-ph/0506532].

\bibitem{SMvariable}

  M.~Sasaki,
  ``Large Scale Quantum Fluctuations in the Inflationary Universe,''
  Prog.\ Theor.\ Phys.\  {\bf 76}, 1036 (1986).

\bibitem{vernizzi}
 
  D.~Langlois and F.~Vernizzi,
  ``Nonlinear perturbations of cosmological scalar fields,''
  JCAP {\bf 0702}, 017 (2007)
  [arXiv:astro-ph/0610064].

\bibitem{loop_corrections}

  D.~Seery,
  ``One-loop corrections to a scalar field during inflation,''
  JCAP {\bf 0711}, 025 (2007)
  [arXiv:0707.3377 [astro-ph]].

  D.~Seery,
  ``One-loop corrections to the curvature perturbation from inflation,''
  JCAP {\bf 0802}, 006 (2008)
  [arXiv:0707.3378 [astro-ph]].

\bibitem{loops1}

  C.~P.~Burgess, L.~Leblond, R.~Holman and S.~Shandera,
  ``Super-Hubble de Sitter Fluctuations and the Dynamical RG,''
  JCAP {\bf 1003}, 033 (2010)
  [arXiv:0912.1608 [hep-th]].

  C.~P.~Burgess, R.~Holman, L.~Leblond and S.~Shandera,
  ``Breakdown of Semiclassical Methods in de Sitter Space,''
  arXiv:1005.3551 [hep-th].

\bibitem{loops2}

  L.~Senatore and M.~Zaldarriaga,
  ``On Loops in Inflation,''
  arXiv:0912.2734 [hep-th].

\bibitem{loops3}

  S.~B.~Giddings and M.~S.~Sloth,
  ``Semiclassical relations and IR effects in de Sitter and slow-roll
  space-times,''
  arXiv:1005.1056 [hep-th].

\bibitem{M&W}
 
  K.~A.~Malik and D.~Wands,
  ``Evolution of second order cosmological perturbations,''
  Class.\ Quant.\ Grav.\  {\bf 21}, L65 (2004)
  [arXiv:astro-ph/0307055].

\bibitem{L&R}

  D.~H.~Lyth and Y.~Rodriguez,
  ``Non-gaussianity from the second-order cosmological perturbation,''
  Phys.\ Rev.\  D {\bf 71}, 123508 (2005)
  [arXiv:astro-ph/0502578].

\bibitem{Sconserved}

  D.~H.~Lyth, K.~A.~Malik and M.~Sasaki,
  ``A general proof of the conservation of the curvature perturbation,''
  JCAP {\bf 0505}, 004 (2005)
  [arXiv:astro-ph/0411220].

\bibitem{riotto_rev}

   A.~Riotto,
  ``Inflation and the theory of cosmological perturbations,''
  arXiv:hep-ph/0210162.

\bibitem{liddle}

  S.~M.~Leach and A.~R.~Liddle,
  ``Inflationary perturbations near horizon crossing,''
  Phys.\ Rev.\  D {\bf 63}, 043508 (2001)
  [arXiv:astro-ph/0010082].

\bibitem{sasaki2}

  S.~M.~Leach, M.~Sasaki, D.~Wands and A.~R.~Liddle,
  ``Enhancement of superhorizon scale inflationary curvature perturbations,''
  Phys.\ Rev.\  D {\bf 64}, 023512 (2001)
  [arXiv:astro-ph/0101406].


\bibitem{jain2}

  R.~K.~Jain, P.~Chingangbam and L.~Sriramkumar,
  ``Amplification of tachyonic perturbations at super-Hubble scales,''
  JCAP {\bf 0710}, 003 (2007)
  [arXiv:astro-ph/0703762].


\end{thebibliography}
\end{document}